\documentclass[twocolumn,english,aps,prl]{revtex4-1}
\usepackage[latin9]{inputenc}

\usepackage{amsmath}
\usepackage{amssymb}
\usepackage{graphicx}

\makeatletter
\usepackage{babel}
\usepackage{color}

\definecolor{green}{rgb}{0,0.7,0.3}

\newcommand{\changelocaltocdepth}[1]{%
  \addtocontents{toc}{\protect\setcounter{tocdepth}{#1}}%
  \setcounter{tocdepth}{#1}%
}

\usepackage{babel}

\usepackage{babel}

\usepackage{babel}

\makeatother

\usepackage{babel}
\begin{document}

\title{Constructing exact representations of quantum\\
 many-body systems with deep neural networks}

\author{Giuseppe Carleo}

\affiliation{Center for Computational Quantum Physics, Flatiron Institute, 162
5th Avenue, New York, NY 10010, USA}

\affiliation{Institute for Theoretical Physics, ETH Zurich, Wolfgang-Pauli-Str. 27,
8093 Zurich, Switzerland}

\author{Yusuke Nomura}

\affiliation{Department of Applied Physics, The University of Tokyo, 7-3-1 Hongo,
Bunkyo-ku, Tokyo 113-8656, Japan}

\author{Masatoshi Imada}

\affiliation{Department of Applied Physics, The University of Tokyo, 7-3-1 Hongo,
Bunkyo-ku, Tokyo 113-8656, Japan}
\begin{abstract}
We develop a constructive approach to generate artificial neural networks
representing the exact ground states of a large class of many-body
lattice Hamiltonians. It is based on the deep Boltzmann
machine architecture, in which two layers of hidden neurons mediate
quantum correlations among physical degrees of freedom in the visible
layer. The approach reproduces the
exact imaginary-time Hamiltonian evolution, and is completely deterministic. 
In turn, compact and exact network representations for the ground states are obtained without 
stochastic optimization of the network parameters.
The number of neurons grows linearly with the system size and total
imaginary time, respectively. Physical quantities can be measured
by sampling configurations of both physical and neuron degrees of
freedom. 
We provide specific examples for
the transverse-field Ising and Heisenberg models by implementing efficient sampling. 
As a compact, classical representation for many-body
quantum systems, our approach is an alternative to the standard path integral, 
and it is potentially useful also to
systematically improve on numerical approaches based on the restricted Boltzmann machine architecture. 
\end{abstract}
\maketitle

\section*{Introduction}

A tremendous amount of successful developments in quantum physics
builds upon the mapping between many-body quantum systems and effective
classical theories. The probably most well known mapping is due to
Feynman, who introduced an exact representation of many-body quantum
systems in terms of statistical summations over classical particles
trajectories \cite{feynman1948spacetime}. Effective classical representations
of quantum many-body systems are however not unique, and other approaches
rely on different inspiring principles, such as perturbative expansions
\cite{dyson1949thes}, or decomposition of interactions with auxiliary
degrees of freedom \cite{hubbard1959calculation,stratonovich1957ona}.
The classical representations of quantum states allow both for novel
conceptual developments and efficient numerical simulations. On one
hand, perturbative approaches based on the graphical resummation of
classes of diagrams are at the heart of many-body analytical approaches
in various fields of research, ranging from particle to condensed-matter
physics \cite{abrikosov1975methods}. On the other hand, several non-perturbative
numerical methods for many-body quantum systems are also based on
these mappings. Quantum Monte Carlo (QMC) methods are among the most
successful numerical techniques, relying on continuos-space polymer
representations~\cite{binder,takahashi1984,takahashi1984-2,ceperley1995pathintegrals},
world-line lattice path integrals~\cite{suzuki,hirsch}, continuous
time algorithm~\cite{wiese1996}, summation of perturbative diagrams~\cite{sandvik1999stochastic,prokofev2007bolddiagrammatic}.
Effective classical representations are also the building block of
variational methods based on correlated many-body wave-functions~\cite{feynman}.
Several successful variational techniques make extensive use of parametric
representations of quantum states, where the effective parameters
are determined by means of the variational principle~\cite{gros,kashima2001,tahara,becca2017quantum}.
In matrix-product and tensor-network-states the ground-state is expressed
as a classical network~\cite{white,orus}. In general, finding alternative,
efficient classical representations of quantum states can help establishing
novel numerical and analytical techniques to study challenging open
issues.

Recently, an efficient variational representation of many-body systems
in terms of artificial neural networks, which consists of classical
degrees of freedom, has been introduced \cite{carleo2017solving}.
Numerical results have shown that artificial neural networks can represent
many-body states with high accuracy \cite{carleo2017solving,torlai2017manybody,nomura2017restrictedboltzmannmachine,deng2017quantum,rocchetto2017learning,glasser2017neuralnetworks,kaubruegger2017chiraltopological,cai2017approximating,saito2017machine,saito2017solving}.
The majority of the variational approaches adopted so-far are based
on shallow neural networks, called Restricted Boltzmann Machines (RBM),
in which the physical degrees of freedom interact with an ensemble
of hidden degrees of freedom (neurons). While shallow RBM states have
promising features in terms of entanglement capacity \cite{chen2018equivalence,clark2017unifying,deng2017quantum,deng2017machine},
only deep networks are guaranteed to provide a complete and efficient
description of the most general quantum states \cite{gao_efficient_2017,huang2017neuralnetwork}.

In this Paper we introduce a constructive approach to explicitly generate
deep network structures corresponding to exact quantum many-body ground
states. We demonstrate this construction for interacting lattice spin
models, including the transverse-field Ising and Heisenberg models.
Our constructions are fully deterministic, in stark contrast to the
shallow RBM case, in which the numerical optimization of the network
parameters is inevitable. The number of neurons required in the construction
scales only polynomially with the system size, thus the present approach
constitutes a new family of efficient quantum-to-classical mappings
exhibiting a prominent representational flexibility. Given as a simple
set of iterative rules, these constructions can be used both as a
self-standing tool, or to systematically improve results obtained
with variational shallow networks. The latter improves the efficiency
of the method because the numerically optimized shallow RBM states
are already good approximations for ground states. Finally, we discuss
sampling strategies from the generated deep networks and show numerical
results for one-dimensional spin models.

\begin{figure}
\noindent \includegraphics[width=0.9\columnwidth]{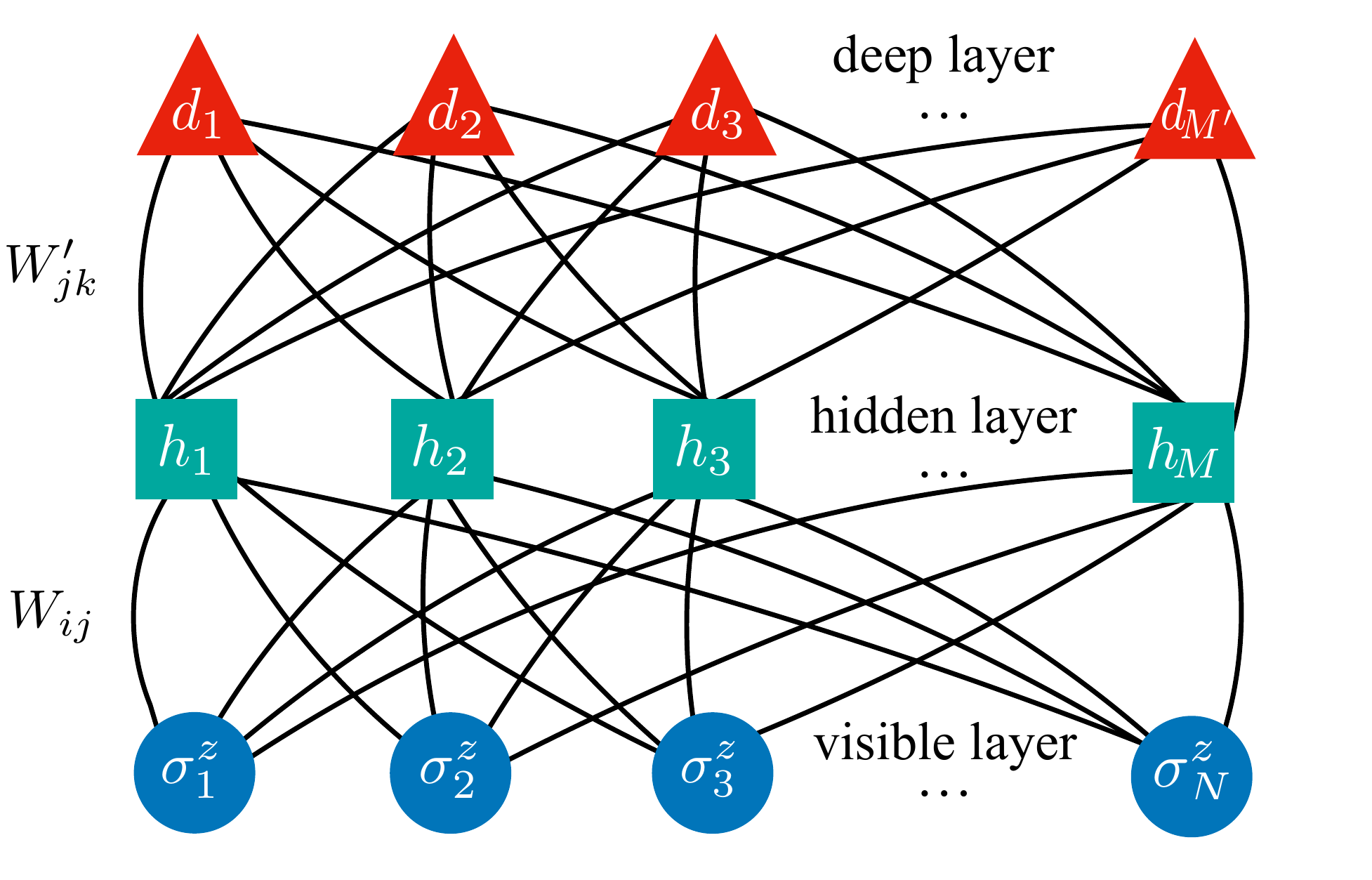} \caption{\label{fig:DBM} \textbf{Structure of deep Boltzmann machine}. Dots,
squares, and triangles represent physical degrees of freedom ($\sigma_{i}^{z}$),
hidden units ($h_{j}$), deep units ($d_{k}$), respectively. Solid
curves represent interlayer couplings ($W_{ij}$ and $W'_{jk}$). }
\end{figure}

\section*{Construction of Deep neural states}

The ground state of a generic Hamiltonian, $\mathcal{H}$, can be
found through imaginary-time evolution, $|\Psi(\tau)\rangle=e^{-\tau\mathcal{H}}|\Psi_{0}\rangle,$
for a sufficiently large $\tau\gg\Delta E^{-1}$. Here $\Delta E$
is the energy gap between the ground and the first excited state,
and $|\Psi_{0}\rangle$ is an arbitrary initial state non-orthogonal
to the exact ground state. For a finite system, the energy gap is
typically finite, and the total propagation time needed to reach the
ground state within an arbitrary given accuracy is expected to grow
at most polynomially with the system size (for systems becoming gapless
in the thermodynamic limit). 

Here, we introduce a representation of
the wave-function coefficients in terms of a deep Boltzmann machine
(DBM) \cite{salakhutdinov2009deepboltzmann}. For the sake of concreteness,
let us consider the case of $N$ spins, described by the quantum numbers
$|\sigma^{z}\rangle=|\sigma_{1}^{z}\dots\sigma_{N}^{z}\rangle$. Then,
we represent generic many-body amplitudes $\langle\sigma_{1}^{z}\dots\sigma_{N}^{z}|\Psi\rangle\equiv\Psi(\sigma^{z})$
in the two-layer DBM form: 
\begin{multline}
\Psi_{\mathcal{W}}(\sigma^{z})=\sum_{\{h,d\}}\exp\left[\sum_{i}a_{i}\sigma_{i}^{z}+\sum_{ij}\sigma_{i}^{z}W_{ij}h_{j}+\right.\\
\left.+\sum_{j}b_{j}h_{j}+\sum_{jk}h_{j}d_{k}W_{jk}^{\prime}+\sum_{k}b_{k}^{\prime}d_{k}\right]\label{eq:psidbm}
\end{multline}
where we have introduced $M$ hidden units $h$, $M^{\prime}$ deep
units $d$, and a set of couplings and bias terms $\mathcal{W}\equiv(a,b,b^{\prime},W,W^{\prime})$.
A sketch of the DBM architecture is shown in Fig. \ref{fig:DBM}.

In the following, we specialize to the case of spin $1/2$, thus all
the units are taken to be $\sigma^{z},h,d=\pm1$. This representation
is the natural deep-network generalization of the shallow RBM, introduced
as variational ansatz in Ref. \cite{carleo2017solving}. As for the
RBM form, also in this case direct connections between variables in
the same layer are not allowed. A crucial difference is however that
the layer of deep variables makes, in general, the evaluation of the
wave-function amplitudes not possible analytically. At variance with
RBM, the DBM form is known to be universal, as proven by Gao and Duan
recently \cite{gao_efficient_2017}. In order to find explicit expressions
for the parameters $\mathcal{W}$ that represent $|\Psi(\tau)\rangle$
for arbitrary imaginary time, we start considering a second-order
Trotter-Suzuki decomposition~\cite{trotter,suzuki}: 
\begin{multline}
|\Psi({\tau})\rangle=\mathcal{G}_{1}(\delta_{\tau}/2)\mathcal{G}_{2}(\delta_{\tau})\dots\mathcal{G}_{1}(\delta_{\tau})\mathcal{G}_{2}(\delta_{\tau})\mathcal{G}_{1}(\delta_{\tau}/2)|\Psi_{0}\rangle,\label{eq:psitau}
\end{multline}
where we have decomposed the Hamiltonian into two non-commuting parts,
$\mathcal{H=\mathcal{H}}_{1}+\mathcal{H}_{2}$, and introduced the
short-time propagators $\mathcal{G}_{\nu}(\delta_{\tau})=e^{-\mathcal{H}_{\nu}\delta_{\tau}}$.
The problem of finding an exact representation for $|\Psi(\tau)\rangle$
then reduces to finding an exact representation for each of the two
type of propagators. As shown in the following concrete examples for
paradigmatic spin models, thanks to the high representability of DBM,
the imaginary time evolution can be tracked exactly by dynamically
modifying the DBM network structure. In practice, this is achieved
either by changing parameters $\mathcal{W}$ at each step of the imaginary
time evolution, or by introducing additional parameters in $\mathcal{W}$,
adding new neurons and creating new connections in the network.

\begin{figure}
\noindent \includegraphics[width=1\columnwidth]{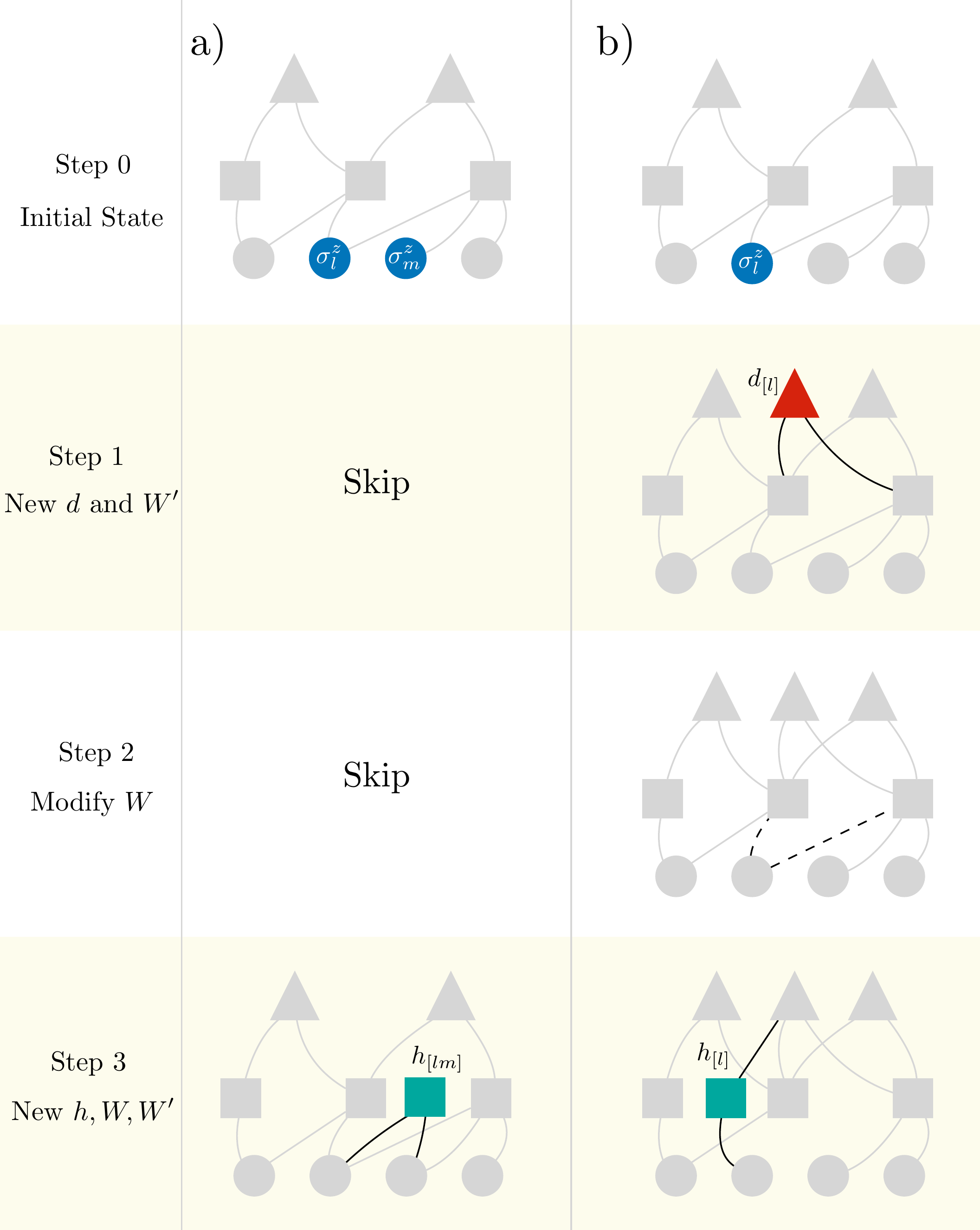}
\caption{\label{fig:Construction-of-exact} \textbf{Construction of exact DBM
representations of transverse-field Ising model}. In this example,
a step of imaginary-time evolution is shown, for the case of the 1-dimensional
transverse-field Ising model. Dots represent physical degrees of freedom
($\sigma_{i}^{z}$), squares represent hidden units ($h_{j}$), triangles
represent deep units ($d_{k}$). In each panel, upper networks are
the initial state with arbitrary network form, and the bottom networks
are the final states, after application of the propagator. Intermediate
steps illustrate how the network is modified, where the relevant modified
couplings at each step are highlighted in black. The highlighted solid
and dashed curves indicate new and vanishing couplings, respectively.
(a) Shows the diagonal (interaction) propagator being applied to the
highlighted blue spins. This introduces a hidden unit (green) connected
only to the two physical spins. In (b) the off-diagonal (transverse-field)
propagator is applied, acting on the blue physical spin. Here, we
then add one deep unit (red triangle), and a hidden unit (green) mediating
visible-deep interactions. }
\end{figure}

\subsection*{Transverse-Field Ising model}

We start considering the transverse-field Ising (TFI) model on an
arbitrary interaction graph. In this case, we decompose the Hamiltonian
into two parts: $\mathcal{H}_{1}=-\sum_{l}\Gamma_{l}\sigma_{l}^{x}$,
and $\mathcal{H}_{2}=\sum_{l<m}V_{lm}\sigma_{l}^{z}\sigma_{m}^{z}$,
where $\sigma$ denote Pauli matrices, $\Gamma_{l}$ ($>0$) are site-dependent
transverse fields, and $V_{lm}$ are arbitrary coupling constants.

In order to implement the mapping to a DBM, we first consider the
action of the diagonal propagator $e^{-\delta_{\tau}V_{lm}\sigma_{l}^{z}\sigma_{m}^{z}}$,
acting on a bond $V_{lm}$. In this case, the goal of finding an exact
DBM representation can be rephrased as finding solutions to 
\begin{eqnarray}
\langle\sigma^{z}|e^{-\delta_{\tau}V_{lm}\sigma_{l}^{z}\sigma_{m}^{z}}|\Psi_{\mathcal{W}}\rangle & = & C\Psi_{\bar{\mathcal{W}}}(\sigma^{z}),\label{eq:diagpropIsing}
\end{eqnarray}
i.e. finding a set of new parameters $\bar{\mathcal{W}}$ that exactly
reproduces the imaginary time evolution on the left hand side. Here
$C$ is an arbitrary finite normalization constant. The diagonal propagator
introduces an interaction between two visible, physical spins, which
is not directly available in the DBM architecture. This interaction
can be mediated by a new hidden unit in the first layer, $h_{[lm]}$
which is only connected to the visible spins on that bond, i.e. $\bar{W}_{l[lm]}$
and $\bar{W}_{m[lm]}$ are finite, but $\bar{W}_{i[lm]}=0,\forall i\neq l,m$
and $\bar{W}_{j[lm]}^{\prime}=0,\forall j$ {[}see Fig. \ref{fig:Construction-of-exact}(a){]}.

More concretely, the new wave function has then the form: 
\begin{eqnarray}
\Psi_{\bar{\mathcal{W}}}(\sigma^{z}) & = & \sum_{h_{[lm]}}e^{\sigma_{l}^{z}W_{l[lm]}h_{[lm]}+\sigma_{m}^{z}W_{m[lm]}h_{[lm]}}\Psi_{\mathcal{W}}(\sigma^{z})\nonumber \\
 & = & 2\cosh\left(\sigma_{l}^{z}W_{l[lm]}+\sigma_{m}^{z}W_{m[lm]}\right)\Psi_{\mathcal{W}}(\sigma^{z}).
\end{eqnarray}
Equation (\ref{eq:diagpropIsing}) is then satisfied if 
\begin{eqnarray}
e^{-\delta_{\tau}V_{lm}\sigma_{l}^{z}\sigma_{m}^{z}}=2C\cosh\left(\sigma_{l}^{z}W_{l[lm]}+\sigma_{m}^{z}W_{m[lm]}\right)\label{eq:sigmasigmarenorm}
\end{eqnarray}
for all the possible values of $\sigma_{l}^{z}$ and $\sigma_{m}^{z}$.
By means of a useful identity {[}Eq.~(\ref{W''1_methods}) in Methods{]},
the new parameters $W_{l[lm]}$ and $W_{m[lm]}$ are given by 
\begin{eqnarray}
W_{l[lm]} & = & \frac{1}{2}\mathrm{arcosh}\left(e^{2|V_{lm}|\delta_{\tau}}\right)\\
W_{m[lm]} & = & -\mathrm{sgn}(V_{lm})\times W_{l[lm]}.
\end{eqnarray}
In this way the classical two-body interaction can, in general, be
represented exactly by the shallow RBM.

Next, to exactly represent the off-diagonal propagator $e^{\delta_{\tau}\Gamma_{l}\sigma_{l}^{x}}|\Psi_{\mathcal{W}}\rangle$,
we must solve: 
\begin{multline}
\cosh(\Gamma_{l}\delta_{\tau})\Psi_{\mathcal{W}}(\sigma^{z})+\sinh(\Gamma_{l}\delta_{\tau})\Psi_{\mathcal{W}}(\sigma_{l}^{z}\rightarrow-\sigma_{l}^{z})=\\
=C\Psi_{\bar{\mathcal{W}}}(\sigma^{z})\label{eq:sigmaxprop}
\end{multline}
for the new weights $\bar{\mathcal{W}}$, and for an appropriate finite
normalization constant $C$. In this case, one possible solution is
obtained by adding one deep $d_{[l]}$ and one hidden $h_{[l]}$ neurons.
For $d_{[l]}$, we create new couplings $W_{j[l]}^{\prime}$ to the
existing hidden neurons $h_{j}$ which are connected to $\sigma_{l}^{z}$.
We simultaneously allow for changes in the existing parameters. By
the procedure given in Methods, after applying the off-diagonal propagator
for the site $l$, a solution of Eq.(\ref{eq:sigmaxprop}) is found
by the matching condition of the hidden unit interactions on the left
and the right hand sides of Eq.(\ref{eq:sigmaxprop}). Overall, the
solution results in a three-step process {[}Fig. \ref{fig:Construction-of-exact}(b){]}:
First, the hidden units attached to $\sigma_{l}^{z}$ are connected
to the newly introduced deep unit $d_{[l]}$ as 
\begin{eqnarray}
W_{j[l]}^{\prime} & = & -W_{lj}\label{eq:wprimex}
\end{eqnarray}
(see Eq.(\ref{eq:DeltaW1})). Second, all the hidden units previously
connected to the spin $\sigma_{l}^{z}$ lose their connection, i.e.,
$\bar{W}_{lj}=0,\forall j$. Third, the spin $\sigma_{l}^{z}$ and
the deep unit $d_{[l]}$ are connected to the new hidden unit, $h_{[l]}$,
through the interaction$W_{l[l]}$ and $W'_{[l][l]}$, respectively
as 
\begin{eqnarray}
W_{l[l]} & = & \frac{1}{2}\mathrm{arcosh}\left(\frac{1}{\tanh(\Gamma_{l}\delta_{\tau})}\right),\label{eq:wl=00003D00003D00003D00003D00003D00005Bx=00003D00003D00003D00003D00003D00005D}\\
W_{[l][l]}^{\prime} & = & -W_{l[l]}.\label{eq:W'll}
\end{eqnarray}

Using the given expressions for the parameters $\bar{\mathcal{W}}$
we can then exactly implement a single step of imaginary-time evolution.
The full imaginary-time evolution is achieved by applying the above
procedure for $\mathcal{H}_{1}$ and $\mathcal{H}_{2}$ alternately
and repeatedly. Example applications of these rules, for both the
diagonal and the off-diagonal propagators are shown in Fig. \ref{fig:Construction-of-exact}.
\begin{figure*}[t]
\noindent \includegraphics[width=2\columnwidth]{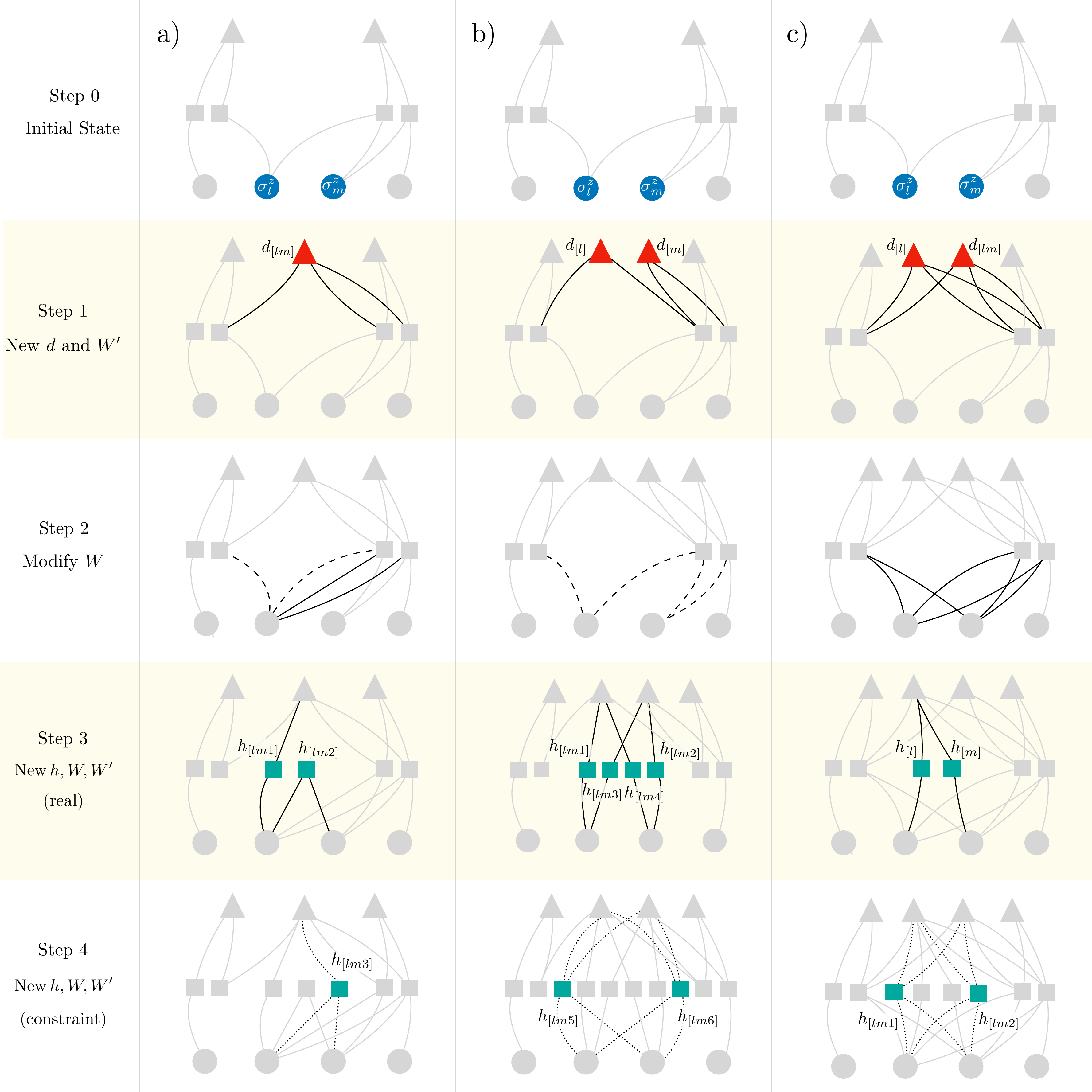}
\caption{\label{fig:Construction-Heis} \textbf{Construction of exact DBM representations
of Heisenberg models}. In this example, a time step of imaginary-time
evolution is shown, for the case of the 1-dimensional antiferromagnetic
Heisenberg model. Dots represent physical degrees of freedom ($\sigma_{i}^{z}$),
squares represent hidden units ($h_{j}$), triangles represent deep
units ($d_{k}$). The three panels (a,b,c) represent different possible
explicit constructions. In each panel, upper networks are the initial
state with arbitrary network form, and the bottom networks are the
final states, after application of the propagator. Intermediate steps
illustrate how the network is modified, where the relevant modified
weights at each step are highlighted in black. In those diagrams,
dashed lines indicate that the corresponding weights are set to zero,
and dotted lines indicate complex-valued weights. The three panels
correspond to the (a) \emph{1 deep, 3 hidden} (1d-3h), (b) \emph{2
deep, 6 hidden} (2d-6h), and (c) \emph{2 deep, 4 hidden} (2d-4h) constructions
(see text for a more detailed explanation of the individual steps
characteristic of each construction). }
\end{figure*}

\subsection*{Heisenberg model}

We now consider the anti-ferromagnetic Heisenberg (AFH) model, on
bipartite lattices. In one dimension, we decompose the Hamiltonian
into odd and even bonds: $\mathcal{H}_{1}=\sum_{\langle l,m\rangle}^{\mathrm{odd}}\mathcal{H}_{lm}^{\mathrm{bond}}$
and $\mathcal{H}_{2}=\sum_{\langle l,m\rangle}^{\mathrm{even}}\mathcal{H}_{lm}^{\mathrm{bond}}$,
with $\mathcal{H}_{lm}^{\mathrm{bond}}=J\left(\sigma_{l}^{x}\sigma_{m}^{x}+\sigma_{l}^{y}\sigma_{m}^{y}+\sigma_{l}^{z}\sigma_{m}^{z}\right)$,
where $\sigma$ denote Pauli matrices. Because the bond Hamiltonian
$\mathcal{H}_{lm}^{\mathrm{bond}}$ is a building block also in higher
dimensional models, construction of an exact DBM representation of
the ground states can be achieved by finding solutions for the bond-propagator
$\langle\sigma^{z}|e^{-\delta_{\tau}{\mathcal{H}}_{lm}^{\mathrm{bond}}}|\Psi_{\mathcal{W}}\rangle=C\langle\sigma^{z}|\Psi_{\bar{\mathcal{W}}}\rangle,$
where the parameters $\bar{\mathcal{W}}$ are such that the previous
equation is satisfied for all the possible $\langle\sigma^{z}|$,
and for an arbitrary finite normalization constant $C$. More explicitly,
we need to satisfy 
\begin{multline}
\delta_{\sigma_{l}^{z},\sigma_{m}^{z}}e^{-J\delta_{\tau}}\Psi_{\mathcal{W}}(\sigma^{z})+(1-\delta_{\sigma_{l}^{z},\sigma_{m}^{z}})e^{J\delta_{\tau}}\cosh(2J\delta_{\tau})\times\\
\left(\Psi_{\mathcal{W}}(\sigma^{z})-\tanh(2J\delta_{\tau})\Psi_{\mathcal{W}}(\sigma_{l}^{z}\leftrightarrow\sigma_{m}^{z})\right)=C\Psi_{\mathcal{\bar{W}}}(\sigma^{z}).\label{eq:Heisenberg0}
\end{multline}
The basic strategy of finding a solution for Eq.(\ref{eq:Heisenberg0})
is similar to that for Eq.(\ref{eq:sigmaxprop}) in the transverse
Ising model. Several possibilities arise when looking for solutions
of the bond-propagator equation, Eq. (\ref{eq:Heisenberg0}). The
existence of non-equivalent solutions prominently shows the non-uniqueness
of DBM structure to represent the very same state and, at the same
time, provides us flexibility in designing DBM architectures. Here,
we show three concrete constructions. See Methods and Supplementary
Information (II-B) for a detailed derivation of the DBM construction
for the Heisenberg model, including anisotropic and bond-disordered
coupling cases.

\subsubsection*{1 deep, 3 hidden}

The first construction is dubbed ``\emph{1 deep, 3 hidden}'' (1d-3h).
It amounts to adding an extra deep neuron, $d_{[lm]}$, and three
more hidden neurons to satisfy Eq. (\ref{eq:Heisenberg0}). A crucial
difference with respect to the TFI model is that the introduced deep
spin $d_{[lm]}$ has a constraint depending on the state of the spins
on the bond: $\sigma_{l}^{z}$ and $\sigma_{m}^{z}$. Specifically,
when $\sigma_{l}^{z}=\sigma_{m}^{z}$ the deep spin is constrained
to be $d_{[lm]}=\sigma_{l}^{z}=\sigma_{m}^{z}$, whereas when $\sigma_{l}^{z}\neq\sigma_{m}^{z}$,
its value is unconstrained. From a pictorial point of view, the action
of the bond propagator is a four-step process {[}see Fig. \ref{fig:Construction-Heis}(a){]}.
Starting from a given initial network (uppermost structures in Fig.
\ref{fig:Construction-Heis}), $d_{[lm]}$ is added and connected,
through $W'_{j[lm]}$ given in Eq. (\ref{1d3h_wp_jlm}), to the existing
hidden units $h_{j}$ connected to $\sigma_{l}^{z}$ and $\sigma_{m}^{z}$.
Second, spin $\sigma_{l}^{z}$ is disconnected to all hidden units
and reconnected to those hidden units the spin $\sigma_{m}^{z}$ is
attached to {[}see Eq. (\ref{1d3h_dw}){]}. Third, two new hidden
units are introduced. One of the hidden units, $h_{[lm1]}$, mediates
the interaction between $\sigma_{l}^{z}$ and $d_{[lm]}$ {[}Eq. (\ref{1d3h_wwp_lm1}){]},
and the other hidden unit $h_{[lm2]}$ mediates a direct spin-spin
interaction between $\sigma_{l}^{z}$ and $\sigma_{m}^{z}$ {[}Eq.
(\ref{1d3h_wwp_lm2}){]}. Fourth, a further hidden unit connected
to $\sigma_{l}^{z}$, $\sigma_{m}^{z}$ and $d_{[lm]}$ is inserted,
in such a way that the constraint previously described is satisfied.
For all but the last step, the DBM weights are real-valued. In the
last step instead the constraint is enforced by introducing imaginary-valued
interactions (dotted lines in Fig.~\ref{fig:Construction-Heis}),
referred to the ``$i\pi/6$\char`\"{} trick, resulting in a sign-problem
free global term $\cos(\pi/6(\sigma_{l}^{z}+\sigma_{m}^{z}-d_{[lm]}))$
after the summation over $\pm1$ for the lastly added hidden unit
$h_{[lm3]}$: $\sum_{h_{[lm3]}=\pm1}\exp[i\pi/6(\sigma_{l}^{z}+\sigma_{m}^{z}-d_{[lm]})h_{[lm3]}]$.
The constraint mentioned above is assured by this cosine term. 

\subsubsection*{2 deep, 6 hidden}

The second construction is dubbed ``\emph{2 deep, 6 hidden''} (2d-6h),
and is more similar to the lattice path-integral formulation. In this
representation, we introduce two auxiliary deep spins per bond, $d_{[l]}$
and $d_{[m]}$ with constraint $d_{[l]}+d_{[m]}=\sigma_{l}^{z}+\sigma_{m}^{z}$,
and six hidden neurons. The action of the bond propagator is schematically
illustrated in Fig. \ref{fig:Construction-Heis}(b): first, two deep
units $d_{[l]}$ and $d_{[m]}$ are introduced, connecting, respectively,
to the hidden units spins $\sigma_{l}^{z}$ and $\sigma_{m}^{z}$
are attached to {[}see Eqs. (\ref{eq:2d6h_wprimex1}), (\ref{eq:2d6h_wprimex2}){]}.
Second, all the connections between spins $\sigma_{l}^{z}$, $\sigma_{m}^{z}$
and hidden units $h_{j}$ are cut off {[}Eqs. (\ref{Eq:2d6h_wlj_change}),
(\ref{Eq:2d6h_wmj_change}){]}. Third, four hidden units $h_{[lm1]},\ldots,h_{[lm4]}$
are introduced, to mediate interactions between the two deep units
and the physical spins $l,m$ {[}Eqs. (\ref{Eq.2d6h_wwp_lm1_lm2}),
(\ref{Eq.2d6h_wwp_lm3_lm4}){]}. Finally, two hidden units $h_{[lm5]}$
and $h_{[lm6]}$ are introduced, connecting both to $d_{[l]},d_{[m]}$
and $\sigma_{l}^{z},\sigma_{m}^{z}$ with imaginary-valued weights.
The last step realizes the constraint $d_{[l]}+d_{[m]}=\sigma_{l}^{z}+\sigma_{m}^{z}$,
through the ``$i\pi/4,\ i\pi/8$\char`\"{} trick discussed in Methods
and Supplementary Information (II-B.2). 

In this representation, if
the hidden neurons are traced out, the imaginary-time evolution becomes
equivalent to that of the path-integral Monte Carlo method. More specifically,
the number of deep neurons introduced at each time slice is exactly
the same as the number of visible spins, and the deep neurons at each
time slice can be regarded as additional classical spin degrees of
freedom in the path-integral. Moreover, the constraint $d_{[l]}+d_{[m]}=\sigma_{l}^{z}+\sigma_{m}^{z}$
ensures that the total magnetization is conserved at each time slice.
Finally, the $W$ and $W'$ interactions reproduce the matrix element
of $\exp(-\delta_{\tau}{\mathcal{H}}_{lm}^{{\rm bond}})$ between
neighboring time slices. See Supplementary Information (II-B.2) for
more detail on this point.

\subsubsection*{2 deep, 4 hidden}

A further possible solution to Eq. (\ref{eq:Heisenberg0}) is dubbed
``\emph{2 deep, 4 hidden}'' (2d-4h) construction. In this case,
we introduce two auxiliary deep variables $d_{[l]}$ and $d_{[lm]}$.
We also introduce four hidden units $h_{[l]}$, $h_{[m]}$, $h_{[lm1]}$,
and $h_{[lm2]}$. Before the imaginary time evolution, $e^{-\delta_{\tau}\mathcal{H}_{lm}^{{\rm bond}}}$,
the physical variables $\sigma_{n}^{z}$ ($n=l$ or $m$) are already
coupled to each hidden variable $h_{j}$ with a coupling $W_{nj}$.
After the time evolution $e^{-\delta_{\tau}\mathcal{H}_{lm}^{{\rm bond}}}$,
as shown schematically in Fig. \ref{fig:Construction-Heis}(c), the
coupling parameters are updated in the following way based on the
old $W_{nj}$: First, the first deep unit $d_{[l]}$ becomes coupled
to the already existing hidden variables $h_{j}$ through the coupling
$W'_{j[l]}$ given in Eq. (\ref{bWZ4}). The second deep unit $d_{[lm]}$
becomes similarly coupled to $h_{j}$ through a term $Z_{lmj}$ given
in Eq. (\ref{bWZ4}). Second, $W_{nj}$ is updated to $\bar{W}_{nj}=W_{nj}+\Delta W_{nj}$
{[}see Eq. (\ref{bWZ3}){]}. Third, newly introduced $h_{[n]}$ ($n=l$
or $m$) gets coupled to $d_{[l]}$ through $W'_{[n][l]}$, and also
to $\sigma_{n}^{z}$ through $W_{n[n]}$ {[}Eqs. (\ref{Wl52}), (\ref{Wl53}){]}.

Within this construction, and as clarified in Methods, we also need to satisfy the constraint
$d_{[l]}d_{[lm]}=\sigma_{l}^{z}\sigma_{m}^{z}.$ Such a constraint
is represented in DBM form as 
\begin{multline}
\sum_{h_{[lm1]},h_{[lm2]}}\exp[\frac{i\pi}{4}(h_{[lm1]}+h_{[lm2]})(\sigma_{l}^{z}+\sigma_{m}^{z}+d_{[l]}+d_{[lm]})],
\end{multline}
which ensures $d_{[l]}d_{[lm]}=\sigma_{l}^{z}\sigma_{m}^{z}$ after
explicit summation of $h_{[lm1]}$ and $h_{[lm2]}$. 

Finally, we remark that the three constructions presented here have different intrinsic
network topologies. In particular, 2d-6h gives rise to a local topology
(because of the equivalence with the path-integral contruction), 1d-3h
has a local structure in the first layer and non-local in the second
one, and 2d-4h is purely non-local in both layers (see Supplementary Information II.B). 
\begin{figure*}[t]
\noindent \includegraphics[width=2\columnwidth]{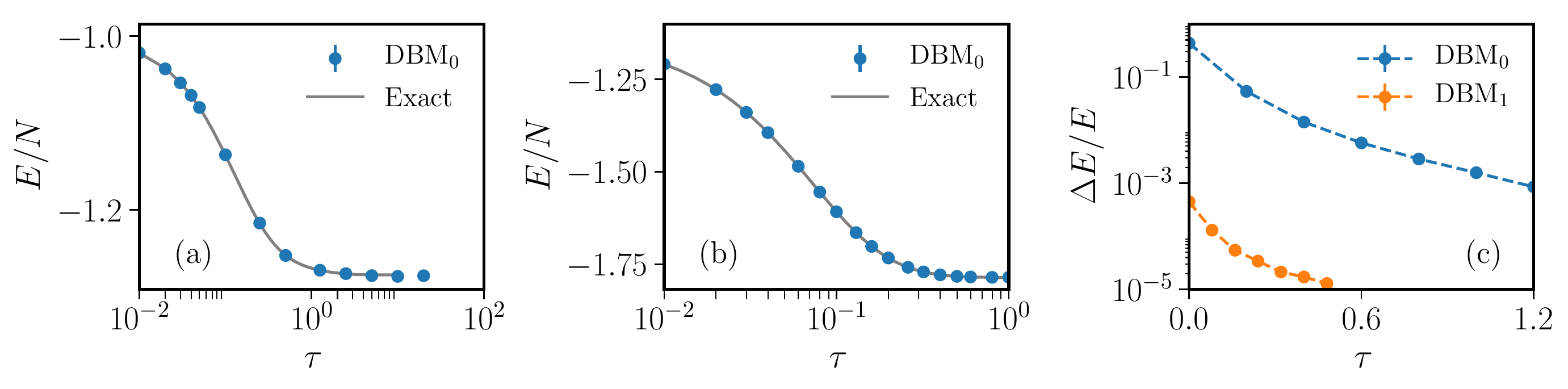}
\caption{\label{fig:results}\textbf{Imaginary-time evolution with a DBM for
1D spin models}. (a) Expectation value of energy of the transverse-field
Ising Hamiltonian in the exact imaginary-time evolution (continuous
line) is compared to the stochastic result obtained with a DBM ($J\delta_{\tau}=0.01$).
We consider the critical point ($\Gamma_{l}=1$), periodic boundary
conditions, and $N=20$ sites. (b) Expectation value of the isotropic
antiferromagnetic Heisenberg Hamiltonian (AFHM) in the exact imaginary-time
evolution (continuous line) is compared to the stochastic result obtained
with a DBM ($J\delta_{\tau}=0.01$) following 2d-6h construction.
We consider periodic boundary conditions, $N=16$ sites. (c) Relative
error on the ground-state energy for the 1D AFHM as a function of
the imaginary time. Here we consider periodic boundary conditions,
$N=80$ sites, and $J\delta_{\tau}=0.01$. The subscript $\alpha$
in $\mathrm{DBM}_{\alpha}$ in panels (a,b,c) specifies a different
initial state $|\Psi_{0}\rangle$: $\alpha=1$ means that the initial
state is an RBM state with hidden-unit density $M/N=1$, whereas when
$\alpha=0$ the initial state is the empty-network state ($M=0$). }
\end{figure*}

\section*{Sampling strategies}

With network structures explicitly determined, we now focus on the
problem of extracting meaningful physical quantities from them. To
this end, it is convenient to decompose the DBM weight into two parts,
such that 
\begin{equation}
\Psi_{\mathcal{W}}(\sigma^{z})=\sum_{\{h,d\}}P_{1}(\sigma^{z},h)P_{2}(h,d),\label{eq:PsiP1P2}
\end{equation}
where $P_{1}(\sigma^{z},h)=e^{\sigma^{z}\cdot a+\sigma^{z}\cdot W\cdot h+h\cdot b}$,
and $P_{2}(h,d)=e^{h\cdot W^{\prime}\cdot d+d\cdot b^{\prime}}.$
The expectation value of an arbitrary (few-body) operator $\mathcal{O}$
can then be computed through the expression 
\begin{eqnarray}
\langle\mathcal{O}\rangle & = & \frac{\sum_{\{\sigma^{z},h,h^{\prime}d,d^{\prime}\}}\Pi(\sigma^{z},h,h^{\prime},d,d^{\prime})O_{\mathrm{loc}}(\sigma^{z},h,h^{\prime})}{\sum_{\{\sigma^{z},h,h^{\prime}d,d^{\prime}\}}\Pi(\sigma^{z},h,h^{\prime},d,d^{\prime})},\label{eq:expdbm}
\end{eqnarray}
where we have introduced the pseudo-probability density $\Pi(\sigma^{z}\!,h,h^{\prime},d,d^{\prime})\equiv P_{1}(\sigma^{z}\!,h)P_{2}(h,d)P_{1}^{\star}(\sigma^{z}\!,h^{\prime})P_{2}^{\star}(h^{\prime},d^{\prime})$,
and the ``local'' estimator $O_{\mathrm{loc}}(\sigma^{z},h,h^{\prime})=\frac{1}{2}\sum_{\sigma^{\prime z}}\left\langle \sigma^{z}\right|\mathcal{O}\left|\sigma^{\prime z}\right\rangle \left(\frac{P_{1}(\sigma^{\prime z},h)}{P_{1}(\sigma^{z},h)}+\frac{P_{1}(\sigma^{\prime z},h^{\prime})^{\star}}{P_{1}(\sigma^{z},h^{\prime})^{\star}}\right)$.

For the sampling over the $\Pi$ distribution, a block Gibbs sampling
analogous to what performed in standard DBM architectures can be performed
\cite{salakhutdinov2009deepboltzmann,salakhutdinov2012anefficient}.
Alternatively, it is possible to devise a set of Metropolis local
updates sampling the exactly known marginals $\tilde{\Pi}(\sigma^{z},h,h^{\prime})=\sum_{\{d,d^{\prime}\}}\Pi(\sigma^{z},h,h^{\prime},d,d^{\prime})$
or $\tilde{\Pi}'(\sigma^{z},d,d^{\prime})=\sum_{\{h,h^{\prime}\}}\Pi(\sigma^{z},h,h^{\prime},d,d^{\prime})$.
We can also employ efficient cluster updates. Sampling is discussed
more in detail in the Supplementary Information (III).

\section*{Numerical results}

We have implemented numerical algorithms to sample and obtain physical
properties from the DBM previously derived. In Fig. \ref{fig:results}
(a) we show results for the one-dimensional TFI model. Specifically,
we show the expectation value of the energy following the imaginary-time
evolution starting from a fully polarized (in the $x$ direction)
initial state. The initial state corresponds to an empty network,
where all the DBM parameters are set to zero. The DBM results closely
match the exact imaginary-time evolution, thus verifying the correctness
of our construction. 

Numerical results for the one-dimensional Heisenberg
model are shown in Figure \ref{fig:results} (b-c). Specifically,
\ref{fig:results}(b) shows the numerical check for the DBM (construction
2d-6h) time evolution for one-dimensional Heisenberg model for $N=16$.
As expected, the DBM results also in this case follow the exact time
evolution. Figure \ref{fig:results}(c) shows the dependence of the
energy from the initial state, for $N=80$ case. Specifically, by
taking a pre-optimized variational RBM as an initial state as an initial
state, we can significantly decrease the time $\tau$ needed to reach
the ground state.

In the case of the TFI model, sampling from the DBM is realized through
the Gibbs scheme previously sketched, in conjunction with a parallel
tempering scheme, to improve ergodicity in the sampling. In the AFH
model with 2d-6h representation, we employ loop update~\cite{PhysRevLett.70.875}
used in the path-integral QMC method, because the imaginary-time evolution
in the 2d-6h representation has a direct correspondence to the path-integral
formulation, allowing for an efficient handling of the constraint
$d_{[l]}+d_{[m]}=\sigma_{l}^{z}+\sigma_{m}^{z}$.

\section*{Discussion}

We have shown how exact ground states of interacting spin Hamiltonians
can be explicitly constructed using artificial neural networks comprising
only two layers of hidden variables. In contrast to approaches based
on one-layer RBMs, the constructions we have derived here do not require
further variational optimization of the network parameters. In the
case of the Heisenberg model, all of the explicit algorithms presented
here give rise to sign-problem-free representations, if the lattice
is bipartite. The DBM representation has an intrinsic conceptual value,
as an alternative to the path-integral representation. Notably, the
additional deep hidden layer in the DBM plays a similar role as an
additional dimension in statistical mechanics. Whereas a single layer
(RBM) is enough to describe exactly the state of a classical system
{[}see Eq. (\ref{eq.classical_decomposition}){]}, a second layer
is necessary to describe exactly quantum mechanical states. 

DBM-based
schemes can be further used to systematically improve upon existing
RBM variational results. More generally, the initial state for the
present DBM scheme can be generic variational states or even combinations
of RBMs and more conventional wave functions~\cite{nomura2017restrictedboltzmannmachine,clark2017unifying}.
We have shown that, by starting the DBM construction from a pre-optimized
variational state, a fast convergence to the exact ground state is
observed. In conjunction with very accurate initial RBM states, this
kind of scheme opens the possibility of characterizing the ground
state even in the case of non-bipartite lattices with frustration
effects, exploiting the transient regime in which the sign problem
can be still efficiently handled numerically, as for example discussed
in Ref.~\cite{ceperley-alder}.

\section*{Methods}

\subsection*{Useful identities}

{\small{}It is useful to introduce several identities, which can be
used when more complicated interactions between the visible spins
$\sigma^{z}$, hidden variables $h$ and deep variables $d$ beyond
the standard form Eq. (\ref{eq:psidbm}) are needed. The first identity
reads 
\begin{multline}
e^{s_{1}s_{2}V}=C\sum_{s_{3}=\pm1}e^{s_{1}s_{3}\tilde{V}_{1}+s_{2}s_{3}\tilde{V}_{2}}=2C\cosh(s_{1}\tilde{V}_{1}+s_{2}\tilde{V}_{2}).\label{W''1_methods}
\end{multline}
with 
\begin{eqnarray}
C & = & \frac{1}{2}e^{-|V|}\\
\tilde{V}_{1} & = & \frac{1}{2}{\rm arcosh}(e^{2|V|})\\
\tilde{V}_{2} & = & {\rm sgn}(V)\times\tilde{V}_{1}
\end{eqnarray}
for Ising variables $s_{1}$ and $s_{2}$, and a real interaction
$V$. This is a gadget for decomposing two-body interactions, and
can be proven by examining all the cases of $s_{1}$ and $s_{2}$.}{\small \par}

{\small{}By taking $s_{1}$ and $s_{2}$ as visible (physical) variables
$\sigma^{z}$ and $s_{3}$ as a hidden variable $h$, the direct classical
two-body interaction between physical variables {[}the leftmost part
in Eq. (\ref{W''1_methods}){]} is cut and instead mediated by the hidden
neuron $h$. Furthermore, a direct interaction between $\sigma^{z}$
and $d$ can also be decomposed: In the following derivations for
the DBM wave constructions, for convenience, we sometimes introduce
the direct interaction between $\sigma^{z}$ and $d$, which is not
allowed in the DBM structure. However, by taking $s_{1}$ as a visible
spin $\sigma^{z}$, $s_{2}$ as a deep variable $d$, and $s_{3}$
as a hidden variable $h$ in Eq.~(\ref{W''1_methods}), one can eliminate
the direct interaction between $\sigma^{z}$ and $d$ and decompose
it into the interaction mediated only by $h$ with trade-off of the
summation over the hidden variable $h$. With this trick, one can
recover the standard DBM form in Eq.~(\ref{eq:psidbm}).}{\small \par}

{\small{}Another identity (decomposition of four-body interaction)
is 
\begin{eqnarray}
e^{s_{1}s_{2}s_{3}s_{4}V} & = & \frac{1}{4}\sum_{s_{5},s_{6},s_{7}}\exp\left[i\frac{\pi}{4}(s_{5}+s_{6})(s_{1}+s_{2}+s_{3}+s_{7})\right]\nonumber \\
 & \times & \exp(s_{4}s_{7}V)\nonumber \\
 & = & \sum_{s_{7}}\cos^{2}\left[\frac{\pi}{4}(s_{1}+s_{2}+s_{3}+s_{7})\right]\exp(s_{4}s_{7}V)\nonumber \\
\label{DHS0_methods}
\end{eqnarray}
for Ising variables $s_{i}$ with $i=1,\cdots,4$. Although we have
introduced complex couplings in the first line, each term in the summation
in the second line of Eq. (\ref{DHS0}) is positive definite if $V$
is real. The second line remains nonzero only if $s_{1}s_{2}=s_{3}s_{7}$,
which proves the identity. This identity with $s_{1}$ and $s_{2}$
as physical variables, $s_{4}$, $s_{5}$, and $s_{6}$ as hidden
variables, and $s_{3}$ and $s_{7}$ as deep variables, reads 
\begin{eqnarray}
e^{\sigma_{1}\sigma_{2}d_{1}h_{1}V} & = & \frac{1}{4}\sum_{h_{2},h_{3},d_{2}}\exp\left[i\frac{\pi}{4}(h_{2}+h_{3})(\sigma_{1}+\sigma_{2}+d_{1}+d_{2})\right]\nonumber \\
 &  & \times\exp(h_{1}d_{2}V),\label{four_int_gadget_methods}
\end{eqnarray}
Note that the right hand side fits the DBM structure.}{\small \par}

{\small{}General three-body and two-body interactions can also be
represented by the two-body form just by putting some of $s_{1},\cdots s_{4}$
as constants in Eq.(\ref{DHS0_methods}). These could be used instead of Eq.
(\ref{W''1_methods}), although we employ Eq. (\ref{W''1_methods}) in the formalism
below for the decoupling of the two-body interaction.}{\small \par}

{\small{}Finally, we discuss the gadgets for decomposing general $N$-body
classical interactions using complex bias term $b_{j}$ in addition
to the couplings $W$ and $W'$, whereas the gadgets Eqs. (\ref{W''1_methods})
and (\ref{four_int_gadget_methods}) are represented only by $W$ and $W'$
interactions. The gadget reads 
\begin{eqnarray}
e^{\sigma_{1}\sigma_{2}\ldots\sigma_{N}V} & = & C\cos^{2}\left(b+\frac{\pi}{4}\sum_{i=1}^{N}\sigma_{i}\right)\\
 & = & \frac{C}{4}\sum_{h_{1},h_{2}}e^{ib(h_{1}+h_{2})}e^{i\frac{\pi}{4}(h_{1}+h_{2})(\sigma_{1}+\sigma_{2}+\ldots+\sigma_{N})}\nonumber \\
\label{eq.classical_decomposition}
\end{eqnarray}
with 
\begin{eqnarray}
b & = & \mathrm{arctan}\left(e^{-V}\right)-\frac{\pi}{4}\mathrm{mod}(N,4),\\
C & = & \frac{1}{\cos\left(\mathrm{arctan}\left(e^{-V}\right)\right)\times\sin\left(\mathrm{arctan}\left(e^{-V}\right)\right).}
\end{eqnarray}
This fact suggests that any classical partition function defined for
Ising spins can be written exactly in terms of an RBM. Although the
RBM is shown to be powerful in representing also the quantum states,
there is no analytical way to map quantum states to the RBM and one
must rely on numerical optimizations to get the RBM parameters. In
the present study, we show analytical mappings from quantum states
to the DBM, which has additional hidden layer. In the statistical
mechanics, it is known that quantum systems with $D$ dimension can
be mapped on $(D+1)$-dimensional classical systems. Therefore, having
additional hidden layer in neural network language is equivalent to
acquiring additional dimension in statistical mechanics.}{\small \par}

\subsection*{Transverse-Field Ising model}

{\small{}The solution of Eq.(\ref{eq:sigmaxprop}) is found in the
following way. The left hand side of Eq.(\ref{eq:sigmaxprop}) can
be rewritten by using the notation Eq.(\ref{eq:PsiP1P2}) as 
\begin{eqnarray}
 &  & \sum_{\{h,d\}}P_{1}(\sigma^{z},h)P_{2}(h,d)\left[1+\tanh(\Gamma_{l}\delta_{\tau})e^{-2\sigma_{l}^{z}\sum_{j}h_{j}W_{lj}}\right]\nonumber \\
 & = & C\Psi_{\bar{\mathcal{W}}}(\sigma^{z}).\label{eq:Psidbmsigmax}
\end{eqnarray}
We look for a solution by adding one deep neuron $d_{[l]}$ and creating
new couplings $W_{j[l]}^{\prime}$ to the existing hidden neurons
$h_{j}$ which are connected to $\sigma_{l}^{z}$. We also allow for
changes in the existing interaction parameters. In particular we set
the new couplings to be $\bar{W}_{lj}=W_{lj}+\Delta W_{lj}$, (with
$\Delta W_{lj}$ to be determined). Moreover, we introduce one hidden
neuron $h_{[l]}$ coupled to $\sigma_{l}^{z}$ and $d_{[l]}$ through
the interactions $W_{l[l]}$ and $W'_{[l][l]}$, respectively. If
we trace out $h_{[l]}$, the hidden neuron $h_{[l]}$ mediates the
interaction between $\sigma_{l}^{z}$ and $d_{[l]}$ (denoted as $W''_{l[l]}$).}{\small \par}

{\small{}With this choice, we have (in the representation where $h_{[l]}$
is traced out): 
\begin{eqnarray}
\Psi_{\bar{\mathcal{W}}}(\sigma^{z}) & = & \sum_{\{h,d\}}\sum_{d_{[l]}}P_{1}(\sigma^{z},h)P_{2}(h,d)\nonumber \\
 &  & e^{\sigma_{l}^{z}\sum_{j}\Delta W_{lj}h_{j}+d_{[l]}\sum_{j}h_{j}W_{j[l]}^{\prime}+\sigma_{l}^{z}d_{[l]}W_{l[l]}^{\prime\prime}}.\label{Psi_barW_1d-3h}
\end{eqnarray}
The equations to be verified are obtained considering the two possible
values of $\sigma_{l}^{z}=\pm1$: 
\begin{eqnarray}
 &  & e^{\sum_{j}h_{j}\left(\Delta W_{lj}+W_{j[l]}^{\prime}\right)+W_{l[l]}^{\prime\prime}}+e^{\sum_{j}h_{j}\left(\Delta W_{lj}-W_{j[l]}^{\prime}\right)-W_{l[l]}^{\prime\prime}}\nonumber \\
 & = & C\times\left(1+\tanh(\Gamma_{l}\delta_{\tau})e^{-2\sum_{j}h_{j}W_{lj}}\right)\\
 &  & e^{\sum_{j}h_{j}\left(-\Delta W_{lj}+W_{j[l]}^{\prime}\right)-W_{l[l]}^{\prime\prime}}+e^{\sum_{j}h_{j}\left(-\Delta W_{lj}-W_{j[l]}^{\prime}\right)+W_{l[l]}^{\prime\prime}}\nonumber \\
 & = & C\times\left(1+\tanh(\Gamma_{l}\delta_{\tau})e^{2\sum_{j}h_{j}W_{lj}}\right).
\end{eqnarray}
This equation has a solution from the requirement that the hidden
unit interactions on the left and right hand sides match, thus we
require 
\begin{eqnarray}
\Delta W_{lj}+W_{j[l]}^{\prime} & = & -2W_{lj}\label{eq:DeltaW1}\\
\Delta W_{lj}-W_{j[l]}^{\prime} & = & 0,\label{eq:DeltaW2}
\end{eqnarray}
and 
\begin{eqnarray}
W_{l[l]}^{\prime\prime} & = & \frac{\log\tanh(\Gamma_{l}\delta_{\tau})}{2}.\label{eq:Wll''}
\end{eqnarray}
Notice that when $\Gamma_{l}>0$, $W_{l[l]}^{\prime\prime}$ is also
real. By using Eq. (\ref{W''1_methods}) with the following replacement $s_{1}\rightarrow\sigma_{l}^{z}$,
$s_{2}\rightarrow d_{[l]}$, $s_{3}\rightarrow h_{[l]}$, $V\rightarrow W_{l[l]}^{''}$,
$\tilde{V}_{1}\rightarrow W_{l[l]}$ and $\tilde{V}_{2}\rightarrow W_{[l][l]}^{\prime}$,
the last condition determines the real couplings $W_{l[l]}$ and $W_{[l][l]}^{\prime}$
as Eqs.(\ref{eq:wl=00003D00003D00003D00003D00003D00005Bx=00003D00003D00003D00003D00003D00005D})
and (\ref{eq:W'll})}.

\subsection*{Heisenberg model}

{\small{}Here, we show the derivation for the general form of bond
Hamiltonian allowing anisotropy and bond-disorder: $\mathcal{H}_{lm}^{\mathrm{bond}}=J_{lm}^{xy}\left(\sigma_{l}^{x}\sigma_{m}^{x}+\sigma_{l}^{y}\sigma_{m}^{y}\right)+J_{lm}^{z}\sigma_{l}^{z}\sigma_{m}^{z}$.
In the case of the bipartite lattice and the antiferromagnetic exchange
$J_{lm}^{z},J_{lm}^{xy}>0$, we further apply a local gauge transformation
by a $\pi$ rotation around the $z$ axis in the spin space as $\sigma^{x}\rightarrow-\sigma^{x}$
and $\sigma^{y}\rightarrow-\sigma^{y}$ on one of the sublattices,
which gives a $-$ sign for $\sigma_{l}^{x}\sigma_{m}^{x}$ and $\sigma_{l}^{y}\sigma_{m}^{y}$
interactions. This transformation is equivalent to taking 
\begin{equation}
J_{lm}^{xy}\rightarrow-J_{lm}^{xy}.\label{gauge}
\end{equation}
The gauge transformation enables to design a DBM neural network with
real couplings $\{W,W'\}$ except for those to put ``constraint\char`\"{}
on the values of deep neuron spins (see more detail about the constraint
in the following sections). It ensures that the DBM algorithm has
no negative sign problems.}{\small \par}

{\small{}In the case of the antiferromagnetic Heisenberg model after
the gauge transformation on the bipartite lattice, we must solve,
for each bond, 
\begin{multline}
\delta_{\sigma_{l}^{z},\sigma_{m}^{z}}e^{-\delta_{\tau}J_{lm}^{z}}\Psi_{\mathcal{W}}(\sigma^{z})+(1-\delta_{\sigma_{l}^{z},\sigma_{m}^{z}})e^{\delta_{\tau}J_{lm}^{z}}\\
\left(\Psi_{\mathcal{W}}(\sigma^{z})\cosh(2J_{lm}^{xy}\delta_{\tau})+\Psi_{\mathcal{W}}(\sigma_{l}^{z}\leftrightarrow\sigma_{m}^{z})\sinh(2J_{lm}^{xy}\delta_{\tau})\right)\\
=C\langle\sigma^{z}|\Psi_{\bar{\mathcal{W}}}\rangle.\label{Eq.bond_operator}
\end{multline}
It is also useful to explicitly write the expression for the exchange
term in the second line above: 
\begin{multline}
\Psi_{\mathcal{W}}(\sigma^{z})\cosh(2J_{lm}^{xy}\delta_{\tau})+\Psi_{\mathcal{W}}(\sigma_{l}^{z}\leftrightarrow\sigma_{m}^{z})\sinh(2J_{lm}^{xy}\delta_{\tau})\\
=\sum_{\{h,d\}}P_{1}(\sigma^{z},h)P_{2}(h,d)\Bigl[\cosh(2J_{lm}^{xy}\delta_{\tau})+\\
\sinh(2J_{lm}^{xy}\delta_{\tau})e^{(\sigma_{m}^{z}-\sigma_{l}^{z})\sum_{j}h_{j}\left(W_{lj}-W_{mj}\right)}\Bigr].
\end{multline}
In the following derivations, for the antiferromagnetic Hamiltonian
($J_{lm}^{z},J_{lm}^{xy}>0$) after the gauge transformation, we look
for a solution with zero bias terms ($a_{i},\ b_{j},\ b'_{k}=0$,
$\forall i,j,k$). We can also derive a sign-problem free solution
for the imaginary time evolution in the absence of the explicit gauge
transformation by introducing a complex bias term $a_{i}$. Indeed,
in the ``}\emph{\small{}2 deep, 4 hidden}{\small{}\char`\"{} representation,
we will explicitly show that taking a specific set of complex bias
term $a_{i}$ on physical spins is equivalent to the gauge transformation,
making a solution free from the sign problem.}{\small \par}

{\small{}In a way similar to the TFI model, solutions of Eq. (\ref{Eq.bond_operator})
can be found by specifying the structure of the deep Boltzmann machine
and the three examples are the following.}{\small \par}

\subsubsection*{1d-3h construction}

{\small{}We assume the structure of the updated wave function (corresponding
to Eq. (\ref{Psi_barW_1d-3h}) for the TFI model) to be 
\begin{multline}
\Psi_{\bar{\mathcal{W}}}(\sigma^{z})=\sum_{\{h,d\}}\sum_{\substack{d_{[lm]}=\pm1\\
d_{[lm]}=\sigma_{l}^{z}\ {\rm if}\ \sigma_{l}^{z}=\sigma_{m}^{z}
}
}P_{1}(\sigma^{z},h)P_{2}(h,d)\\
e^{\sigma_{l}^{z}\sum_{j}\Delta W_{lj}h_{j}+d_{[lm]}\sum_{j}h_{j}W_{j[lm]}^{\prime}+d_{[lm]}\sigma_{l}^{z}W_{l[lm]}^{\prime\prime}+V_{[lm]}\sigma_{l}^{z}\sigma_{m}^{z}}.\label{new_wf_1d3h}
\end{multline}
Similarly to the case of the TFI model, a solution of Eq.~(\ref{Eq.bond_operator})
is given by 
\begin{eqnarray}
\Delta W_{lj} & = & -W_{lj}+W_{mj}\label{1d3h_dw}\\
W_{j[lm]}^{\prime} & = & W_{lj}-W_{mj}.\label{1d3h_wp_jlm}
\end{eqnarray}
and 
\begin{eqnarray}
W_{l[lm]}^{\prime\prime} & = & -\left(\log\tanh(2J_{lm}^{xy}\delta_{\tau})\right)/2\\
V_{[lm]} & = & -\left(\log\cosh(2J_{lm}^{xy}\delta_{\tau})\right)/2-J_{lm}^{z}\delta_{\tau}
\end{eqnarray}
Notice that the first condition is equivalent to cutting all connections
from spin $l$ to the hidden units and attaching the spin $l$ to
all the hidden units connected to spin $m$, with an interaction $W_{mj}$.}{\small \par}

{\small{}Although the terms proportional to $W_{l[lm]}^{''}$ and
$V_{lm}$ do not satisfy the standard DBM form, they can be transformed
to the DBM form by introducing new hidden neurons $h_{[lm1]}$ and
$h_{[lm2]}$ {[}see the gadget Eq. (\ref{W''1_methods}){]}: 
\begin{multline*}
\!\!\!\!e^{\sigma_{l}^{z}d_{[lm]}W_{l[lm]}^{\prime\prime}}\!=\!C_{[lm1]}\!\!\sum_{h_{[lm1]}}\!\!e^{\sigma_{l}^{z}h_{[lm1]}W_{l[lm1]}+h_{[lm1]}d_{[lm]}W'_{[lm1][lm]}},
\end{multline*}
with 
\begin{eqnarray}
W_{l[lm1]}=W'_{[lm1][lm]}=\frac{1}{2}{\rm arcosh}\left(\frac{1}{\tanh(2J_{lm}^{xy}\delta_{\tau})}\right).\label{1d3h_wwp_lm1}
\end{eqnarray}
Similarly, the coupling $V_{[lm]}$ is decomposed as 
\begin{multline*}
e^{\sigma_{l}^{z}\sigma_{m}^{z}V_{[lm]}}=C_{[lm2]}\sum_{h_{[lm2]}}e^{\sigma_{l}^{z}h_{[lm2]}W_{l[lm2]}+\sigma_{m}^{z}h_{[lm2]}W_{m[lm2]}},
\end{multline*}
with 
\begin{multline}
W_{l[lm2]}=-W_{m[lm2]}=\frac{1}{2}{\rm arcosh}\left(\cosh(2J_{lm}^{xy}\delta_{\tau})e^{2J_{lm}^{z}\delta_{\tau}}\right).\label{1d3h_wwp_lm2}
\end{multline}
Finally, as discussed in the main text, the constraint $d_{[lm]}=\sigma_{l}^{z}$
when $\sigma_{l}^{z}=\sigma_{m}^{z}$ can be satisfied by adding the
third neuron $h_{[lm3]}$, introducing pure complex $i\pi/6$ couplings.}{\small \par}

\subsubsection*{2d-6h construction}

{\small{}{}In this case, the form of the new wave function reads
\begin{multline}
\Psi_{\bar{\mathcal{W}}}(\sigma^{z})=\sum_{\{h,d\}}\sum_{\substack{d_{[l]},d_{[m]}\\
d_{[l]}+d_{[m]}=\sigma_{l}^{z}+\sigma_{m}^{z}
}
}\!\!\!\!P_{1}(\sigma^{z},h)P_{2}(h,d)\\
e^{\sum_{j}\sum_{n=l,m}h_{j}(\Delta W_{nj}\sigma_{n}^{z}+W_{j[n]}^{\prime}d_{[n]})
+\sum_{n=l,m}\sigma_{n}^{z}(W''_{n[l]}d_{[l]}+W''_{n[m]}d_{[m]})}.\label{wf_new_2d6h}
\end{multline}
A solution of Eq.~(\ref{Eq.bond_operator}) is given by 
\begin{eqnarray}
W_{j[l]}^{\prime} & = & W_{lj},\label{eq:2d6h_wprimex1}\\
W_{j[m]}^{\prime} & = & W_{mj},\label{eq:2d6h_wprimex2}\\
\Delta W_{lj} & = & -W_{lj},\label{Eq:2d6h_wlj_change}\\
\Delta W_{mj} & = & -W_{mj},\label{Eq:2d6h_wmj_change}
\end{eqnarray}
and 
\begin{eqnarray}
W_{l[l]}^{\prime\prime}=W_{m[m]}^{\prime\prime} & = & -\frac{J_{lm}^{z}\delta_{\tau}}{2}-\frac{1}{4}\log\sinh(2J_{lm}^{xy}\delta_{\tau}),\label{Eq.Wpp_ll}\\
W_{l[m]}^{\prime\prime}=W_{m[l]}^{\prime\prime} & = & -\frac{J_{lm}^{z}\delta_{\tau}}{2}-\frac{1}{4}\log\cosh(2J_{lm}^{xy}\delta_{\tau}).\label{Eq.Wpp_lm}
\end{eqnarray}
}{\small \par}

{\small{}The direct interactions between $(\sigma_{l}^{z},d_{[l]})$,
$(\sigma_{m}^{z},d_{[m]})$, $(\sigma_{l}^{z},d_{[m]})$, and $(\sigma_{m}^{z},d_{[l]})$,
are mediated by $h_{[lm1]}$, $h_{[lm2]}$, $h_{[lm3]}$, and $h_{[lm4]}$,
respectively, as follows 
\begin{eqnarray*}
e^{\sigma_{l}^{z}d_{[l]}W_{l[l]}^{\prime\prime}} & = & C_{[lm1]}\sum_{h_{[lm1]}}e^{\sigma_{l}^{z}h_{[lm1]}W_{l[lm1]}+h_{[lm1]}d_{[l]}W'_{[lm1][l]}},\\
e^{\sigma_{m}^{z}d_{[m]}W_{m[m]}^{\prime\prime}} & = & C_{[lm2]}\sum_{h_{[lm2]}}e^{\sigma_{m}^{z}h_{[lm2]}W_{m[lm2]}+h_{[lm2]}d_{[m]}W'_{[lm2][m]}},\\
e^{\sigma_{l}^{z}d_{[m]}W_{l[m]}^{\prime\prime}} & = & C_{[lm3]}\sum_{h_{[lm3]}}e^{\sigma_{l}^{z}h_{[lm3]}W_{l[lm3]}+h_{[lm3]}d_{[m]}W'_{[lm3][m]}},\\
e^{\sigma_{m}^{z}d_{[l]}W_{m[l]}^{\prime\prime}} & = & C_{[lm4]}\sum_{h_{[lm4]}}e^{\sigma_{m}^{z}h_{[lm4]}W_{m[lm4]}+h_{[lm4]}d_{[l]}W'_{[lm4][l]}}.
\end{eqnarray*}
By applying the gadget Eq.~(\ref{W''1_methods}), the new $W$ and $W'$
interactions are given by, for small $\delta_{\tau}$ (such that $\frac{e^{-J_{lm}^{z}\delta_{\tau}}}{\sqrt{\sinh(2J_{lm}^{xy}\delta_{\tau})}}>1$),
\begin{multline}
W_{l[lm1]}=W'_{[lm1][l]}=W_{m[lm2]}=W'_{[lm2][m]}\\
=\frac{1}{2}\mathrm{arcosh}\left(\frac{e^{-J_{lm}^{z}\delta_{\tau}}}{\sqrt{\sinh(2J_{lm}^{xy}\delta_{\tau})}}\right)\label{Eq.2d6h_wwp_lm1_lm2}
\end{multline}
and 
\begin{multline}
W_{l[lm3]}=-W'_{[lm3][m]}=W_{m[lm4]}=-W'_{[lm4][l]}\\
=\frac{1}{2}\mathrm{arcosh}\left(\sqrt{\cosh(2J_{lm}^{xy}\delta_{\tau})}\times e^{J_{lm}^{z}\delta_{\tau}}\right).\label{Eq.2d6h_wwp_lm3_lm4}
\end{multline}
Finally, the constraint $d_{[l]}+d_{[m]}=\sigma_{l}^{z}+\sigma_{m}^{z}$
can be put by introducing additionally two hidden neurons $h_{[lm5]}$
and $h_{[lm6]}$, and by introducing complex couplings 
\begin{multline}
\sum_{h_{[lm5]},h_{[lm6]}}e^{i\frac{\pi}{4}\left((\sigma_{l}^{z}+\sigma_{m}^{z})h_{[lm5]}-h_{[lm5]}(d_{[l]}+d_{[m]})\right)}\\
\times e^{i\frac{\pi}{8}\left((\sigma_{l}^{z}+\sigma_{m}^{z})h_{[lm6]}-h_{[lm6]}(d_{[l]}+d_{[m]})\right)}
\end{multline}
This term gives interactions among $d_{[l]}$, $d_{[m]}$, $\sigma_{l}^{z}$
and $\sigma_{m}^{z}$: $4\cos\left(\frac{\pi}{4}(\sigma_{l}^{z}+\sigma_{m}^{z}-d_{[l]}-d_{[m]})\right)\cos\left(\frac{\pi}{8}(\sigma_{l}^{z}+\sigma_{m}^{z}-d_{[l]}-d_{[m]})\right)$,
which realize the constraint.}{\small \par}

\subsubsection*{2d-4h construction}

{\small{}For this construction, we assume the following structure
for the wave-function after the propagator: 
\begin{multline*}
\Psi_{\bar{\mathcal{W}}}(\sigma^{z})=\sum_{\{h,d\}}\sum_{d_{[l]}}P_{1}(\sigma^{z},h)P_{2}(h,d)e^{\sum_{j,n=l,m}\sigma_{n}^{z}h_{j}\Delta W_{nj}}\\
\times e^{\sum_{j}h_{j}d_{[l]}W_{j[l]}^{\prime}+\sum_{n=l,m}\sigma_{n}^{z}d_{[l]}W_{n[l]}^{''}+\sum_{j}\sigma_{l}^{z}\sigma_{m}^{z}h_{j}d_{[l]}Z_{lmj}}.
\end{multline*}
In this case, we also look for a solution for the bond operator without
the gauge transformation. This shows that the introduction of a complex
bias term $a_{i}$ can play the same role as the gauge transformation.
Then, we need to solve: 
\begin{multline}
\delta_{\sigma_{l}^{z},\sigma_{m}^{z}}e^{-\delta_{\tau}J_{lm}^{z}}\Psi_{\mathcal{W}}(\sigma^{z})+(1-\delta_{\sigma_{l}^{z},\sigma_{m}^{z}})e^{\delta_{\tau}J_{lm}^{z}}\\
\left(\Psi_{\mathcal{W}}(\sigma^{z})\cosh(2J_{lm}^{xy}\delta_{\tau})-\Psi_{\mathcal{W}}(\sigma_{l}^{z}\leftrightarrow\sigma_{m}^{z})\sinh(2J_{lm}^{xy}\delta_{\tau})\right)\\
=C\langle\sigma^{z}|\Psi_{\bar{\mathcal{W}}}\rangle.\label{Eq.bond_operator2}
\end{multline}
Note that the sign for $\Psi_{\mathcal{W}}(\sigma_{l}^{z}\leftrightarrow\sigma_{m}^{z})\sinh(2J_{lm}^{xy}\delta_{\tau})$
term is different from that in Eq. (\ref{Eq.bond_operator}).}{\small \par}

{\small{}A solution of Eq. (\ref{Eq.bond_operator2}) is obtained
as 
\begin{eqnarray}
\Delta W_{lj} & = & -\Delta W_{mj}=-\frac{1}{2}(W_{lj}-W_{mj}),\label{bWZ3}
\end{eqnarray}
where $W_{nj}$ $(n=l,m)$ is updated to $\bar{W}_{nj}$ with the
increment $\Delta W_{nj}$ as $\bar{W}_{nj}=W_{nj}+\Delta W_{nj}$.
The new couplings $W_{j[l]}^{\prime}$, $Z_{lmj}$ and $W''_{n[l]}$
are also given by 
\begin{eqnarray}
W_{j[l]}^{\prime}=-Z_{lmj}=-\frac{1}{2}(W_{lj}-W_{mj})\label{bWZ4}
\end{eqnarray}
and 
\begin{eqnarray}
W''_{l[l]} & = & \frac{1}{4}\Biggl[\log\left[-e^{-2a_{l-m}}\tanh(2J_{lm}^{xy}\delta_{\tau})\right]\nonumber \\
 & + & 2{\rm arcosh}\biggl[\frac{e^{-2J_{lm}^{z}\delta_{\tau}}}{\sqrt{-2e^{-2a_{l-m}}\sinh(4J_{lm}^{xy}\delta_{\tau})}}\biggr]\Biggr]\label{W''l}\\
W''_{m[l]} & = & \frac{1}{4}\Biggl[-\log\left[-e^{
-2a_{l-m}}\tanh(2J_{lm}^{xy}\delta_{\tau})\right]\nonumber \\
 & + & 2{\rm arcosh}\biggl[\frac{e^{-2J_{lm}^{z}\delta_{\tau}}}{\sqrt{-2e^{-2a_{l-m}}\sinh(4J_{lm}^{xy}\delta_{\tau})}}\biggr]\Biggr]\label{W''m}
\end{eqnarray}
with $a_{l-m}=a_{l}-a_{m}$. On a bipartite lattice, to avoid the
negative sign (or complex phase) problem we need to keep $W''_{l[l]}$
and $W''_{m[l]}$ real. This can be achieved by choosing $a_{l}=0$
for any $l$ if $J_{lm}<0$ (ferromagnetic case). For $J_{lm}>0$
(antiferromagnetic case), $a_{l}=n\pi i$ with an arbitrary integer
$n$ if the site $l$ belongs to the sub-lattice A and $a_{l}=(n+1/2)\pi i$
if $l$ belongs to the sub-lattice B. This local gauge for $J_{lm}>0$
is equivalent to the transformation $J_{lm}^{xy}\rightarrow-J_{lm}^{xy}$
and $a_{l}=0$ for any site $l$. We further notice that $W''_{m[l]}$
can be taken positive if we take a sufficiently small $\delta_{\tau}$
in Eq (\ref{W''m}), with the leading order term $-\log(2J_{lm}^{xy}\delta_{\tau})/2$.
On the other hand, in Eq.~(\ref{W''l}), the leading order term is
negative ($=-J_{lm}\delta_{\tau}$).}{\small \par}

{\small{}To recover the original form of the DBM, we first use Eq.
(\ref{W''1_methods}) with the replacement $s_{1}\rightarrow\sigma_{n}^{z}$,
$s_{2}\rightarrow d_{[l]}$, $s_{3}\rightarrow h_{[n]}$, $C\rightarrow D_{n}$,
$V\rightarrow W''_{n[l]}$ $\tilde{V}_{1}\rightarrow W_{n[n]}$ and
$\tilde{V}_{2}\rightarrow W'_{[n][l]}$ for $n=l,m$. Then a solution
for $D_{n}$, $W_{n[n]}$, and $W'_{[n][l]}$ are represented by using
$W_{n[l]}^{''}$ as 
\begin{eqnarray}
D_{n} & = & \frac{1}{2}\exp[-W_{n[l]}^{''}]\\
W_{n[n]} & = & W'_{[n][l]}=\frac{1}{2}{\rm arcosh}(\exp[2W_{n[l]}^{''}]),
\label{Wl52}
\end{eqnarray}
for positive $W_{n[l]}^{''}$ and 
\begin{eqnarray}
D_{n} & = & \frac{1}{2}\exp[W_{n[l]}^{''}]\\
W_{n[n]} & = & -W'_{[n][l]}=\frac{1}{2}{\rm arcosh}(\exp[-2W_{n[l]}^{''}]),\label{Wl53}
\end{eqnarray}
for negative $W_{n[l]}^{''}$ to give real $W_{n[n]}$ and $W'_{[n][l]}$.}{\small \par}

{\small{}To completely recover the original DBM form, we next use
Eq. (\ref{four_int_gadget_methods}) by replacing $\sigma_{1}$ with $\sigma_{l}^{z}$,
$\sigma_{2}$ with $\sigma_{m}^{z}$, $d_{1}$ with $d_{[l]}$, $d_{2}$
with $d_{[lm]}$, $h_{1}$ with $h_{j}$, $h_{2}$ with $h_{[lm1]}$,
$h_{3}$ with $h_{[lm2]}$, and $V$ with $Z_{lmj}$.}{\small \par}

{\small{}With these solutions, by ignoring the trivial constant factors
including $D_{l}$ and $D_{m}$, the evolution is described by introducing
two deep and four hidden additional variables $d_{[l]}$, $d_{[lm]}$,
$h_{[l]}$, $h_{[m]}$, $h_{[lm1]}$, and $h_{[lm2]}$ as 
\begin{multline}
\Psi_{\bar{\mathcal{W}}}(\sigma^{z})=\sum_{\{\bar{h},\bar{d}\}}P_{1}(\sigma^{z},h)P_{2}(h,d)\exp\biggl[\sum_{j,n=l,m}\sigma_{n}^{z}h_{j}\Delta W_{nj}\\
+\sum_{j}h_{j}d_{[l]}W_{j[l]}^{\prime}+\sum_{n=l,m}h_{[n]}(\sigma_{n}^{z}W_{n[n]}+d_{[l]}W_{[n][l]}^{'})\\
+d_{[lm]}\sum_{j}h_{j}Z_{lmj}+\frac{i\pi}{4}(h_{[lm1]}+h_{[lm2]})(\sigma_{l}^{z}+\sigma_{m}^{z}+d_{[l]}+d_{[lm]})\biggr],\label{DBM_summary}
\end{multline}
where $\{\bar{h},\bar{d}\}$ is a set consisting of the existing and
new neurons.}{\small \par}

\section*{Acknowledgements}

{\small{}G.C. acknowledges useful discussions with Xun Gao, and Markus
Heyl. Y.N. and M.I. are grateful for the useful discussions with Youhei
Yamaji and Andrew S. Darmawan. Y.N. was financially supported by Grant-in-Aids
for Scientific Research (JSPS KAKENHI) (No. 17K14336). M.I. and Y.N.
were financially supported by a Grant-in- Aid for Scientific Research
(No. 16H06345) from Ministry of Education, Culture, Sports, Science
and Technology, Japan. Part of the calculations were done at Supercomputer
Center, Institute for Solid State Physics, University of Tokyo. This
work was also supported in part by MEXT as a social and scientific
priority issue (Creation of new functional devices and high-performance
materials to support next-generation industries CDMSI) to be tackled
by using post-K computer. We also thank the support provided by the
RIKEN Advanced Institute for Computational Science through the HPCI
System Research project (hp170263) supported by Ministry of Education,
Culture, Sports, Science, and Technology, Japan. }{\small \par}

\changelocaltocdepth{1}
\onecolumngrid
\clearpage

\part*{\centering Supplementary Information}

\section{Deep Boltzmann Machines}

\noindent
{\bf Deep Boltzmann machine representation of quantum states.}
In the main text we have considered a representation of the many-body
wave-function in terms of a two-layers deep Boltzmann Machine (DBM).
In the following we specialize to the case of $N$ spin $1/2$ particles,
described by the quantum numbers $|\sigma^z \rangle=|\sigma_{1}^{z}\dots\sigma_{N}^{z}\rangle$
with $\sigma_{i}^z=\pm1$.
Then, we represent the amplitudes $\langle\sigma_{1}^{z}\dots\sigma_{N}^{z}|\Psi\rangle\equiv\Psi(\sigma^z)$
in the DBM form:
\begin{eqnarray}
\Psi_{\mathcal{W}}(\sigma^z) & = & \sum_{\{h\}}e^{\sum_{i}a_{i}\sigma_{i}^{z}}e^{\sum_{ij}\sigma_{i}^{z}h_{j}W_{ij}+\sum_{j}b_{j}h_{j}}\sum_{\{d\}}e^{\sum_{jk}h_{j}d_{k}W_{jk}^{\prime}+\sum_{k}b'_{k}d_{k}}.\label{eq:PsiDeep}
\end{eqnarray}
Here, we have introduced $M$ hidden units $h_{j}$, $M^{\prime}$
deep units $d_{k}$, and a set of couplings and bias terms $\mathcal{W}\equiv(a,b,b',W,W^{\prime})$.
Hereafter, we call the neurons in the 1st hidden layer just hidden neurons
 and distinguish them from the neurons in the 2nd hidden layer, which are called deep neurons.

All those parameters, in general, must be taken complex-valued to
represent a generic many-body state. The hidden and deep units are
taken here to be of spin $1/2$, i.e. $h_{j}=\pm1$, $d_{k}=\pm1$,
and the summations are over all the possible values of those variables.
From a pictorial point of view, the DBM architecture features direct
connections (interactions) between nearest-neighboring layers. In
particular, the visible layer of physical degrees of freedom ($\sigma_{1}^{z}\dots\sigma_{N}^{z}$)
is connected only to the first layer of hidden variables $(h_{1}\dots h_{M})$,
whereas the first layer is connected both to the visible spins and
to the deep spins $(d_{1}\dots d_{M^{\prime}})$. 

For the following derivations, it is useful to write the DBM amplitudes
as: 
\begin{eqnarray}
\Psi_{\mathcal{W}}(\sigma^z) & = & \sum_{\{h,d\}}P_{1}(\sigma^z,h)P_{2}(h,d),
\end{eqnarray}
where we have introduced the two quantities: 
\begin{eqnarray}
P_{1}(\sigma^z,h) & = & e^{\sum_{i}a_{i}\sigma_{i}^{z}}e^{\sum_{ij}\sigma_{i}^{z}h_{j}W_{ij}+\sum_{j}b_{j}h_{j}}\\
P_{2}(h,d) & = & e^{\sum_{jk}h_{j}d_{k}W_{jk}^{\prime}+\sum_{k}b'_{k}d_{k}}.
\end{eqnarray}
Notice that, in general, those weights are complex-valued, and cannot
be interpreted as genuine Boltzmann weights. From these expressions,
it is also straightforward to see that the Restricted Boltzmann Machine
(RBM) expression for the wave-function is recovered when $M^{\prime}=0$,
i.e. taking 
\begin{eqnarray}
\Psi_{\mathcal{W}}^{\mathrm{RBM}}(\sigma^z) & = & \sum_{\{h\}}P_{1}(\sigma^z,h)\\
 & = & e^{\sum_{i}a_{i}\sigma_{i}^{z}}\Pi_{j}^{M}2\cosh\left(\sum_{i}^{N}\sigma_{i}^{z}h_{j}W_{ij}+b_{j}\right),
\end{eqnarray}
where we have explicitly performed the summation of the hidden variables.
At variance with the RBM case, in the more general case when $M^{\prime}>0$, it is not possible to analytically obtain the DBM amplitudes.
\\

\noindent
{\bf Useful gadgets in constructing DBM neural network.}
In the Methods we have discussed several useful identities to decompose spin interactions. In particular, those identities are very useful if we need more complicated interactions between the visible spins $\sigma^z$, hidden variables $h$ and deep variables $d$ beyond the standard form Eq. (\ref{eq:PsiDeep}).
For the sake of completeness of this Supplementary Information, we reproduce here the identities for decomposing two-body, three-body, and four-body interactions.

The first identity reads 
\begin{eqnarray}
e^{s_1 s_2 V  }   =  C\sum_{s_3  = \pm1 } e^{s_1  s_3  \tilde{V}_1  + s_2 s_3 \tilde{V}_2} = 2C \cosh (s_1 \tilde{V}_1 + s_2\tilde{V}_2   ) .
\label{W''1}
\end{eqnarray}
with 
\begin{eqnarray}
  C &=&  \frac{1}{2} e^{ - | V | } \\ 
  \tilde{V}_1 &=&  \frac{1}{2} {\rm arcosh} ( e^{2|V|}) \\ 
  \tilde{V}_2 &=&  {\rm sgn} (V ) \times \tilde{V}_1  
\end{eqnarray}
for Ising variables $s_1$ and $s_2$, and a real interaction $V$.
This is the gadget for decomposing two-body interactions discussed in Methods. 
In the following, we will use this identity to decompose either interactions between visible (physical spins) (in that case $s_1$ and $s_2$ are both $\sigma^z$ variables), or to decompose direct interactions between a $\sigma^z$ spin and a deep unit $d$.

Another identity (decomposition of four-body interaction) is
\begin{eqnarray}
e^{s_1s_2s_3 s_4 V} &=& 
\frac{1}{4}
\sum_{s_5,s_6,s_7} \exp \left[i \frac{\pi}{4}( s_5+s_6 )(s_1+s_2+s_3 + s_7) \right ] \exp  (s_4 s_7 V) 
 \nonumber \\
 &=& \sum_{s_7}  \cos^2 \left [\frac{\pi}{4}(s_1+s_2+s_3+s_7) \right ]  \exp  (s_4 s_7 V)
 \label{DHS0}
\end{eqnarray}
for Ising variables $s_i$ with $i=1, \cdots, 4$.
Although we have introduced complex couplings in the first line, 
each term in the summation in the second line of Eq. (\ref{DHS0}) is positive definite if $V$ is real. 
The second line remains nonzero only for 
$s_7=1$ if $s_1s_2s_3=1$ and only for $s_7=-1$ if $s_1s_2s_3=-1$, 
which proves the identity. 
This identity with $s_1$ and $s_2$ as physical variables, $s_4$, $s_5$, and $s_6$ as hidden variables, and $s_3$ and $s_7$ as deep variables, which reads
\begin{eqnarray}
 e^{\sigma_1 \sigma_2 d_1 h_1 V} &=&\frac{1}{4}  \sum_{h_2,h_3, d_2} \exp \left[i \frac{\pi}{4}( h_2+h_3 )(\sigma_1+\sigma_2+ d_1 + d_2) \right ] \exp  (h_1 d_2 V), 
 \label{four_int_gadget}
\end{eqnarray}
will be used in Sec.~\ref{section-two-four-rep}. 
Note that the right hand side fits the DBM structure.

Although identities for decomposing three-body interactions
are not used in the following derivation, it is nonetheless useful to show them: 
\begin{eqnarray}
e^{s_1s_2s_3 V} &=& 
\frac{1}{4}
\sum_{s_4,s_5,s_6} \exp \left[i \frac{\pi}{4}( s_4+s_5 )(s_1+s_2+s_3 + s_6) \right ] \exp  ( s_6 V) 
 \nonumber \\
 &=& \sum_{s_6}  \cos^2 \left [\frac{\pi}{4}(s_1+s_2+s_3+s_6) \right ]  \exp  (s_6 V). 
 \label{three-body-gadget1}
\end{eqnarray}
This gadget for three-body interactions is obtained by fixing $s_4=1$ in Eq. (\ref{DHS0}) (and changing variables).
 Alternative form is obtained by replacing $s_3$ with $1$ in 
Eq. (\ref{DHS0}), which gives, 
\begin{eqnarray}
e ^{ s_1 s_2 s_3 V} 
   &=& \frac{1}{4} \sum_{s_4, s_5, s_6} \exp \left [ i \frac{ \pi}{4}  (s_4 + s_5)    ( s_1 + s_2 + s_6 +1 )  \right]    \exp (  s_3 s_6 V )  
   \nonumber \\
   &=& \sum_{s_6}  \cos^2 \left [ \frac{\pi}{4} ( s_1 + s_2 + s_6 +1 )  \right]    \exp(  s_3 s_6 V ).   
\label{three-body-gadget2}
\end{eqnarray}
As we see, the gadgets for three-body interactions [Eqs. (\ref{three-body-gadget1}) and (\ref{three-body-gadget2})] have been derived from the gadget for four-body interactions [Eq. (\ref{DHS0})] trivially.  

Gadgets for two-body interactions which are different from Eq. (\ref{W''1}) can also be obtained from Eq. (\ref{DHS0})
by fixing two variables out of $s_1$, $s_2$, $s_3$, $s_4$ to be 1. 
These could be used instead of (\ref{W''1}), 
although we employ (\ref{W''1}) in the formalism below for the decoupling of the two-body interaction.

\section{Representing Ground-States}

As discussed in the main text, our goal is to construct explicit DBM
representations of ground-states of local Hamiltonians. This goal
is achieved 
by finding a representation of the imaginary-time evolved
state: 
\begin{eqnarray}
|\Psi(\tau)\rangle & = & e^{-\tau {\mathcal H}}|\Psi_{0}\rangle,
\end{eqnarray}
where $ |\Psi_{0}\rangle$ is empty RBM ($\langle\sigma^z |\Psi_{0}\rangle=\mathrm{const.}$) or 
pre-optimized RBM state, converging
to the exact ground-state for large enough $\tau$. To achieve this
goal we first consider a second-order Trotter-Suzuki decomposition:
\begin{eqnarray}
|\Psi ({\tau}) \rangle & = & \mathcal{G}_{1}(\delta_{\tau}/2)\mathcal{G}_{2}(\delta_{\tau})\dots\mathcal{G}_{1}(\delta_{\tau})\mathcal{G}_{2}(\delta_{\tau})\mathcal{G}_{1}(\delta_{\tau}/2)|\Psi_{0}\rangle,\label{eq:psitau}
\end{eqnarray}
where $\delta_{\tau}$ is a small time step, the Hamiltonian is decomposed
into two non-commuting parts, $\mathcal{H=\mathcal{H}}_{1}+\mathcal{H}_{2}$,
and $\mathcal{G}_{\nu}(\delta_\tau)=e^{-\mathcal{H}_{\nu}\delta_\tau}$ are
short-time propagators. For given Hamiltonian, we then need to find
specific rules to apply the short-time propagators to a generic DBM,
and obtain a new (time-evolved) DBM, possibly with a larger total
number of hidden and deep neurons. 
In the following, we show concrete examples for the transverse-field Ising and Heisenberg models. 

\subsection{Transverse-Field Ising model}

Let us start with the case of the transverse-field Ising model. We
consider a Trotter-Suzuki decomposition of the imaginary-time propagator,
into two parts: ${\mathcal H}_{1}=-\sum_{i}\Gamma_{i}\sigma_{i}^{x}$, and ${\mathcal H}_{2}=\sum_{l<m}V_{lm}\sigma_{l}^{z}\sigma_{m}^{z}$.
In the following derivation, we assume that $\Gamma_i $ is positive ($\Gamma_i >0$).
In this case, we look for a solution with zero bias terms: $ a_i = b_j = b'_k = 0$, $\forall i,j,k$. 
The case of negative $\Gamma_i$ can also be treated, and is discussed more in detail at the end of this section. 
\\

\noindent
{\bf Interaction propagator.}
The interaction propagator $e^{-\delta_{\tau}V_{lm}\sigma_{l}^{z}\sigma_{m}^{z}}$
is diagonal in the $\sigma^{z}$ basis, and applying it to
 a DBM will lead to a modification in the
DBM parameters. 
In particular, the goal is to satisfy the equation:
\begin{eqnarray}
\langle\sigma^z  |e^{-\delta_{\tau}V_{lm}\sigma_{l}^{z}\sigma_{m}^{z}}|\Psi_{\mathcal{W}}\rangle & = & C\langle\sigma^z |\Psi_{\bar{\mathcal{W}}}\rangle,\label{eq:diagpropIsing}
\end{eqnarray}
i.e. to explicitly find a set of parameters $\bar{\mathcal{W}}$ that
satisfies the previous equation for all the possible $\langle\sigma^z |$,
and for an arbitrary constant $C$. 

We can achieve this goal adding a hidden unit in the first layer,
$h_{[lm]}$ such that it is only connected to the visible spins: $W_{[lm]k}^{\prime}=0, \forall k$.
The new wave function has then the form: 
\begin{eqnarray}
\Psi_{\bar{\mathcal{W}}}(\sigma^z) & = & \sum_{\{h,d\}}\sum_{ h_{[lm]} } P_{1}(\sigma^z,h)P_{2}(h,d)e^{\sigma_{l}^z W_{l[lm]}h_{[lm]}+\sigma_{m}^z W_{m[lm]}h_{[lm]}}\\
 & = & 2\cosh\left(\sigma_{l}^z W_{l[lm]}+\sigma_{m}^z W_{m[lm]}\right)\Psi_{\mathcal{W}}(\sigma^z).
\end{eqnarray}
Equation (\ref{eq:diagpropIsing}) is then satisfied if
\begin{eqnarray}
e^{-\delta_{\tau}V_{lm}\sigma_{l}^{z}\sigma_{m}^{z}} = 2C
\cosh\left(\sigma_{l}^z W_{l[lm]}+\sigma_{m}^z W_{m[lm]}\right) 
\label{eq:sigmasigmarenorm}
\end{eqnarray}
for all the possible values of $\sigma_{l}^{z}$ and $\sigma_{m}^{z}$.
By using the gadget Eq.~(\ref{W''1}), the new parameters $W_{l[lm]}$ and $W_{m[lm]}$ are given by 
\begin{eqnarray}
W_{l[lm]} & = & \frac{1}{2}\mathrm{arcosh}\left(e^{2|V_{lm}|\delta_{\tau}}\right)\\
W_{m[lm]} & = & -\mathrm{sgn}(V_{lm}) \times W_{l[lm]}.
\end{eqnarray}
\\

\noindent
{\bf Transverse-field propagator.}
The propagator involving the transverse-field $e^{\delta_{\tau}\Gamma_{l}\sigma_{l}^{x}}$
is off-diagonal in $\sigma^z$ basis.
For this off-diagonal part, we must solve a slightly more involved
equation:
\begin{eqnarray}
\langle\sigma^z |e^{\delta_{\tau}\Gamma_{l}\sigma_{l}^{x}}|\Psi_{\mathcal{W}}\rangle & = & \Psi_{\mathcal{W}}(\sigma^z )\times\cosh(\Gamma_{l}\delta_{\tau})+\Psi_{\mathcal{W}}(\sigma_{1}^{z},\dots-\sigma_{l}^{z},\dots, \sigma_{N}^{z})\times\sinh(\Gamma_{l}\delta_{\tau})\\
 & = & C\langle\sigma^z |\Psi_{\bar{\mathcal{W}}}\rangle,
\end{eqnarray}
for the new parameters $\bar{\mathcal{W}}$, and for an arbitrary finite
normalization constant $C$. In turn, this equation is equivalent to: 
\begin{eqnarray}
\sum_{\{h,d\}}P_{1}(\sigma^z,h)P_{2}(h,d)\left[1+\tanh(\Gamma_{l}\delta_{\tau})e^{-2\sigma_{l}^{z}\sum_{j}h_{j}W_{lj}}\right] & = & C\Psi_{\bar{\mathcal{W}}}(\sigma^z).\label{eq:Psidbmsigmax}
\end{eqnarray}
We look for a solution
by adding one deep neuron $d_{[l]}$ 
and creating new couplings $W_{j[l]}^{\prime}$ to the existing hidden neurons $h_j$ 
which are connected to $\sigma_l^z$. 
We also allow for 
changes in the existing interaction parameters. 
In particular we set the new couplings to be $\bar{W}_{lj}=W_{lj}+\Delta W_{lj}$,
(with $\Delta W_{lj}$ to be determined).

Moreover, we introduce one hidden neuron $h_{[l]}$ coupled to $\sigma_l^z$ and $d_{[l]}$
through the interactions $W_{l[l]}$ and $W'_{[l][l]}$, respectively. 
If we trace out $h_{[l]}$, the hidden neuron $h_{[l]}$ mediates the interaction between $\sigma_l^z$ and $d_{[l]}$ (denoted as $W''_{l[l]}$).   

With this choice, we have (in the representation where $h_{[l]}$ is traced out): 
\begin{eqnarray}
\Psi_{\bar{\mathcal{W}}}(\sigma^z) & = & \sum_{\{h,d\}}\sum_{ d_{[l]} }P_{1}(\sigma^z,h)P_{2}(h,d)e^{\sigma^z_{l}\sum_{j}\Delta W_{lj}h_{j}+d_{[l]}\sum_{j}h_{j}W_{j[l]}^{\prime}+\sigma^z_{l}d_{[l]}W_{l[l]}^{\prime\prime}}.
\end{eqnarray}

The equations to be verified are obtained considering the two
possible values of $\sigma^z_{l}=\pm1$: 
\begin{eqnarray}
e^{\sum_{j}h_{j}\left(\Delta W_{lj}+W_{j[l]}^{\prime}\right)+W_{l[l]}^{\prime\prime}}+e^{\sum_{j}h_{j}\left(\Delta W_{lj}-W_{j[l]}^{\prime}\right)-W_{l[l]}^{\prime\prime}} & = & C\times\left(1+\tanh(\Gamma_{l}\delta_{\tau})e^{-2\sum_{j}h_{j}W_{lj}}\right)\\
e^{\sum_{j}h_{j}\left(-\Delta W_{lj}+W_{j[l]}^{\prime}\right)-W_{l[l]}^{\prime\prime}}+e^{\sum_{j}h_{j}\left(-\Delta W_{lj}-W_{j[l]}^{\prime}\right)+W_{l[l]}^{\prime\prime}} & = & C\times\left(1+\tanh(\Gamma_{l}\delta_{\tau})e^{2\sum_{j}h_{j}W_{lj}}\right).
\end{eqnarray}
This equation has a solution if the hidden unit interactions on the
l.h.s. and on the r.h.s match, i.e. when: 
\begin{eqnarray}
\Delta W_{lj}+W_{j[l]}^{\prime} & = & -2W_{lj}\\
\Delta W_{lj}-W_{j[l]}^{\prime} & = & 0,
\end{eqnarray}
which in turn are verified when 
\begin{eqnarray}
W_{j[l]}^{\prime} & = & -W_{lj}\label{eq:wprimex}\\
\Delta W_{lj} & = & -W_{lj},\label{eq:deltawx}
\end{eqnarray}
and if 
\begin{eqnarray}
W_{l[l]}^{\prime\prime} & = & \frac{\log\tanh(\Gamma_{l}\delta_{\tau})}{2}.
\end{eqnarray}

When $\Gamma_l > 0$, $W_{l[l]}^{\prime\prime}$ is real. 
By using Eq. (\ref{W''1}) with the following replacement $s_1 \rightarrow \sigma^z_{l}$, $s_2  \rightarrow d_{[l]}$, $s_3  \rightarrow h_{[l]}$, $V  \rightarrow W_{l[l]}^{''}$, $\tilde{V}_1  \rightarrow W_{l[l]}$ and $\tilde{V}_2  \rightarrow W_{[l][l]}^{\prime}$,
the last condition determines the real couplings $W_{l[l]}$ and
$W_{[l][l]}^{\prime}$,  
which read 
\begin{eqnarray}
W_{l[l]} & = & \frac{1}{2}\mathrm{arcosh}\left(\frac{1}{\tanh(\Gamma_{l}\delta_{\tau})}\right)\label{eq:wl=00005Bx=00005D}\\
W_{[l][l]}^{\prime} & = & -W_{l[l]}.
\end{eqnarray}
Notice that because of condition (\ref{eq:deltawx}), after applying
the off-diagonal propagator all the interactions $W_{lj}$ between
spin $l$ and hidden units $h_{j}$ are set to zero. However, because
of condition (\ref{eq:wl=00005Bx=00005D}), the spin $l$ is reconnected
to the new hidden unit $h_{[l]}$ with the $W_{l[l]}$ interaction. 
\\

\noindent
{\bf Negative transverse field.}
When $\Gamma_i < 0$, it is still possible to recover a DBM representation with purely real interaction weights $W$ and $W^\prime$. In order to do so, we apply the gauge transformation $\sigma_i^x \rightarrow -\sigma_i^x$ 
and $\sigma_i^y \rightarrow -\sigma_i^y$ ($\pi$ spin rotation around the $z$ axis), which maps onto the Hamiltonian with positive $\Gamma_i$. 
This gauge transformation can be achieved by taking a finite bias terms $a_i$ in Eq.~(\ref{eq:PsiDeep}) as $a_i = i\pi/2$
and fix them during the imaginary time evolution. 
With this complex bias term $a_i = i\pi/2$, $| \! \uparrow \rangle$ ($|\! \downarrow \rangle$) state at the $i$th site 
acquires a phase as follows $| \! \uparrow \rangle  \rightarrow e^{i\frac{\pi}{2} }| \! \uparrow \rangle = i | \!  \uparrow \rangle$
($| \! \downarrow \rangle  \rightarrow e^{-i\frac{\pi}{2} }| \! \downarrow \rangle = -i | \! \downarrow \rangle  $), 
which is equivalent to a $\pi$ spin rotation around the $z$ axis.
In the case when $\Gamma_i $ is originally positive, we can set all the bias terms $\{ a, b, b' \}$ to be zero. 

\subsection{Heisenberg Model}
\label{section_dbm_Heisenberg} 

We now consider the case of the Heisenberg model, whose Hamiltonian reads
\begin{eqnarray}
{\mathcal H} &=&\sum_{\langle lm \rangle} {\mathcal H}_{lm}\\
{\mathcal H}_{lm}&=& {\mathcal H}_{lm}^z+{\mathcal H}_{lm}^{xy} \\
{\mathcal H}_{lm}^z&=&   J_{lm}^z\sigma_{l}^{z}\sigma_{m}^{z}  \\
{\mathcal H}_{lm}^{xy}&=& J_{lm}^{xy} (\sigma_{l}^{x}\sigma_{m}^{x} + \sigma_{l}^{y}\sigma_{m}^{y}) =  2J_{lm}^{xy}(\sigma_{l}^{+}\sigma_{m}^{-}+\sigma_{l}^{-}\sigma_{m}^{+})
\label{Heisenberg}
\end{eqnarray} 
with $J_{lm}^z=J_{lm}^{xy}=J$. 
We write the Hamiltonian in a general form because the following DBM algorithm can be straightforwardly extended to the more general case of
anisotropic/disordered bonds.
As a starting point for our construction, we decompose the Hamiltonian into pieces by a Trotter-Suzuki decomposition of the imaginary-time
propagator: $e^{-\delta_{\tau} {\mathcal H}}\sim \prod_{\langle lm \rangle}e^{-\delta_{\tau} {\mathcal H}_{lm}}+O({\delta_\tau}^2)$.
Then in this Section, we represent $e^{-\delta_{\tau} {\mathcal H}_{lm}}$ by using the DBM in three different forms, which are all exact. By taking $\delta_{\tau}$ small enough and operating $e^{-\delta_{\tau} {\mathcal H}_{lm}}$ many times, those constructions ensure that the ground state is obtained with any controlled accuracy.

For $e^{-\delta_{\tau}{\mathcal H}_{lm}}$, 
and the antiferromagnetic exchange $J_{lm}^z,J_{lm}^{xy}>0$,
if the lattice is bipartite, we further apply a local gauge transformation by $\pi$ rotation around $z$ axis in the spin space as
\begin{equation}
\sigma^x \rightarrow -\sigma^x
\ \ \ \  {\rm and} \ \ \ \ \sigma^y \rightarrow -\sigma^y
\label{gauge0}
\end{equation}
 on one of the sublattices, which gives a $-$ sign for the $\sigma_{l}^{x}\sigma_{m}^{x}$ and $\sigma_{l}^{y}\sigma_{m}^{y}$ interactions. It is equivalent to the following transformation in the couplings:
\begin{equation}
J_{lm}^{xy}\rightarrow -J_{lm}^{xy}.
\label{gauge}
\end{equation}
The gauge transformation enables to design a DBM neural network with real couplings $\{ W, W' \}$ 
except for those necessary to enforce local constraints on the values of deep neuron spins (see more detail about the constraint in the following sections). Overall, we show in the following that the 3 different DBM constructions have no negative sign problem.  

On the bipartite lattice, the Suzuki-Trotter decomposition is frequently expressed by decomposing the Hamiltonian 
${\mathcal H}$ into several groups. For instance, on the one dimensional chain, if it is natural to decompose it into odd and even bonds: 
\begin{equation}
{\mathcal H}_{1}=\sum_{\langle l,m\rangle \in {\rm odd} \  {\rm bond}}
{\mathcal H}_{lm}
, \ \ \ 
{\mathcal H}_{2}=\sum_{\langle l,m\rangle \in {\rm even} \ {\rm bond}}{\mathcal H}_{lm}, 
\end{equation}
further decompositions  
$e^{-\delta_{\tau}{\mathcal H}_1}=\prod_{\langle l,m \rangle \in \ {\rm odd \ bond}  }e^{-\delta_{\tau} {\mathcal H}_{lm}}$
and 
$e^{-\delta_{\tau}{\mathcal H}_2}=\prod_{\langle l,m \rangle \in \ {\rm even \ bond}  }e^{-\delta_{\tau} {\mathcal H}_{lm}}$
contain commuting elements and are therefore exact. For the square lattice, a similar procedure requires the decomposition of the Hamiltonian into 4 parts, in a checkerboard fashion.
In all cases, the fundamental ingredient to represent the ground-state as a DBM is to find an exact expression for the bond propagator, $e^{-\delta_{\tau} {\mathcal H}_{lm}}$, when applied to an existing DBM state. 

In the case of antiferromagnetic Heisenberg model after the gauge transformation on the bipartite lattice, we must solve, 
for each bond, 
\begin{multline}
\langle \sigma^z | e^{\delta_{\tau}J_{lm}^{xy}\left(\sigma_{l}^{x}\sigma_{m}^{x}+\sigma_{l}^{y}\sigma_{m}^{y}\right)
-\delta_{\tau}J_{lm}^z\sigma_{l}^{z}\sigma_{m}^{z}}
| \Psi_{\mathcal{W}}\rangle \\
= \delta_{\sigma_{l}^{z},\sigma_{m}^{z}}e^{-\delta_{\tau}J_{lm}^z}\Psi_{\mathcal{W}}(\sigma^z) 
+(1-\delta_{\sigma_{l}^{z},\sigma_{m}^{z}})e^{\delta_{\tau}J_{lm}^z}\left(\Psi_{\mathcal{W}}(\sigma^z)\times\cosh(2J_{lm}^{xy}\delta_{\tau})+\Psi_{\mathcal{W}}(\sigma_{l}^{z}\leftrightarrow\sigma_{m}^{z})\times\sinh(2J_{lm}^{xy}\delta_{\tau})\right)\\
= C\langle\sigma^z|\Psi_{\bar{\mathcal{W}}}\rangle.
\label{Eq.bond_operator}
\end{multline}
It is also useful to explicitly write the expression for the exchange
term in the second line above: 
\begin{multline}
\Psi_{\mathcal{W}}(\sigma^z )\times\cosh(2J_{lm}^{xy}\delta_{\tau})+\Psi_{\mathcal{W}}(\sigma_{l}^{z}\leftrightarrow\sigma_{m}^{z})\times\sinh(2J_{lm}^{xy}\delta_{\tau})\\
=\sum_{\{h,d\}}P_{1}(\sigma^z,h)P_{2}(h,d)\left[\cosh(2J_{lm}^{xy}\delta_{\tau})+\sinh(2J_{lm}^{xy}\delta_{\tau})e^{(\sigma_{m}^{z}-\sigma_{l}^{z})\sum_{j}h_{j}\left(W_{lj}-W_{mj}\right)}\right].
\end{multline}
In the following derivations, for the antiferromagnetic Hamilonian ($J_{lm}^z, J_{lm}^{xy} >0$)
after the gauge transformation, we look for a solution with zero bias terms ($a_i,\  b_j, \ b'_k = 0$, $\forall i, j, k$). 
We can also derive a sign-problem free solution for the imaginary time evolution in the absence of the explicit gauge transformation by introducing complex bias term $a_i$. 
Indeed, in the ``\emph{2 deep, 4 hidden}" representation in Sec.~\ref{section-two-four-rep}, 
we will explicitly show that taking a specific set of complex bias term 
$a_i$ on physical spins is equivalent to 
the gauge transformation, making a solution free from the sign problem. 

\subsubsection{\emph{1 deep, 3 hidden} (1d-3h) representation}
\label{section-one-three-rep}
\noindent
{\bf Strategy.}
The first representation we propose is obtained adding one deep neuron
$d_{[lm]}$, 
which gives new couplings $W'_{j[lm]}$ to the hidden units $h_j$ connected to $\sigma_l^z$ and $\sigma_m^z$. 
We also allow for changes
in the existing DBM parameters. 
In particular we set the new couplings to be
$\bar{W}_{lj}=W_{lj}+\Delta W_{lj}$, (with $\Delta W_{lj}$ to be
determined). 
We introduce a coupling $W_{l[lm]}^{\prime\prime}$
between
$\sigma_l^z$
and 
$d_{[lm]}$,  
and a coupling $V_{[lm]}$
between $\sigma_l^z$ and $\sigma_m^z$, which are both not allowed in the DBM architecture. 
By using the gadget Eq.~(\ref{W''1}), 
these interactions can be mediated by hidden neurons 
$h_{[lm1]}$ and $h_{[lm2]}$, respectively, and the DBM form is recovered. 
Furthermore, we look for a solution with a constraint: 
$d_{[lm]}=\sigma^z_{l}$  when $\sigma^z_{l}=\sigma^z_{m}$
(when $\sigma^z_{l}\neq \sigma^z_{m}$, the $d_{[lm]}$ value is not constrained).
Imposing the constraint on the value of the deep unit 
is a crucial difference from the DBM solution for the TFI model.
We will show that this constraint can be achieved by adding additional hidden neuron $h_{[lm3]}$
and introducing complex couplings (``$i\pi/6$" trick).  
We discuss this trick in more detail later. 

In total, we introduce one deep and three hidden neurons. 
After tracing out the three hidden neurons $h_{[lm1]}$, $h_{[lm2]}$, and $h_{[lm3]}$, 
the new wave function reads 
\begin{eqnarray}
\Psi_{\bar{\mathcal{W}}}(\sigma^z) & = & \sum_{\{h,d\}}\sum_{ 
\substack{d_{[lm]}  = \pm1 \\ d_{[lm]}= \sigma_l^z \ {\rm if } \ \sigma_l^z = \sigma_m^z } 
}P_{1}(\sigma^z,h)P_{2}(h,d)e^{\sigma^z_{l}\sum_{j}\Delta W_{lj}h_{j}+d_{[lm]}\sum_{j}h_{j}W_{j[lm]}^{\prime}+d_{[lm]}\sigma^z_{l}W_{l[lm]}^{\prime\prime}+V_{[lm]}\sigma^z_{l}\sigma^z_{m}}.
\label{new_wf_1d3h}
\end{eqnarray}
\\

\noindent
{\bf Derivation for the update of parameters.}
The equations to be verified are then obtained considering 
all the
possible values of $\sigma^z_{l}=\pm1$ and $\sigma^z_{m}=\pm1$, in addition
to the constraints on $d_{[lm]}$ previously introduced. 
We then have two equations for 
$\sigma^z_{l} = \sigma^z_{m} = \pm1$: 
\begin{eqnarray}
e^{\sum_{j}\left(\Delta W_{lj}+W_{j[lm]}^{\prime}\right)h_{j}+W_{l[lm]}^{\prime\prime}+V_{[lm]}} & = & C\times\exp(-J_{lm}^z\delta_{\tau})\\
e^{\sum_{j}\left(-\Delta W_{lj}-W_{j[lm]}^{\prime}\right)h_{j}+W_{l[lm]}^{\prime\prime}+V_{[lm]}} & = & C\times\exp(-J_{lm}^z\delta_{\tau}),
\end{eqnarray}
and the other two equations for 
$\sigma^z_{l} = -\sigma^z_{m} = \pm1$:
\begin{multline}
e^{\sum_{j}\left(\Delta W_{lj}+W_{j[lm]}^{\prime}\right)h_{j}+W_{l[lm]}^{\prime\prime}-V_{[lm]}}+e^{\sum_{j}\left(\Delta W_{lj}-W_{j[lm]}^{\prime}\right)h_{j}-W_{l[lm]}^{\prime\prime}-V_{[lm]}}\\
=C\times\exp(J_{lm}^z\delta_{\tau})\left(\cosh(2J_{lm}^{xy}\delta_{\tau})+\sinh(2J_{lm}^{xy}\delta_{\tau})e^{-2\sum_{j}h_{j}\left(W_{lj}-W_{mj}\right)}\right),
\end{multline}
\begin{multline}
e^{\sum_{j}\left(-\Delta W_{lj}+W_{j[lm]}^{\prime}\right)h_{j}-W_{l[lm]}^{\prime\prime}-V_{[lm]}}+e^{\sum_{j}\left(-\Delta W_{lj}-W_{j[lm]}^{\prime}\right)h_{j}+W_{l[lm]}^{\prime\prime}-V_{[lm]}} \\
=C\times\exp(J_{lm}^z\delta_{\tau})\left(\cosh(2J_{lm}^{xy}\delta_{\tau})+\sinh(2J_{lm}^{xy}\delta_{\tau})e^{2\sum_{j}h_{j}\left(W_{lj}-W_{mj}\right)}\right).
\end{multline}
These equations have a solution if the hidden unit interactions on
the l.h.s. and on the r.h.s match, i.e. when:
\begin{eqnarray}
\Delta W_{lj}+W_{j[lm]}^{\prime} & = & 0\\
\Delta W_{lj}-W_{j[lm]}^{\prime} & = & -2(W_{lj}-W_{mj})
\end{eqnarray}
which implies 
\begin{eqnarray}
\Delta W_{lj} & = & -W_{lj}+W_{mj}\\
W_{j[lm]}^{\prime} & = & W_{lj}-W_{mj}.
\end{eqnarray}
Notice that the first condition 
gives $\bar{W}_{lj} = W_{lj} + \Delta W_{lj} = W_{mj}$, 
which is equivalent to cutting all connections
from spin $l$ to the hidden units and attaching the spin $l$ to
all the hidden units connected to spin $m$, with an interaction $W_{mj}$. 

In order to match the coefficients we must also have: 
\begin{eqnarray}
W_{l[lm]}^{\prime\prime}+V_{[lm]} & = & \log C-J_{lm}^z\delta_{\tau}\\
W_{l[lm]}^{\prime\prime}-V_{[lm]} & = & \log C+\log\cosh(2J_{lm}^{xy}\delta_{\tau})+J_{lm}^z\delta_{\tau}\\
-W_{l[lm]}^{\prime\prime}-V_{[lm]} & = & \log C+\log\sinh(2J_{lm}^{xy}\delta_{\tau})+J_{lm}^z\delta_{\tau},
\end{eqnarray}
which has the solution:
\begin{eqnarray}
W_{l[lm]}^{\prime\prime} & = & -\left(\log\tanh(2J_{lm}^{xy}\delta_{\tau})\right)/2\\
V_{[lm]} & = & -\left(\log\cosh(2J_{lm}^{xy}\delta_{\tau})\right)/2-J_{lm}^{z}\delta_{\tau}
\end{eqnarray}
\\

\noindent
{\bf Recovery of standard DBM}.
The coupling $W_{l[lm]}^{\prime\prime}$ between the deep unit $d_{[lm]}$ 
and the visible spin $\sigma_l^z$ is mediated by the hidden unit $h_{[lm1]}$ coupled to 
$\sigma_l^z$ by $W_{l[lm1]}$ and 
$d_{[lm]}$ by $W'_{[lm1][lm]}$: 
\begin{eqnarray} 
\exp( \sigma_l^z d_{[lm]} W_{l[lm]}^{\prime\prime} ) = 
C_{[lm1]} \sum_{h_{[lm1]} } \exp( \sigma_l^z h_{[lm1]} W_{l[lm1]} + h_{[lm1]}  d_{[lm]}W'_{[lm1][lm]} ).
\end{eqnarray}
By using Eq. (\ref{W''1}) with the following replacement $s_1 \rightarrow \sigma^z_{l}$, $s_2  \rightarrow d_{[lm]}$, $s_3  \rightarrow h_{[lm1]}$, $V  \rightarrow W''_{l[lm]}$, $\tilde{V}_1  \rightarrow W_{l[lm1]}$ and $\tilde{V}_2  \rightarrow W'_{[lm1][lm]}$,
$W_{l[lm1]}$ and $W'_{[lm1][lm]}$ are given by 
\begin{eqnarray}
  W_{l[lm1]} = W'_{[lm1][lm]} = \frac{1}{2} {\rm arcosh}  \left(  \frac{1}{\tanh(2J_{lm}^{xy}\delta_{\tau})} \right).
\end{eqnarray}

Similarly, the coupling $V_{[lm]}$ between visible spins $\sigma_l^z$ and $\sigma_m^z$ 
is mediated by the hidden unit $h_{[lm2]}$ coupled to 
$\sigma_l^z$ by $W_{l[lm2]}$ and 
$\sigma_m^z$ by $W_{m[lm2]}$: 
\begin{eqnarray}
 \exp( \sigma_l^z \sigma_m^z V_{[lm]} ) = 
 C_{[lm2]} \sum_{h_{[lm2]} } \exp( \sigma_l^z h_{[lm2]} W_{l[lm2]} + \sigma_m^z h_{[lm2]} W_{m[lm2]} ).
\end{eqnarray}
By using Eq. (\ref{W''1}) with the following replacement $s_1 \rightarrow \sigma^z_{l}$, $s_2  \rightarrow \sigma^z_m$, $s_3  \rightarrow h_{[lm2]}$, $V  \rightarrow V_{lm}$, $\tilde{V}_1  \rightarrow W_{l[lm2]}$ and $\tilde{V}_2  \rightarrow W_{m[lm2]}$,
$W_{l[lm2]}$ and $W_{m[lm2]}$ are given by 
\begin{eqnarray}
  W_{l[lm2]} =  - W_{m[lm2]} = \frac{1}{2} {\rm arcosh}  \left( \cosh(2J_{lm}^{xy}\delta_{\tau})  e^{2J_{lm}^z \delta_\tau} \right).
\end{eqnarray}
\\

\noindent
{\bf How to enforce the constraint \mbox{\boldmath $d_{[lm]} = \sigma^z_l$} when \mbox{\boldmath $\sigma^z_l = \sigma^z_m$}
 (``\mbox{\boldmath $i\pi/6$}" trick).}
The constraint $d_{[lm]} = \sigma^z_l$ when $\sigma^z_l = \sigma^z_m$
can be exactly satisfied by introducing pure complex connections. 
We can replace the sum with the constraint in Eq.~(\ref{new_wf_1d3h}) as follows
(we ignore trivial constant factor): 
\begin{eqnarray}
 \sum_{  \substack{d_{[lm]}  = \pm1 \\ d_{[lm]}= \sigma_l^z \ {\rm if } \ \sigma_l^z = \sigma_m^z }}
\longrightarrow   \ \ 
\sum_{ d_{[lm]}  } \sum_{h_{[lm3]}  } 
e^{  i \frac{\pi}{6}    \left ( (\sigma^z_{l} + \sigma^z_{m} )   h_{[lm3]}   -   h_{[lm3]}  d_{[lm]}    \right)} 
= \sum_{ d_{[lm]} } 
2 \cos \left ( \frac{\pi}{6}     \left (\sigma^z_{l} + \sigma^z_{m} -  d_{[lm]}  \right )   \right )  
  \end{eqnarray}
One can easily see that the cosine term in the rightmost part gives nonzero value only when $d_{[lm]} = \sigma^z_l$ if $\sigma^z_l = \sigma^z_m$.
On the other hand, if $\sigma^z_l \neq \sigma^z_m$, both $d_{[lm]} \pm 1 $ contributions survive. 
\\

\begin{figure}[tb]
\begin{center}
\includegraphics[width=0.55\columnwidth]{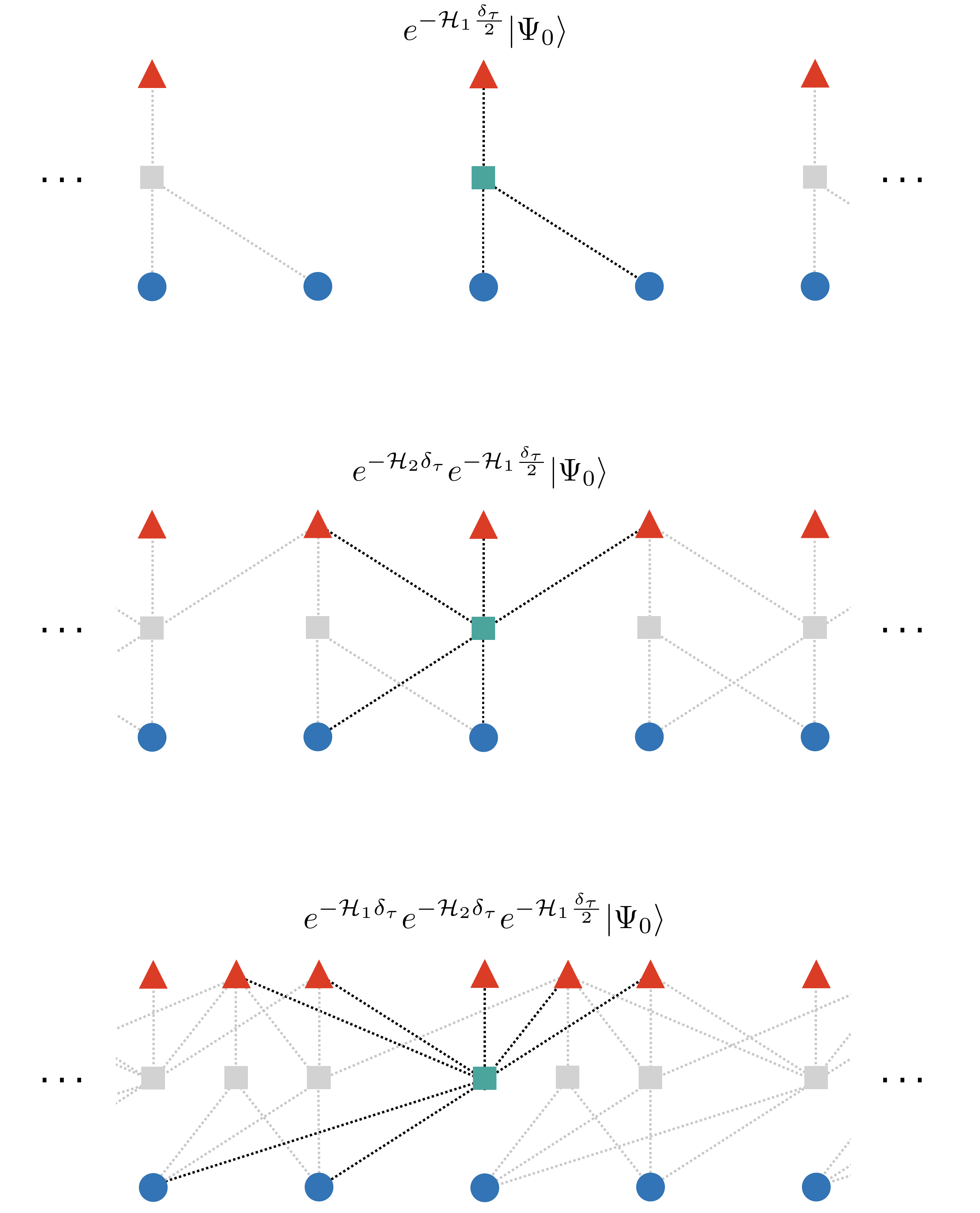}
\caption{
{\bf Imaginary-time evolution of complex couplings in 1d-3h construction for one-dimensional Heisenberg model.}
The figure shows how the complex couplings with weight $\pm i \pi/6$ evolve from an empty RBM ($\langle \sigma^z | \Psi_0 \rangle = \rm{const.}$). 
Dots, squares, triangles indicate physical spins $\sigma_i^z$, hidden neurons $h_j$, and deep neurons $d_k$, respectively. 
For visibility, only hidden neurons having complex couplings and the associated complex couplings are shown. 
Therefore, at each imaginary-time evolution, one hidden neuron (called $h_{[lm3]}$ in the text) appears for each bond.  
One hidden neuron (green) and the associated couplings (black) are highlighted. 
As discussed in Step 2 in Fig. 3, at each evolution on $\sigma_l^z$ and $\sigma_m^z$, the $W$ couplings to $\sigma_l^z$ are cut and $\sigma_l^z$ is reconnected to the hidden neuron coupled to $\sigma_m^z$. By this ``cut and reconnect" procedure, the positions of nonzero $W$ couplings from a specific hidden neuron move, however, the $W$ couplings stay local.
On the other hand, the number of nonzero $W'$ couplings increases by imaginary-time evolution, resulting in non-local structure of $W'$ couplings.        
For the same reason, the real $W$ couplings stay local, whereas the real $W'$ couplings become nonlocal. 
}
\label{fig_timeevolution_one_deep}
\end{center}
\end{figure}

\noindent
{\bf Proof of no negative sign}.
Here, we show that the marginal probability density  
$\tilde{\Pi}'(\sigma^{z},d,d^{\prime})=\sum_{h,h^{\prime}}\Pi(\sigma^{z},h,h^{\prime},d,d^{\prime})$
obtained by tracing out the hidden unit 
is non-negative definite. 
Therefore, we can perform the Metropolis sampling using $\tilde{\Pi}'$ density without suffering from the negative signs
(see more detail on the sampling scheme in Sec.~\ref{sec_Metropolis_sampling}).
To prove this, it is sufficient to show 
\begin{eqnarray}
\sum_{ \{h \} }  P_1 (\sigma^z, h) P_2 (h, d) \geq 0.
\label{eq_no_nega}
\end{eqnarray} 
for all possible $\sigma^z$ and $d$ configurations. 

In the 1d-3h representation, $i \pi / 6 $ complex couplings are originally introduced to put the constraint locally. 
However, as time evolves, these complex couplings become non-local
(see Fig.~\ref{fig_timeevolution_one_deep}).
Because the pure complex couplings give cosine terms after tracing out hidden variables, 
they have a potential to give negative signs. 
Here, we prove that this is not the case. 

We assume that Eq. (\ref{eq_no_nega}) is satisfied for all possible $\sigma^z$ and $d$ after several steps of the imaginary time evolution.
Then, we apply the bond propagator $e^{-{\mathcal H}_{lm} \delta_\tau}$ to obtain the new wave function. 
In the case when $\sigma^z_l = -\sigma^z_m = 1$, the solution in the 1d-3h representation can be rewritten as  
\begin{eqnarray}
   \sum_{ h_{[lm1]} , h_{[lm2]} , h_{[lm3]}} 
 \bar{P}_1 (\sigma^z,  \bar{h} ) \bar{P}_2 (  \bar{h}  , d, d_{[lm]} = 1)  &=&  P_1 (\sigma^z, h) P_2 (h, d ) \times ({\rm positive \ constant}) \\ 
  \sum_{   h_{[lm1]} , h_{[lm2]} , h_{[lm3]}   } 
 \bar{P}_1 (\sigma^z,   \bar{h}) \bar{P}_2 ( \bar{h} , d, d_{[lm]} = -1 )  &=&  P_1 (\sigma_{l}^{z}\leftrightarrow\sigma_{m}^{z}, h) P_2 (h, d ) \times  ({\rm positive \ constant}) 
\end{eqnarray}
where $\bar{P}_1 \times \bar{P}_2$ on the left hand side is the new weight after the imaginary time evolution, 
and $\{ \bar{h} \}$ consists of the existing hidden neurons $\{h \}$ and 
the newly introduced hidden neurons $h_{[lm1]}$, $h_{[lm2]}$, and $h_{[lm3]}$. 
By taking the summation on the existing hidden variables on both sides, we get 
\begin{eqnarray}
     \sum_{ \{ \bar{h} \}  } \bar{P}_1 (\sigma^z,  \bar{h}) \bar{P}_2 ( \bar{h}, d, d_{[lm]} = 1)  &=&  \sum_{ \{ h \}} 
     P_1 (\sigma^z,  h) P_2 (h, d ) \times ({\rm positive \ constant}) \geq  0  \\ 
  \sum_{\{ \bar{h} \}  }  \bar{P}_1 (\sigma^z, \bar{h} ) \bar{P}_2 ( \bar{h} , d, d_{[lm]} = -1)  &=&  \sum_{ \{ h \} } P_1 (\sigma_{l}^{z}\leftrightarrow\sigma_{m}^{z}, h) P_2 (h, d ) \times ({\rm positive \ constant})   \geq 0  
\end{eqnarray}
Here, we used Eq. (\ref{eq_no_nega}) to obtain the rightmost inequality. 
It proves that the new weight with the hidden variables being traced out is also non-negative. 
In the same way, we can show the non-negativeness of the new weight for $\sigma_l = -\sigma_m = -1$.

Next we consider the case $\sigma^z_l = \sigma^z_m = 1$. 
In this case, 
\begin{eqnarray}
 \sum_{   h_{[lm1]} , h_{[lm2]} , h_{[lm3]}   }
\bar{P}_1 (\sigma^z,  \bar{h}) \bar{P}_2 ( \bar{h}, d, d_{[lm]} = 1)  &=&  P_1 (\sigma^z, h) P_2 (h, d ) \times ({\rm positive \ constant}), \\ 
 \sum_{   h_{[lm1]} , h_{[lm2]} , h_{[lm3]}   }
 \bar{P}_1 (\sigma^z,  \bar{h} ) \bar{P}_2 ( \bar{h}, d, d_{[lm]} = -1)  &=& 0.
\end{eqnarray}
By taking the summation on the existing hidden variables on both sides, we obtain
\begin{eqnarray}
  \sum_{ \{ \bar{h}  \} } \bar{P}_1 (\sigma^z,  \bar{h} ) \bar{P}_2 ( \bar{h} , d, d_{[lm]} = 1)  &=&  \sum_h P_1 (\sigma^z, h) P_2 (h, d ) \times ({\rm positive \ constant}) \geq  0  \\ 
  \sum_{ \{ \bar{h} \}  }  \bar{P}_1 (\sigma^z,  \bar{h} ) \bar{P}_2 ( \bar{h} , d, d_{[lm]} = -1)  &=&  0.
  \end{eqnarray}
Therefore, the non-negativeness of the weight is ensured. 
The proof for $\sigma^z_l = \sigma^z_m = -1$ case can be done in an analogous way.  

We have proven that the new weight after applying the bond propagator $e^{-{\mathcal H}_{lm} \delta_\tau}$ 
is non negative for all the possible $\sigma^z$ and $\bar{d}$ configurations:
\begin{eqnarray}
 \sum_{ \{ \bar{h} \} }  \bar{P}_1 (\sigma^z,  \bar{h} ) \bar{P}_2 ( \bar{h}, \bar{d} ) \geq 0
\end{eqnarray}
with $\{ \bar{d} \}$ consisting of $\{ d \}$ and $d_{[lm]}$. 
It ensures the non-negativeness of the weight at any time during the imaginary time evolution.   
\\

\noindent
{\bf Summary of 1d-3h representation}.
The action the bond propagator is summarized as follows. 
First, the new deep neuron $d_{[lm]}$ is attached to the existing hidden neurons connected to $\sigma_{l}^{z}$ and $\sigma_{m}^{z}$. 
Second, $\sigma_{l}^{z}$ is disconnected to all hidden units 
and reconnected to the hidden units having finite couplings to $\sigma_{m}^{z}$ 
($\bar{W}_{lj} =  W_{mj}$).
Third, four couplings are inserted, involving new hidden neurons $h_{[lm1]}$ and 
$h_{[lm2]}$: $\sigma_l^z \leftrightarrow h_{[lm1]}$,  
$h_{[lm1]} \leftrightarrow d_{[l]}$, $\sigma_l^z \leftrightarrow h_{[lm2]}$ and $\sigma_m^z \leftrightarrow h_{[lm2]}$,
Finally, the new hidden neuron $h_{[lm3]}$ puts the constraint on the $d_{[l]}$ sum by the imaginary couplings to 
$\sigma_l^z$, $\sigma_m^z$, and $d_{[l]}$.

By successively applying the imaginary-time evolutions, the $W'$ couplings become nonlocal or long ranged.  
On the other hand, the $W$ couplings stay local (see Fig. \ref{fig_timeevolution_one_deep}). 

\subsubsection{\emph{2 deep, 6 hidden} (2d-6h) representation}
\label{section-two-six-rep}
\noindent
{\bf Strategy.}
We look for a solution where we add two deep neurons $d_{[l]}$ and $d_{[m]}$,
giving new couplings $W'_{j[l]},\ W'_{j[m]}$ to the existing hidden spins $h_j$ connected to $\sigma_l^z$ and $\sigma_m^z$. 
We also allow for changes in the existing  
$W$ parameters:  
We set the new couplings to be $\bar{W}_{lj} =W_{lj}+\Delta W_{lj}$ and 
$\bar{W}_{mj} =W_{mj}+\Delta W_{mj}$
(with $\Delta W_{lj}$,   $\Delta W_{mj}$ to be determined).
Furthermore, we add four hidden neurons $h_{[lm1]}$, $h_{[lm2]}$, $h_{[lm3]}$, and $h_{[lm4]}$ to mediate the interactions between $(\sigma^z_l, d_{[l]})$, $(\sigma^z_m, d_{[m]})$, $(\sigma^z_l, d_{[m]})$, and $(\sigma^z_m, d_{[l]})$, respectively.   
We solve the equation with the constraint $\sigma^z_l + \sigma^z_m = d_{[l]} + d_{[m]}$. 
This constraint can be achieved, for example, by adding two further hidden neurons ($h_{[lm5]}$ and $h_{[lm6]}$, respectively) and introducing complex connections  (``$i \pi/4, i\pi/8$" trick).
This trick will be discussed in detail later.  

In total, we add two deep neurons ($d_{[l]}$ and $d_{[m]}$) and six hidden neurons ($h_{[lm1]}, \ldots, h_{[lm6]}$).
In the following, to make equations simple, we employ a representation in which the new hidden neurons are analytically traced out. 
The interactions between $(\sigma_l^z, d_{[l]})$, $(\sigma_m^z, d_{[m]})$, $(\sigma_l^z, d_{[m]})$, and $(\sigma_m^z, d_{[l]})$, 
which are mediated by 1st to 4th hidden neurons, will be denoted as $W''_{l[l]}$, $W''_{m[m]}$, $W''_{l[m]}$, and $W''_{m[l]}$, respectively. 
The 5th and 6th hidden neurons filter out $\sigma^z_l + \sigma^z_m \neq d_{[l]} + d_{[m]}$ contributions. 
With this setting, the new wave function is represented as 
\begin{eqnarray}
\Psi_{\bar{\mathcal{W}}}(\sigma^z) =  \sum_{\{h,d\}}\sum_{\substack{d_{[l]}, d_{[m]} \\ d_{[l]} + d_{[m]} = \sigma^z_l + \sigma^z_m  } }
\! \! \! \!  &&
P_{1}(\sigma^z,h)P_{2}(h,d) \
e^{  \sum_{j}   h_{j}  (  \Delta W_{lj}   \sigma^z_{l}   + W_{j[l]}^{\prime} d_{[l]} ) 
+ \sum_{j}   h_{j}  (   \Delta W_{mj}  \sigma^z_{m}  + W_{j[m]}^{\prime} d_{[m]} ) }  \nonumber \\
&& \times \ e^{  \sigma^z_{l} (W''_{l[l]} d_{[l]} + W''_{l[m]} d_{[m]}) 
+   \sigma^z_{m} (W''_{m[l]} d_{[l]} +  W''_{m[m]} d_{[m]})  }.
\label{wf_new_2d6h}
\end{eqnarray}
\\

\noindent
{\bf Derivation for the update of parameters.}
When the $l$th and $m$th physical spins are anti-parallel ($\sigma^z_{l}  = - \sigma^z_{m} = \pm1$), 
$d_{[l]}  =  - d_{[m]}  = \pm 1 $ contributions survive in the sum over $d_{[l]}$ and $d_{[m]}$ variables
in Eq.~(\ref{wf_new_2d6h}), and thus the equations to be satisfied are
\begin{multline}
 e^{\sum_{j}h_{j}\left(\Delta W_{l-m,j}+W_{j[l]}^{\prime} - W_{j[m]}^{\prime} \right) +  W_{l-m,[l]}^{\prime\prime} - W_{l-m,[m]}^{\prime\prime} }  
 + e^{\sum_{j}h_{j}\left(\Delta W_{l-m,j}-W_{j[l]}^{\prime} + W_{j[m]}^{\prime}\right)-W_{l-m,[l]}^{\prime\prime} + W_{l-m,[m]}^{\prime\prime}  }
 \\  =   
 C e^{J_{lm}^z \delta_{\tau}} \left( \cosh( 2 J_{lm}^{xy} \delta_\tau )  + \sinh( 2 J_{lm}^{xy} \delta_\tau )  e^{-2\sum_{j}h_{j}W_{l-m,j}}\right)
\label{Eq:anti1} 
\end{multline}
for $\sigma^z_{l}  = - \sigma^z_{m} = 1$ and 
\begin{multline}
e^{\sum_{j}h_{j}\left( - \Delta W_{l-m,j}+W_{j[l]}^{\prime} - W_{j[m]}^{\prime} \right) -   W_{l-m,[l]}^{\prime\prime} + W_{l-m,[m]}^{\prime\prime}   }  
 + e^{\sum_{j}h_{j}\left(- \Delta W_{l-m,j}-W_{j[l]}^{\prime} + W_{j[m]}^{\prime}\right) + W_{l-m,[l]}^{\prime\prime} - W_{l-m,[m]}^{\prime\prime} }
 \\ = 
C e^{J_{lm}^z \delta_{\tau}} \left( \cosh( 2 J_{lm}^{xy} \delta_\tau )  + \sinh( 2 J_{lm}^{xy} \delta_\tau )  e^{2\sum_{j}h_{j}W_{l-m,j}}\right) 
\label{Eq:anti2} 
\end{multline}
for $\sigma^z_{l}  = - \sigma^z_{m} = -1$, respectively. 
Here, $W_{l-m,[\alpha]} = W_{l[\alpha]} - W_{m[\alpha]} $,  $\Delta W_{l-m,[\alpha]} = \Delta W_{l[\alpha]} - \Delta W_{m[\alpha]} $, $W_{l-m,[\alpha]}^{\prime\prime} = W_{l[\alpha]}^{\prime\prime} - W_{m[\alpha]}^{\prime\prime}$ with $\alpha = l, m$.

When the $l$th and $m$th physical spins are parallel ($\sigma^z_{l}  = \sigma^z_{m} = \pm1$), 
only $d_{[l]} \! = \! d_{[m]} \! = \! \sigma^z_{l}  \! = \! \sigma^z_{m}$ contribution survives in the sum over $d_{[l]}$ and $d_{[m]}$ variables in Eq.~(\ref{wf_new_2d6h}), and thus the equations to be satisfied are
\begin{eqnarray}
 e^{\sum_{j}h_{j}\left(\Delta W_{l+m,j}+W_{j[l]}^{\prime} + W_{j[m]}^{\prime} \right) +  W_{l+m,[l]}^{\prime\prime} + W_{l+m,[m]}^{\prime\prime}  }  =  C e^{- J_{lm}^z  \delta_\tau}
 \label{Eq:para1} 
 \end{eqnarray}
 for $\sigma^z_{l}  = \sigma^z_{m} =  1$
 \begin{eqnarray}
 e^{\sum_{j}h_{j}\left(-\Delta W_{l+m,j} - W_{j[l]}^{\prime}  - W_{j[m]}^{\prime} \right) +  W_{l+m,[l]}^{\prime\prime} + W_{l+m,[m]}^{\prime\prime}  }  
 =  C e^{- J_{lm}^z  \delta_\tau}
 \label{Eq:para2}  
 \end{eqnarray}
 for $\sigma^z_{l}  = \sigma^z_{m} =  -1$, respectively. 
 Here, $W_{l+m,[\alpha]} = W_{l[\alpha]} + W_{m[\alpha ]} $,  $\Delta W_{l+m,[\alpha]} = \Delta W_{l[\alpha]} + \Delta W_{m[\alpha]} $, $W_{l+m,[\alpha]}^{\prime\prime} = W_{l[\alpha]}^{\prime\prime} + W_{m[\alpha]}^{\prime\prime}$ with $\alpha =l, m$. 

The equations (\ref{Eq:anti1}),  (\ref{Eq:anti2}),  (\ref{Eq:para1}), and (\ref{Eq:para2}) are satisfied if 
\begin{eqnarray}
\Delta W_{l-m,j} - W_{j[l]}^{\prime} + W_{j[m]}^{\prime} & = & -2W_{l-m,j},\\
\Delta W_{l-m,j} + W_{j[l]}^{\prime}  - W_{j[m]}^{\prime}& = & 0, \\ 
\Delta W_{l+m,j} + W_{j[l]}^{\prime} + W_{j[m]}^{\prime} & = & 0,
\end{eqnarray}
and  
\begin{eqnarray}
  W_{l-m,[l]}^{\prime\prime}   - W_{l-m,[m]}^{\prime\prime}  & = & \log C +  J_{lm}^z \delta_\tau +  \log     \cosh(2J_{lm}^{xy}  \delta_\tau  ), \\ 
 - W_{l-m,[l]}^{\prime\prime}  + W_{l-m,[m]}^{\prime\prime}  & = & \log C +  J_{lm}^z \delta_\tau +  \log    \sinh(2J_{lm}^{xy}  \delta_\tau) , \\ 
W_{l+m,[l]}^{\prime\prime} + W_{l+m,[m]}^{\prime\prime} & = & \log C -J_{lm}^z  \delta_\tau.
\end{eqnarray}
These conditions give
\begin{eqnarray}
W_{j[l]}^{\prime} & = & W_{lj}, \label{eq:wprimex1}\\
W_{j[m]}^{\prime} & = & W_{mj}, \label{eq:wprimex2}\\
\Delta W_{lj} & = & - W_{lj}, \label{Eq:wlj_change}\\
\Delta W_{mj} & = &  - W_{mj},
\label{Eq:wmj_change}
\end{eqnarray}
and 
\begin{eqnarray}
W_{l [l]}^{\prime\prime}  =    W_{m [m] }^{\prime\prime}  &=& - \frac{J_{lm}^z \delta_\tau}{2}   -  \frac{1}{4} \log \sinh(2J_{lm}^{xy} \delta_\tau ),
\label{Eq.Wpp_ll} \\ 
W_{l [m]}^{\prime\prime}  =    W_{m[l] }^{\prime\prime}  &=&   -\frac{J_{lm}^z  \delta_\tau}{2} - \frac{1}{4} \log \cosh(2J_{lm}^{xy} \delta_\tau ). 
\label{Eq.Wpp_lm}
\end{eqnarray}
\\

\noindent
{\bf Recovery of standard DBM}. The direct interactions between $(\sigma_l^z, d_{[l]})$, $(\sigma_m^z, d_{[m]})$, $(\sigma_l^z, d_{[m]})$, and $(\sigma_m^z, d_{[l]})$, 
are mediated by $h_{[lm1]}$,  $h_{[lm2]}$,  $h_{[lm3]}$,  and $h_{[lm4]}$, respectively, as follows 
\begin{eqnarray} 
\exp( \sigma_l^z d_{[l]} W_{l [l]}^{\prime\prime}  ) &=& 
C_{[lm1]} \sum_{h_{[lm1]} } \exp( \sigma_l^z h_{[lm1]} W_{l[lm1]} + h_{[lm1]}  d_{[l]}W'_{[lm1][l]} ), \\
\exp( \sigma_m^z d_{[m]} W_{m [m]}^{\prime\prime}  ) &=& 
C_{[lm2]} \sum_{h_{[lm2]} } \exp( \sigma_m^z h_{[lm2]} W_{m[lm2]} + h_{[lm2]}  d_{[m]}W'_{[lm2][m]} ), \\
\exp( \sigma_l^z d_{[m]} W_{l [m]}^{\prime\prime}  ) &=& 
C_{[lm3]} \sum_{h_{[lm3]} } \exp( \sigma_l^z h_{[lm3]} W_{l[lm3]} + h_{[lm3]}  d_{[m]}W'_{[lm3][m]} ), \\
\exp( \sigma_m^z d_{[l]} W_{m [l]}^{\prime\prime}  ) &=& 
C_{[lm4]} \sum_{h_{[lm4]} } \exp( \sigma_m^z h_{[lm4]} W_{m[lm4]} + h_{[lm4]}  d_{[l]}W'_{[lm4][l]} ). 
\end{eqnarray}
By applying the gadget Eq.~(\ref{W''1}), the new $W$, $W'$ interactions are given by, for small $\delta_\tau$
(such that $\frac{e^{-J_{lm}^z \delta_\tau}  } { \sqrt{\sinh(2J_{lm}^{xy}\delta_{\tau}) }} > 1 $): 
\begin{eqnarray}
&  W_{l[lm1]} = W'_{[lm1][l]} = W_{m[lm2]} = W'_{[lm2][m]} = \frac{1}{2} {\rm arcosh}  \left( \frac{e^{-J_{lm}^z \delta_\tau}  } { \sqrt{\sinh(2J_{lm}^{xy}\delta_{\tau}) }} \right), \\ 
&   W_{l[lm3]} = - W'_{[lm3][m]} = W_{m[lm4]} = - W'_{[lm4][l]} = \frac{1}{2} {\rm arcosh}  \left( \sqrt{ \cosh(2J_{lm}^{xy}\delta_{\tau})  } \times  e^{J_{lm}^z \delta_\tau}     \right).
\end{eqnarray}
\\

\noindent
{\bf How to enforce the constraint  \mbox{\boldmath $\sigma^z_l + \sigma^z_m = d_{[l]} + d_{[m]}$} (``\mbox{\boldmath $i\pi/4, i\pi/8$}" trick)}. 
Here, we discuss how to design the network to satisfy the constraint $\sigma^z_l + \sigma^z_m = d_{[l]} + d_{[m]}$. 
We rewrite the sum with the constraint in Eq.~(\ref{wf_new_2d6h}) as follows (we ignore trivial constant factor):
\begin{multline}
\sum_{\substack{d_{[l]}, d_{[m]} \\ d_{[l]} + d_{[m]} = \sigma^z_l + \sigma^z_m  } }
\longrightarrow 
\sum_{  d_{[l]} , d_{[m]} }
\sum_{  h_{[lm5]} , h_{[lm6]} } 
   e^{  i \frac{\pi}{4}    \left ( (\sigma^z_{l} + \sigma^z_{m} )   h_{[lm5]}   -   h_{[lm5]} ( d_{[l]} +  d_{[m]} )   \right)}   \times e^{  i \frac{\pi}{8}    \left ( (\sigma^z_{l} + \sigma^z_{m} )   h_{[lm6]}   -   h_{[lm6]} ( d_{[l]} +  d_{[m]} )   \right)}  \\
 = \sum_{  d_{[l]} , d_{[m]} } 2 \cos \left (  \frac{\pi}{4}    (\sigma^z_{l} + \sigma^z_{m} -  d_{[l]}  - d_{[m]} )  \right) \times 
 2\cos \left (  \frac{\pi}{8}    (\sigma^z_{l} + \sigma^z_{m} -  d_{[l]}  - d_{[m]} )  \right) 
\end{multline}
One can easily see that the second line of the equation 
gives nonzero contribution only when  $d_{[l]} + d_{[m]} = \sigma^z_l + \sigma^z_m $. 
\\


\noindent
{\bf Summary of the 2d-6h  representation.}
The network changes induced by the bond propagator at each imaginary time step are summarized as follows. 
Eqs. (\ref{Eq:wlj_change}) and (\ref{Eq:wmj_change}) imply that $\bar{W}_{lj} = W_{lj} + \Delta W_{lj} = 0$
and  $\bar{W}_{mj}  = W_{mj} + \Delta W_{mj} = 0$, i.e., all the existing connections between physical spins and hidden neurons vanish.
Then, the $l$th and $m$th physical spins will be connected to the new hidden neurons $h_{[lm1]}, \ldots, h_{[lm6]} $, 
The new deep neurons $d_{[l]}$ and $d_{[m]}$ are also connected to 
$h_{[lm1]}, \ldots, h_{[lm6]} $.
In total, we have 16 new connections in the deep Boltzmann network. 

\begin{figure}[h!]
\begin{center}
\includegraphics[width=\textwidth]{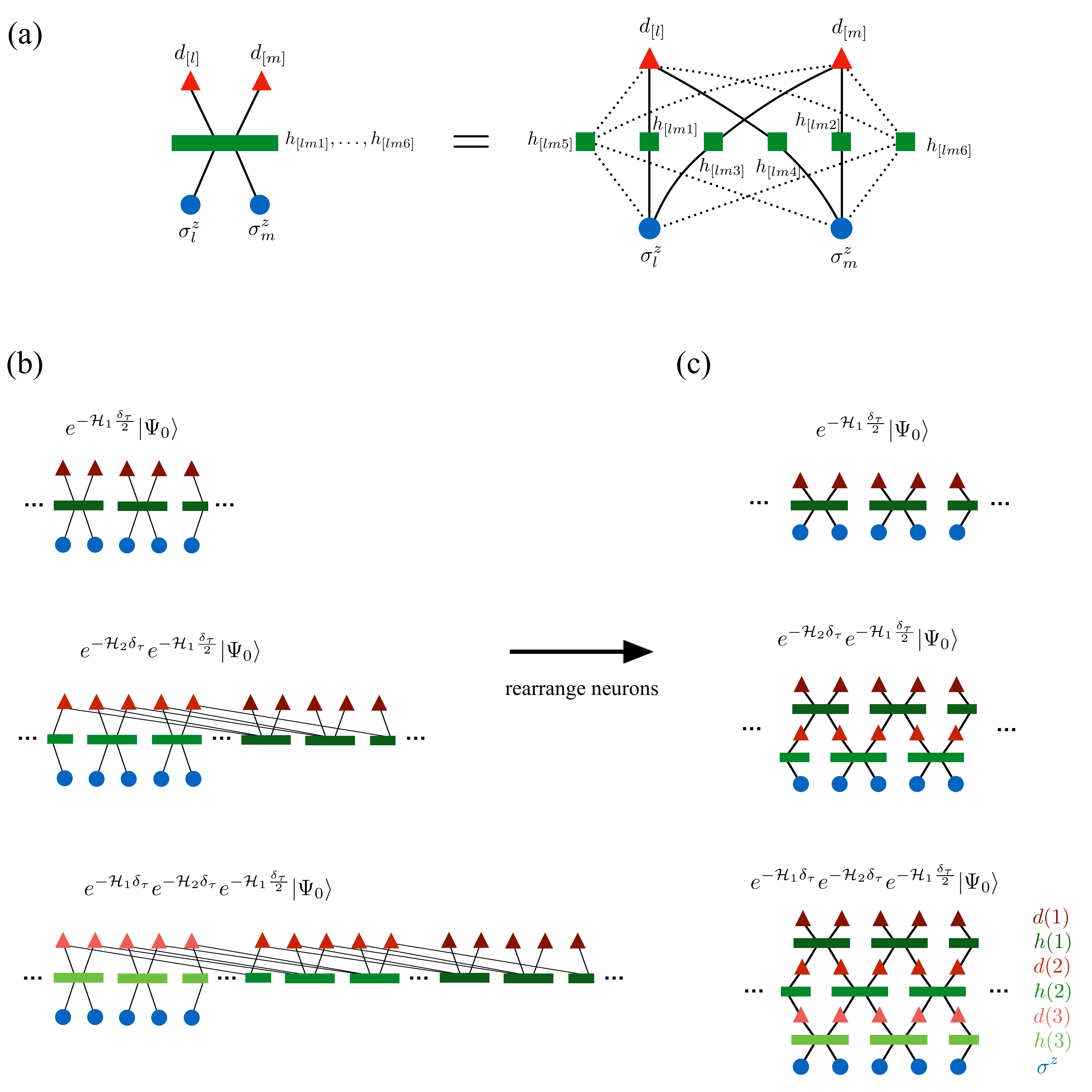}
\caption{
{\bf Schematic picture for imaginary time evolution of DBM neural network in the 2d-6h construction.}
Dots, squares, and triangles indicate physical spins $\sigma_i^z$, hidden neurons $h_j$, and deep neurons $d_k$, respectively.
A set of six hidden neurons are depicted as rectangles. 
(a) Building block of the imaginary-time evolution. 
The left part is a simplified picture of the complete figure in the right part. 
This simplified picture is used in the panels (b) and (c) for the sake of visibility.  
(b) The imaginary time evolution of the network starting from an empty RBM ($\langle \sigma^z | \Psi_0 \rangle = {\rm const.}$).
The hidden neurons introduced at $t$-th step ($h(t)$'s)  lose their connections to physical spins at $(t\!+\!1)$-th step, and instead they get connections to $(t\!+\!1)$-th deep neurons  ($d(t\!+\!1)$'s).  
(c) When we rearrange the neurons, one can see a clear correspondence between the 2d-6h representation and the path-integral formulation (see the text for detail). 
}
\label{fig_timeevolution_two_deep}
\end{center}
\end{figure}

By continuing the imaginary time evolution, the neural network grows as in Fig.~\ref{fig_timeevolution_two_deep}. 
The number of neurons increases linearly with the number $N_{\rm slice}$ of Suzuki-Trotter time slice. 
For example, in the case of the one-dimensional Heisenberg model, the total number of deep and hidden neurons are $N_{\rm site} (2  N_{\rm slice}  +1)$ and $3 N_{\rm site} (2 N_{\rm slice} +1)$, respectively. 
The number of nonzero connections in the network is $8 N_{\rm site} (2  N_{\rm slice}  +1)$. 
The origin of $2 N_{\rm slice} +1$ is coming from the fact that we apply ${\mathcal G}$ propagators $2 N_{\rm slice} +1$ times when we apply the second-order Suzuki-Trotter decomposition.  
The ``$i\pi/4, i\pi/8$" trick plays a role to preserve the total magnetization for deep spins at each imaginary-time step, i.e.,   $\sum_k d_k(t \!+\!1) = \sum_k d_k(t)$,  
where $d(t \!+\!1)$ [$d(t)$] are the deep neurons introduced at $(t\!+\!1)$-th [$t$-th] step.  
\\

\noindent
{\bf Relationship between the 2d-6h representation and the path-integral quantum Monte Carlo method.} 
In the final part of this section, we discuss the similarity between the 2d-6h representation and the imaginary-time path-integral quantum Monte Carlo method~\cite{doi:10.1143/PTP.58.1377}.  
We will show that, in the 2d-6h representation, the deep neurons can be regard as the additional degrees of freedom along the imaginary time in the path-integral formulation. 

In the quantum Monte Carlo simulations using Suzuki-Trotter decomposition~\cite{suzuki,trotter},  the partition function $Z$ is evaluated as 
\begin{eqnarray}
Z  &=&   \langle  \sigma^z (0) |  e^{- \beta {\mathcal H}}  |  \sigma^z(0) \rangle  \nonumber \\ 
&\simeq&  \sum_{\sigma^z(0),\ldots, \sigma^z(2{N_{\rm slice}}-1) }  
 \langle  \sigma^z(0) |   e^{-{\mathcal H}_2 \delta_\tau}   | \sigma^z(2N_{\rm slice} \! - \! 1)  \rangle
 \langle  \sigma^z(2N_{\rm slice} \!-\! 1) |  e^{-{\mathcal H}_1 \delta_\tau }   | \sigma^z (2N_{\rm slice} \!-\!2) \rangle 
 \dots  \nonumber  \\ 
&& \phantom{ \sum_{\sigma^z(0),\ldots, \sigma^z(2{N_{\rm slice}}-1) }   } \dots 
\langle \sigma^z (4)  |  e^{-{\mathcal H}_2 \delta_\tau}  | \sigma^z(3)  \rangle
 \langle \sigma(3)  |
 e^{-{\mathcal H}_1 \delta_\tau}   | \sigma^z(2)  \rangle
  \langle \sigma^z (2) | 
 e^{-{\mathcal H}_2 \delta_\tau}
  | \sigma^z (1) \rangle
  \langle \sigma^z (1) | 
 e^{-{\mathcal H}_1 \delta_\tau } 
 | \sigma^z (0) \rangle \nonumber \\ 
\label{eq:psi_tau_full_2}
\end{eqnarray}
In the evaluation of the matrix element of  $\langle  \sigma^z (t \!+ \!1) |  e^{-{\mathcal H}_\nu \delta_\tau}  | \sigma^z (t) \rangle$ ($\nu=1$ or 2), 
in the case of one-dimensional Heisenberg model, it is sufficient to consider one specific bond, 
$\langle  \sigma^z_l (t \!+ \!1) \sigma^z_m (t \!+ \!1) |  e^{-{\mathcal H}_{lm}  \delta_\tau}  | \sigma^z_l(t) \sigma^z_m (t) \rangle$. The matrix elements are given by
\begin{eqnarray}
\langle  \sigma^z_l (t \!+ \!1) \sigma^z_m (t \!+ \!1) |  e^{-{\mathcal H}_{lm}  \delta_\tau}  | \sigma^z_l(t) \sigma^z_m (t) \rangle = 
e^{ J_{lm}^z \delta_\tau}
\begin{pmatrix} 
 e^{ - 2J_{lm}^z \delta_\tau}   & 0 & 0 & 0 \\
 0 & \cosh(2  J_{lm}^{xy}  \delta_\tau) & \sinh (2  J_{lm}^{xy}  \delta_\tau) & 0 \\
 0 &  \sinh (2  J_{lm}^{xy}  \delta_\tau)   &  \cosh(2  J_{lm}^{xy}  \delta_\tau)  & 0 \\
 0  & 0 & 0 & e^{ - 2J_{lm}^z \delta_\tau} \\ 
\end{pmatrix}
\label{eq_mat_elem}
\end{eqnarray}
in the basis $\{  | \! \!  \uparrow \uparrow \rangle,    | \! \! \uparrow \downarrow \rangle,   |  \! \! \downarrow \uparrow \rangle,   | \!  \! \downarrow \downarrow \rangle \}$. 

On the other hand,
the imaginary time evolution in Eq. (2) in the main text [or equivalently, Eq.~(\ref{eq:psitau})] can be rewritten as 
\begin{eqnarray}
\langle \sigma^z |\Psi (\tau)\rangle & = & \! \! \! \! \sum_{\sigma^z(1),\ldots, \sigma^z(2{N_{\rm slice}} +1) } \!  \!
\langle  \sigma^z |   e^{-  {\mathcal H}_1 \frac{ \delta_\tau} {2}}   |  \sigma^z(2N_{\rm slice} \!+\!1) \rangle
\langle  \sigma^z(2N_{\rm slice} \!+\! 1) |  e^{-{\mathcal H}_2 \delta_\tau}  | \sigma^z (2N_{\rm slice}) \rangle 
 \dots  \nonumber  \\ 
&& 
\phantom {  \sum_{d(1),\ldots, d(2{N_{\rm slice}} \!+\!1) } }
 \dots 
  \langle \sigma^z(4)  | 
 e^{-{\mathcal H}_1 \delta_\tau}
  | \sigma^z(3) \rangle
 \langle \sigma^z (3)  | 
 e^{-{\mathcal H}_2 \delta_\tau}
  | \sigma^z(2 ) \rangle
  \langle  \sigma^z  (2)  | 
 e^{-{\mathcal H}_1 \frac {\delta_\tau }{2}  } 
 | \sigma^z (1) \rangle 
 \langle  \sigma^z (1)  | \Psi_{0} \rangle 
\label{eq:psi_tau_full_1}
\end{eqnarray}
by inserting complete basis sets at each time slice.
The matrix element used here is exactly the same as that of QMC in Eq. (\ref{eq_mat_elem}).
Here, the $D$ dimensional quantum spin system is mapped on the $D+1$ dimensional classical system as in the case of the 
path integral quantum Monte Carlo method.  
Because the neuron spins are defined as the classical Ising-type spins, we can represent the summation over $\sigma^z(1),\ldots, \sigma^z(2{N_{\rm slice}} \!+\!1)$ by the summation over $N_{\rm site} (2  N_{\rm slice}  +1)$ neuron spins.  
Assuming that these $N_{\rm site} (2  N_{\rm slice}  +1)$ neuron spins are in the deep layer,
the imaginary time evolution in Eq.~(\ref{eq:psi_tau_full_1}) reads
\begin{eqnarray}
\langle \sigma^z |\Psi (\tau)\rangle & = & \! \! \! \! \sum_{d(1),\ldots, d(2{N_{\rm slice}} + 1) } \! \! 
\langle  \sigma^z |   e^{-{\mathcal H}_1 \frac{\delta_\tau} {2}}   | d(2N_{\rm slice} \!+\!1) \rangle
\langle  d(2N_{\rm slice} \!+\! 1) |  e^{-{\mathcal H}_2 \delta_\tau}  | d(2N_{\rm slice}) \rangle 
 \dots  \nonumber  \\ 
&& 
\phantom {  \sum_{d(1),\ldots, d(2{N_{\rm slice}} \!+\!1) } }
 \dots 
  \langle  d(4)  | 
 e^{-{\mathcal H}_1 \delta_\tau}
  | d (3) \rangle
 \langle d(3)  | 
 e^{-{\mathcal H}_2 \delta_\tau}
  | d (2 ) \rangle
  \langle d (2)  | 
 e^{-{\mathcal H}_1 \frac {\delta_\tau }{2}  } 
 | d(1) \rangle 
 \langle  d(1)  | \Psi_{0} \rangle. 
\label{eq:psi_tau_full_3}
\end{eqnarray}
The matrix element $\langle  d_l (t \!+ \!1) d_m (t \!+ \!1) |  e^{-{\mathcal H}_{lm} \delta_\tau}  | d_l(t) d_m(t) \rangle$ can be reproduced, for example, by 
the following interaction 
\begin{eqnarray}
   && e^{W''_1(  d_l (t + 1)  d_l (t) +  d_m (t + 1) d_m (t) )}
e^{W''_2   (  d_l (t + 1)  d_m (t) +  d_m (t + 1) d_l (t) ) }   \nonumber \\
 &&  \times \cos   \left (    \frac{\pi}{4}  \left ( d_l (t  \! +  \!1)  + d_m (t \! + \!1) -     d_l (t) - d_m (t)  \right ) \right )  
\cos   \left (   \frac{\pi}{8}  \left ( d_l (t \! +\! 1)  + d_m (t \! + \!1) -     d_l (t) - d_m (t)   \right ) \right ) 
\end{eqnarray}
with 
\begin{eqnarray}
W''_1  &=& - \frac{J_{lm}^z\delta_\tau}{2}   -  \frac{1}{4} \log \sinh(2J_{lm}^{xy} \delta_\tau ) , 
\label{Eq.wpp1}
\\ 
W''_2  &=&   -\frac{J_{lm}^z \delta_\tau}{2} - \frac{1}{4} \log \cosh(2J_{lm}^{xy} \delta_\tau ),
\label{Eq.wpp2} 
\end{eqnarray}
This interaction can be mediated by adding hidden neurons and mediating the interactions between $d (t + 1)$ and $d(t)$. 
Then, Eq.~(\ref{eq:psi_tau_full_2})  can be mapped onto the DBM representation. 

Indeed, the 2d-6h representation presented in this section correspond to this specific DBM construction: 
In the 2d-6h representation, two deep neurons are introduced for each bond at each imaginary time evolution. 
Because each imaginary time evolution acts on either even or odd bonds, the number of deep neurons introduced at one step is exactly same as the number of physical spins. 
In this case, the deep neurons can be considered as the spin degrees of freedom in the imaginary time layers $d(1),\ldots, d(2{N_{\rm slice}} \!+\!1)$. 
The interactions in Eqs.~(\ref{Eq.wpp1}) and (\ref{Eq.wpp2}) are equivalent to those in Eqs.~(\ref{Eq.Wpp_ll}) and (\ref{Eq.Wpp_lm}).
The ``$i\pi/4, i\pi/8$" trick appears to put constraint to conserve the total magnetization at each layer. 
Therefore, the 2d-6h representation is equivalent to the path-integral formulation.
Indeed, if we rearrange the neurons in this DBM construction (Fig.~\ref{fig_timeevolution_two_deep}), one can see a clear correspondence between the DBM network and the path-integral formulation. 
The extended systems including physical spins and deep neurons can be regard as the $D+1$ dimensional classical spin systems mapped from $D$ dimensional quantum systems.

\subsubsection{\emph{2 deep, 4 hidden} (2d-4h) representation}
\label{section-two-four-rep}

\noindent
{\bf Strategy.} 
We first extend DBM in the following way:
\begin{eqnarray}
\Psi_{\bar{\mathcal{W}}}(\sigma^z)
& = & \sum_{ \{h ,  d \} } \sum_{d_{[l]} } P_{1}(\sigma^z,h)P_{2}(h,d)
e^{\sum_{j,n=l,m}\sigma^z_{n}h_{j}\Delta W_{nj} + \sum_{j}h_{j}d_{[l]}W_{j[l]}^{\prime} 
  + \sum_{n=l,m}\sigma^z_nd_{[l]}W^{''}_{n[l]}+\sum_{j}\sigma^z_l \sigma^z_mh_jd_{[l]}Z_{lmj} }. 
 \label{DBM_extended} \nonumber \\
\end{eqnarray}
Here, we have introduced terms which break the standard DBM form, in particular the terms proportional to $W^{''}_{n[l]}$ and $Z_{lmj}$ with $n=l,m$. Those are essential for this construction, and their reduction to the pure DBM will be shown later.  
Also notice that the sum over $j$ runs through all the hidden neuron sites coupled to $\sigma_l^z$ and $\sigma_m^z$,
 thus it incorporates nonlocal couplings between hidden variables ($h$), physical ($\sigma^z$) and deep ($d$) variables.
The term proportional to $a_i$ in $P_1(\sigma^z,h)$ is a local site-dependent magnetic-field term in the DBM acting on the physical variables $\sigma^z$, which can also flexibly represent any local gauge transformation, if $a_i$ is taken complex. Here we fix $a_{i}$ to be site-dependent constants, which stay unchanged through the imaginary time evolution.
We later use the fact that the gauge transformation $\sigma^x \rightarrow -\sigma^x$ and $\sigma^y \rightarrow -\sigma^y$ on one of the sublattices (or $J^{xy}\rightarrow -J^{xy}$) on a bipartite lattice as in Eq. (\ref{gauge}) is equivalent to $a_i= i\pi/2$ if $i$ is on this sublattice and $a_i=0$ on the other sublattice as a special choice of $a_i$.

In the imaginary time evolution of ${\mathcal H}_{lm}$, we update 
$\bar{W}_{nj}$ ($n=l,m$) 
with the increment $\Delta W_{nj}$, in such a way that $\bar{W}_{nj}=W_{nj}+\Delta W_{nj}$. In addition to the deep variable $d_{[l]}$, we further introduce one additional deep variable $d_{[lm]}$ to recover the standard DBM by transforming the term proportional to $W^{''}$ and $Z$, with supplementary four hidden variables. 
\\

\noindent
{\bf Derivation for the update of parameters.} 
For $\sigma^z_l\sigma^z_m=-1$, the imaginary time evolution of the bond ${\mathcal H}_{lm}$ is given as
\begin{eqnarray}
\langle\sigma^z | e^{-\delta_{\tau}(J_{lm}^z\sigma_l^z\sigma_m^z+2J_{lm}^{xy}(\sigma_{l}^{+}\sigma_{m}^{-}+\sigma_{l}^{-}\sigma_{m}^{+}))}|\Psi_{\mathcal{W}}\rangle & = & \Psi_{\mathcal{W}}(\sigma^z) e^{J_{lm}^z\delta_{\tau}} \cosh(2J_{lm}^{xy}\delta_{\tau}) \nonumber \\
&-&\Psi_{\mathcal{W}}(\sigma_{1}^{z},\dots-\sigma_{l}^{z},\cdots -\sigma_{m}^{z}\dots) e^{J_{lm}^z\delta_{\tau}}\sinh(2J_{lm}^{xy}\delta_{\tau})\\
 & = & C'\langle\sigma^z |\Psi_{\bar{\mathcal{W}}}\rangle,
\end{eqnarray}
which is equivalent to 
\begin{eqnarray}
\sum_{ \{ h,d \} }\Psi_{\mathcal{W}}\left[1-\tanh(2J_{lm}^{xy}\delta_{\tau})e^{-2\sum_{n=l,m} (\sigma_{n}^{z}\sum_{j}h_{j}W_{nj}+a_n\sigma_n^z ) }\right] & = & C\Psi_{\bar{\mathcal{W}}} \label{eq:Psidbmsigmax_offD_Heis}
\end{eqnarray}
and $C=(e^{-J_{lm}^z\delta_{\tau}}/\cosh( 2J_{lm}^{xy}\delta_{\tau}) )C'$.
Notice that, here, we keep the bias term $a_n$ in Eq. (\ref{eq:PsiDeep}) instead of 
applying the gauge transformation in Eq. (\ref{gauge}). 

For  $\sigma^z_l\sigma^z_m=1$, we obtain
\begin{eqnarray}
\sum_{ \{ h,d \} }\Psi_{\mathcal{W}}e^{-2J_{lm}^z\delta_{\tau}} /\cosh ( 2J_{lm}^{xy}\delta_{\tau} )& = & C\Psi_{\bar{\mathcal{W}}}.\label{eq:Psidbmsigmax_offD_Heis2}
\end{eqnarray}

To make these imaginary time evolutions exact, 
$W_{nj}$ $(n=l,m)$
is updated to $\bar{W}_{nj}$ with the increment $\Delta W_{nj}$ as $\bar{W}_{nj}=W_{nj}+\Delta W_{nj}$ 
with
\begin{eqnarray}
\Delta W_{lj}&=&-\Delta W_{mj}= -\frac{1}{2}(W_{lj}-W_{mj}).
\label{bWZ3}
\end{eqnarray}
The new couplings $W_{j[l]}^{\prime}$, $Z_{lmj}$ and $W''_{n[l]}$ are also
given by
\begin{eqnarray}
W_{j[l]}^{\prime}  = -Z_{lmj}= -\frac{1}{2}(W_{lj}-W_{mj})
\label{bWZ4}
\end{eqnarray}
and from 
\begin{eqnarray}
2 ( W_{l[l]}^{\prime\prime}  -   W_{m[l]}^{\prime\prime}  )
& = & \log[-e^{
-2a_{l-m}}\tanh (2J_{lm}^{xy}\delta_{\tau})]
\label{Wlx}
\end{eqnarray}
and
\begin{eqnarray}
2\cosh(W^{''}_{l[l]} + W^{''}_{m[l]}) =\frac{e^{-2J_{lm}^z\delta_{\tau}-W^{''}_{l-m}}}{\cosh(2J_{lm}^{xy}\delta_{\tau})},
\label{bWZ1}
\end{eqnarray}
we obtain
\begin{eqnarray}
W''_{l[l]}&=&\frac{1}{4} \left [\log \left [-e^{-2a_{l-m}}\tanh (2J_{lm}^{xy}\delta_{\tau}) \right ]
+2{\rm arcosh} \left [\frac{e^{-2J_{lm}^z\delta_{\tau}}}{\sqrt{-2e^{-2a_{l-m}}\sinh (4J_{lm}^{xy}\delta_{\tau})}} \right] \right ] \label{W''l}\\
W''_{m[l]}&=&\frac{1}{4} \left [-\log \left [-e^{
-2a_{l-m}}\tanh (2J_{lm}^{xy}\delta_{\tau}) \right ]
+2{\rm arcosh} \left [\frac{e^{-2J_{lm}^z\delta_{\tau}}}{\sqrt{-2e^{-2a_{l-m}}\sinh (4J_{lm}^{xy}\delta_{\tau})}} \right] \right ]
\label{W''m}
\end{eqnarray}
with $a_{l-m} = a_l -a_m$.
On a bipartite lattice, to avoid the negative sign (or complex phase) problem  we need to keep  
$W''_{l[l]}$ and $W''_{m[l]}$ real.

This can be achieved by choosing $a_{l}=0$ for any $l$ if $J_{lm}<0$ (ferromagnetic case). For $J_{lm}>0$ (antiferromagnetic case), $a_l =n\pi i$ with an arbitrary integer $n$ if the site $l$ belongs to the sublattice A and $a_l =(n+1/2)\pi i$ if $l$ belongs to the sublattice B.
This local gauge for $J_{lm}>0$ is equivalent to take $J_{lm}^{xy}\rightarrow -J_{lm}^{xy}$ and 
$a_{l}=0$ 
for any site $l$ as is formulated in Eq.(\ref{gauge}).
We further note that $W''_{m[l]}$ can be taken positive if we take sufficiently small $\delta_\tau$ in Eq. (\ref{W''m}), with the leading order term $-\log (2J^{xy}_{lm} \delta_\tau )/2$. On the other hand, 
in Eq.~(\ref{W''l}), the leading order term is negative ($=-J_{lm} \delta_{\tau}$).
\\

\noindent
{\bf Recovery of the standard DBM form.} 
To recover the original form of the DBM, we first use Eq. (\ref{W''1})
with the replacement $s_1 \rightarrow \sigma^z_n$, $s_2 \rightarrow  d_{[l]}$, 
$s_3 \rightarrow  h_{[n]}$, 
$C \rightarrow  D_n$,
$V \rightarrow  W''_{n[l]}$
$\tilde{V}_1 \rightarrow  W_{n[n]}$ and $\tilde{V}_2 \rightarrow  W'_{[n][l]}$
for $n=l,m$. 
We have added here two hidden variables $h_{[l]}$ and $h_{[m]}$.
Then a solution for $D_n$,
$W_{n[n]}$, and $W'_{[n][l]}$ are represented by using $W^{''}_{n[l]}$ as 
\begin{eqnarray}
D_n &=& \frac{1}{2}\exp[-W^{''}_{n[l]}] \\
W_{n[n]} 
& = & W'_{[n][l]}=\frac{1}{2}{\rm arcosh}(\exp[2W^{''}_{n[l]}]), \label{Wl52}
\end{eqnarray}
if $W^{''}_{n[l]}$ is positive (as in the case of $W^{''}_{m[l]}$ for small $\delta_{\tau}$), which gives real $W_{n[n]}$ and $W'_{[n][l]}$. On the other hand, if  $W^{''}_{n[l]}$ is negative (as in the case of $W^{''}_{l[l]}$ for small $\delta_{\tau}$),
we should take 
\begin{eqnarray}
D_n &=& \frac{1}{2}\exp[W^{''}_{n[l]}] \\
W_{n[n]} 
& = & -W'_{[n][l]}=\frac{1}{2}{\rm arcosh}(\exp[-2W^{''}_{n[l]}]), \label{Wl53}
\end{eqnarray}
to give real $W_{n[n]}$ and $W_{[n][l]}$.

To completely recover the original DBM form, we next use Eq. (\ref{four_int_gadget}) by replacing
$\sigma_1$ with $\sigma^z_l$, $\sigma_2$ with $\sigma^z_m$, $d_1$ with $d_{[l]}$, $d_2$ with $d_{[lm]}$, $h_1$ with $h_j$, $h_2$ with $h_{[lm1]}$, $h_3$ with $h_{[lm2]}$, and $V$ with $Z_{lmj}$.

With these solutions, 
by ignoring the trivial constant factors including $D_l$ and $D_m$, 
the evolution is described by introducing two deep and four hidden additional variables $d_{[l]}$, $d_{[lm]}$, $h_{[l]}$, $h_{[m]}$, $h_{[lm1]}$, and $h_{[lm2]}$ as
\begin{eqnarray}
\Psi_{\bar{\mathcal{W}}}(\sigma^z)
& = & \sum_{ \{ \bar{h} ,  \bar{d} \} } P_{1}(\sigma^z,h)P_{2}(h,d) 
\exp \Bigl [
\sum_{j,n=l,m}\sigma^z_{n}h_{j}\Delta W_{nj}
+\sum_{j}h_{j}d_{[l]}W_{j[l]}^{\prime}
\nonumber \\ 
&+&\sum_{n=l,m}h_{[n]}(\sigma^z_nW_{n[n]}+d_{[l]}W^{'}_{[n][l]}) +d_{[lm]}\sum_{j}h_jZ_{lmj} 
\nonumber \\
&+&\frac{i\pi}{4}(h_{[lm1]}+h_{[lm2]})(\sigma^z_l+\sigma^z_m+d_{[l]}+d_{[lm]})  \Bigr ],
 \label{DBM_summary}
\end{eqnarray}
where $\{ \bar{h} , \bar{d} \}$ is a set consisting of the existing and new neurons.  
Equation (\ref{DBM_summary}) recovers the standard form of deep Boltzmann machine, where the physical spins $\sigma^z$ as well as the deep variables $d$ are not interacting each other and couples only to the hidden variables $h$. 
\\

\begin{figure}[tb]
\vspace{0cm}
\begin{center}
\includegraphics[width=0.5\columnwidth]{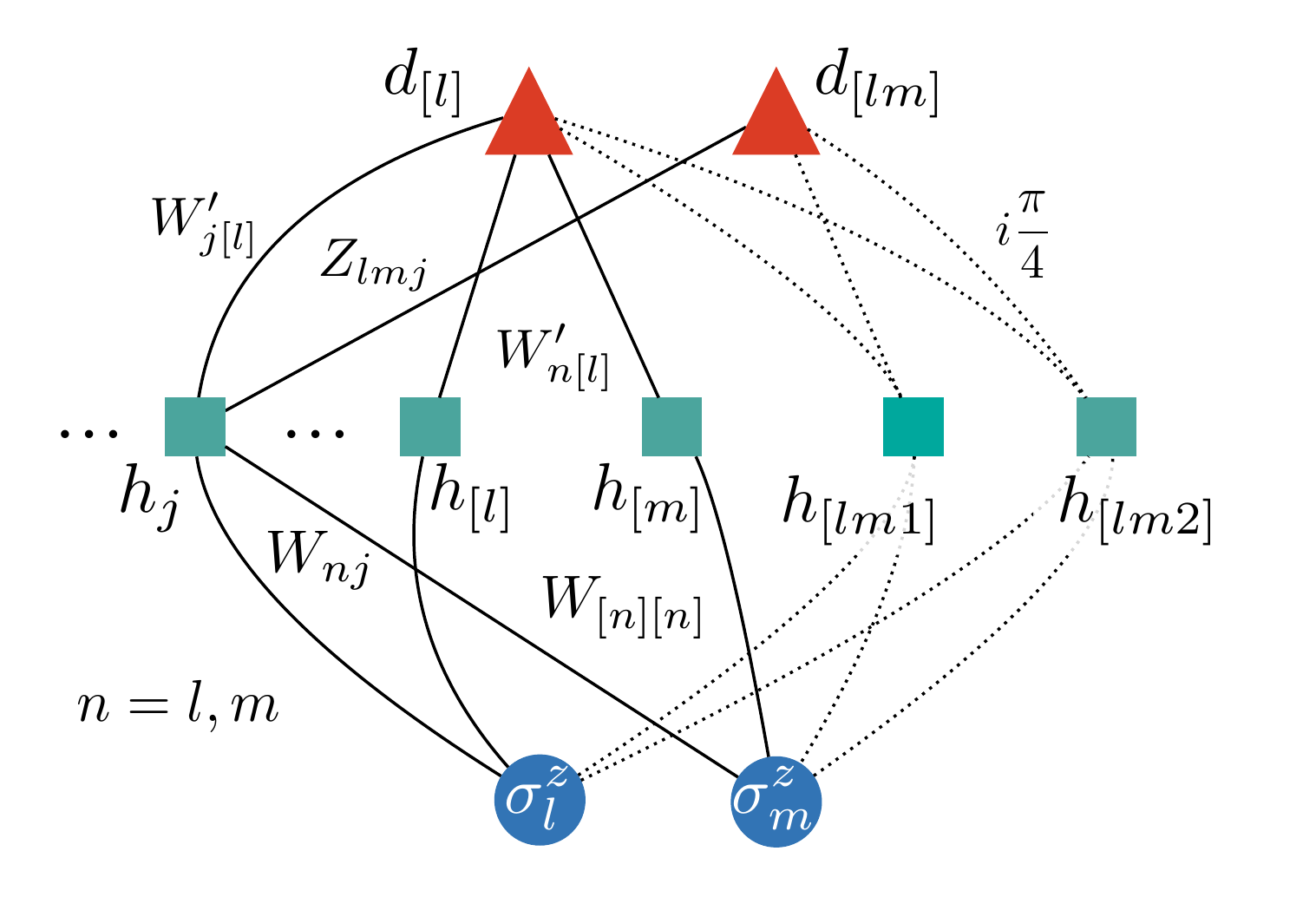}
\caption{
 {\bf Schematic picture for 2d-4h DBM network}. 
Dots, squares, triangles represent physical ($\sigma_i^z$), hidden ($h_j$), deep ($d_k$) variables. 
In 2d-4h construction, both $W$ and $W'$ couplings become nonlocal.
}
\label{fig_evolution_twodeep_fourhidden}
\end{center}
\end{figure}

\noindent
{\bf Summary.}
After summing over $\{ \bar{h} \}$,
we reach
\begin{eqnarray}
\Psi_{\bar{\mathcal{W}}}(\sigma)
& = &
 \sum_{ \{ \bar{d} \} } 
\exp  [ \sum_{n=l,m}a_{n}\sigma^z_{n}]\prod_{j}
\bigl [2\cosh[\sum_{i} \sigma^z_iW_{ij}+\sum_{k}W^{'}_{jk}d_{k}
+d_{[l]}W_{j[l]}^{\prime}+d_{[lm]}Z_{lmj} ] \bigr ]
\nonumber \\ 
&\times&
\prod_{n=l,m} \bigl ( 2\cosh [\sigma^z_{n}W_{n[n]}+d_{[l]}W^{'}_{[n][l]}]  \bigr ) 
\bigl [2 \cos[\frac{\pi}{4}(\sigma^z_l+\sigma^z_m+d_{[lm]}+d_{[lm]})] \bigr ]^2
 \label{DBM_summary2}
\end{eqnarray}
where the parameters $W, W'$ and $Z$ are given 
in Eqs. (\ref{bWZ3}), (\ref{bWZ4}), (\ref{W''l}), (\ref{W''m}), and (\ref{Wl52}) (or (\ref{Wl53})). 

We have introduced 2 deep and 4 hidden variables. Among them, $h_{[lm1]}$ and $h_{[lm2]}$ are simply to relate  $d_{[l]}$ and $d_{[lm]}$ to $\sigma^z_l$ and $\sigma^z_m$. 
With this trick, one can constrain $d_{[l]}=d_{[lm]}$ for $\sigma^z_{l}=\sigma^z_{m}$ and $d_{[l]}=-d_{[lm]}$ for $\sigma^z_{l}=-\sigma^z_{m}$.
After repeatedly operating Eq. (\ref{DBM_summary}), for all the combinations of $l,m$, 
the DBM structure becomes nonlocal as we see in Fig.~\ref{fig_evolution_twodeep_fourhidden}.
After the sufficiently long imaginary-time evolution,  with the analytical sum on $\{ h \}$ and the Monte sampling over $\{ d \}$, 
 one can obtain the ground state wave function. 

\newpage
\section{Sampling}

Once we have determined specific rules to obtain the parameters of the
DBM, the remaining question to be addressed is how to compute expectation
values of physical quantities. Consider a quantum operator $\mathcal{O}$,
then its expectation value over the DBM is given by the expression
\begin{eqnarray}
\langle\mathcal{O}\rangle & = & \frac{\sum_{ \{ \sigma^z,h,h^{\prime}d,d^{\prime} \} }\Pi(\sigma^z,h,h^{\prime},d,d^{\prime})O_{\mathrm{loc}}(\sigma^z,h,h^{\prime})}{\sum_{\{ \sigma^z,h,h^{\prime}d,d^{\prime} \} }\Pi(\sigma^z,h,h^{\prime},d,d^{\prime})},\label{eq:expdbm}
\end{eqnarray}
where we have introduced the pseudo-probability density $\Pi(\sigma^z,h,h^{\prime},d,d^{\prime})\equiv P_{1}(\sigma^z,h)P_{2}(h,d)P_{1}^{\star}(\sigma^z,h^{\prime})P_{2}^{\star}(h^{\prime},d^{\prime})$,
and the ``local'' estimator $O_{\mathrm{loc}}(\sigma^z,h,h^{\prime})=\frac{1}{2}\sum_{\{\sigma^{\prime z} \}}\left\langle \sigma^z\right|\mathcal{O}\left|\sigma^{\prime z}\right\rangle \left(\frac{P_{1}(\sigma^{\prime z},h)}{P_{1}(\sigma^z ,h)}+\frac{P_{1}(\sigma^{\prime z},h^{\prime})^{\star}}{P_{1}(\sigma^z ,h^{\prime})^{\star}}\right)$.
For a large number of spins and hidden/deep units, it is not possible
to compute those sums numerically, because of the exponential number
of terms involved. However, there are specific cases in which efficient
sampling strategies can be devised, allowing to stochastically compute
the quantum expectation values. In general, when the DBM weights are
all real $\Pi(\sigma^z,h,h^{\prime},d,d^{\prime})\geq0$, and it can
be interpreted as an (unnormalized) probability density. Thus, Markov-chain
sampling techniques can be applied, similarly to the case of applications
in standard machine learning. In the case of complex-valued weights,
the straightforward probabilistic interpretation breaks down, and
a sign (phase) problem arises. However, there are specific cases in
which one can still recover a properly defined probability density,
and efficiently sample from it. In the following we describe two main
sampling methods based on Markov chain techniques. First, Gibbs sampling,
then Metropolis-Hastings sampling. In both cases we discuss when the
sign problem can be circumvented. 

\subsection{Gibbs sampling }

We start discussing a strategy which is the natural generalization
of what traditionally used in most applications of DBM in machine
learning. The approach is based on Gibbs sampling, a strategy which
amounts to generate samples using the exact conditional probabilities
for block of variables. In practice, we introduce three kind of moves,
which allow to generate a Markov chain of visible, hidden, and deep
variables distributed according to $\Pi(\sigma^z,h,h^{\prime},d,d^{\prime})$.

\subsubsection{Sampling visible spins}

The first kind of move consists in freezing all the hidden and deep
variables, and sampling the visible spins $\sigma^z$. Specifically,
we generate new visible spin configurations according to the conditional
probability: 
\begin{eqnarray*}
\Pi(\sigma^z  |  h,h^{\prime},d,d^{\prime}) & = & \frac{P_{1}(\sigma^z , h )P_{2}(h,d)P_{1}^{\star}(\sigma^z ,h^{\prime})P_{2}^{\star}(h^{\prime},d^{\prime})} 
{\sum_{\{  \tilde{\sigma}^z \} }P_{1}(\tilde {\sigma}^z,h)P_{2}(h,d)P_{1}^{\star}(\tilde{\sigma}^z,h^{\prime}) P_{2}^{\star}(h^{\prime},d^{\prime}) }\\
 & = & \frac{P_{1}(\sigma^z,h)P_{1}^{\star}(\sigma^z,h^{\prime})}{\sum_{\{ \tilde{\sigma}^z \} }P_{1}(\tilde{\sigma}^z,h)P_{1}^{\star}(\tilde{\sigma}^z,h^{\prime})}\\
 & = & \frac{\Pi_{i}^{N}\exp\left\{ \sigma_{i}^z \left[\sum_{j}\left(h_{j}W_{ij}+h_{j}^{\prime}W_{ij}^{\star}\right)+2a_{i}^{\mathrm{r}}\right]\right\} }{\Pi_{i}^{N}2\cosh\left(\sum_{j}\left(h_{j}W_{ij}+h_{j}^{\prime}W_{ij}^{\star}\right)+2a_{i}^{\mathrm{r}}\right)}.
\end{eqnarray*}
Here, $a_{i}^{\mathrm{r}}$ is a real part of $a_i$. 
A particularly appealing aspect of this transition probability is
that each visible spin can be treated independently from the others,
thus we can update in parallel all visible spins at once. The probability
of a given spin to be up for example is: 
\begin{eqnarray}
P(\sigma^z_{i}=1|h,h^{\prime},d,d^{\prime}) & = & \mathrm{Logistic} (2\lambda_{i}^{[\sigma^z]}), \label{eq:Psigmacond}
\end{eqnarray}
with $\lambda_{i}^{[\sigma^z]}=\sum_{j}\left(h_{j}W_{ij}+h_{j}^{\prime}W_{ij}^{\star}\right)+2a_{i}^{\mathrm{r}}$,
and $\mathrm{Logistic}(x)=\frac{1}{1+\exp(-x)}$. Thus, during this
phase we generate $N$ random numbers $\eta_{i}$ uniformly distributed
in $[0,1)$, and set the spin $\sigma_{i}^{z}=1$ if $\eta_{i}<\mathrm{Logistic}(2\lambda_{i}^{[\sigma^z]})$.
For this approach to be feasible, we must have that the $\lambda_{i}^{[\sigma^z]}$
are real. 
Necessary conditions for this condition to be satisfied
are discussed at the end of this section.

\subsubsection{Sampling hidden spins}

The second type of move consists in freezing visible and deep spins,
and sampling hidden variables $h$ and $h^{\prime}$. For example,
to sample $h$ the transition probability reads: 
\begin{eqnarray*}
\Pi(h| \sigma^z,h^{\prime},d,d^{\prime}) & = & \frac{P_{1}(\sigma^z ,h)P_{2}(h,d)}{\sum_{\{ \tilde{h} \}  }P_{1}(\sigma^z,\tilde{h})P_{2}(\tilde{h},d)}\\
 & = & \frac{\Pi_{j}^{M}\exp\left[h_{j}\left(\sum_{i}\sigma^z_{i}W_{ij}+b_{j}+\sum_{k}d_{k}W_{jk}^{\prime}\right)\right]}{\Pi_{j}^{M}2\cosh\left(\sum_{i}\sigma^z_{i}W_{ij}+b_{j}+\sum_{k}d_{k}W_{jk}^{\prime}\right)}.
\end{eqnarray*}
The probability of having $h_{j}=1$ is then: 
\begin{eqnarray}
P(h_{j}=1|\sigma^z,h^{\prime},d,d^{\prime}) & = & \mathrm{Logistic}(2\lambda_{j}^{[h]}),\label{eq:phiddencond}
\end{eqnarray}
with $\lambda_{j}^{[h]}=\sum_{i}\sigma^z_{i}W_{ij}+b_{j}+\sum_{k}d_{k}W_{jk}^{\prime}$.
Again, one can therefore efficiently update all the $M$ hidden spins
at once, without rejection. Analogously, for $h^{\prime}$ we have
$\lambda_{j}^{[h^{\prime}]}=\sum_{i}\sigma^z_{i}W_{ij}^{\star}+b_{j}^{\star}+\sum_{k}d_{k}^{\prime}W_{jk}^{\prime^{\star}}$. 

\subsubsection{Sampling deep spins}

The final set of moves consists in freezing visible and hidden spins,
and sample from deep variables $d$ and $d^{\prime}$. For example,
to sample $d$ the transition probability is: 
\begin{eqnarray*}
\Pi(d|\sigma^z,h,h^{\prime},d^{\prime}) & = & \frac{P_{2}(h,d)}{\sum_{\{  \tilde{d} \} }P_{2}(h,\tilde{d})}\\
 & = & \frac{\Pi_{k}^{M^{\prime}}\exp\left[d_{k}\left(\sum_{j}h_{j}W_{jk}^{\prime}+c_{k}\right)\right]}{\Pi_{k}^{M^{\prime}}2\cosh\left(\sum_{j}h_{j}W_{jk}^{\prime}+c_{k}\right)}.
\end{eqnarray*}
The probability of having $d_{k}=1$ is then: 
\begin{eqnarray}
P(d_{k}=1|\sigma^z,h,h^{\prime},d^{\prime}) & = & \mathrm{Logistic}(2\lambda_{k}^{[d]}),\label{eq:Pdeepcond}
\end{eqnarray}
with $\lambda_{k}^{[d]}=\sum_{j}h_{j}W_{jk}^{\prime}+c_{k}$. Analogously,
we have $\lambda_{k}^{[d^{\prime}]}=\sum_{j}h_{j}^{\prime}W_{jk}^{\prime\star}+c_{k}^{\star}$. 

\subsubsection{Overall scheme: alternate block sampling }

The overall sampling scheme is therefore realized putting together
all those individual Gibbs samplings. In particular, we can devise
a two-step block sampling, which takes into account the conditional
dependence of all the probabilities previously derived. 

The overall sampling scheme then works as follow: 
\begin{enumerate}
\item Sample $h$ and $h^{\prime}$, fixing all the other variables. This
is realized using the probabilities (\ref{eq:phiddencond}) for all
the hidden spins.
\item Sample $\sigma^z$,$d$,$d^{\prime}$ fixing the values of $h$ and
$h^{\prime}$.This is realized using the probabilities (\ref{eq:Psidbmsigmax})
and (\ref{eq:Pdeepcond}) for all the visible and deep spins, respectively.
\item Cycle between 1 and 2.
\end{enumerate}

\subsubsection{Phase problem in the Gibbs scheme}

In order to get a consistent sampling scheme, we must have that all
the quantities $\lambda_{i}^{[\sigma^z]}$,$\lambda_{j}^{[h]}$,$\lambda_{j}^{[h^{\prime}]}$,$\lambda_{k}^{[d]},\lambda_{k}^{[d^{\prime}]}$
are real valued. In the absence of this condition, we have a phase
problem, and we cannot directly use a stochastic approach to sample
from the DBM. Looking more closely at what conditions are needed,
we start noticing that the visible bias can take arbitrary (complex)
values, since only the real parts, $a_{i}^{\mathrm{r}}$, enter $\lambda_{i}^{[\sigma^z]}$.
In general, there might be specific choices of the DBM parameters which still
guarantee absence of phase problem. One possibility is realized, for
example, when fixing the total magnetizations in the three layers,
i.e. the constraints $\sum_{i}\sigma^z_{i}=\sigma^z_{\mathrm{tot}}$,
 $\sum_{j}h_{j}=h_{\mathrm{tot}}$, $\sum_{k}d_{k}=d_{\mathrm{tot}}$.
We further assume that $\mathrm{Im}(W_{ij})=W^{\mathrm{I}}$, a constant,
as well as $\mathrm{Im}(W_{jk}^{\prime})=W^{^{\prime}\mathrm{I}}$.
Then, it is easy to see that the phase problem is avoided when $b_{j}^{\mathrm{I}}=-\sigma^z_{\mathrm{tot}}W^{\mathrm{I}}-d_{\mathrm{tot}}W^{^{\prime}\mathrm{I}}$
and $c_{k}^{\mathrm{I}}=-h_{\mathrm{tot}}W^{\prime\mathrm{I}}.$ Notice
that those are just a specific set of conditions, and less stringent
ones can be found using other sampling schemes. 

When each sample has the imaginary part or negative signs, another possibility of avoiding the phase problem is to take the partial trace summation explicitly so that such partial sum gives always a real nonnegative value.  
We will discuss this point in more detail in the next section.

\subsection{Metropolis sampling}
\label{sec_Metropolis_sampling}

\subsubsection{Marginal probability density}
Because there are no intralayer interactions in the DBM architecture, one can analytically trace out either one of $h, h'$ and $d,d'$. 
Then we get marginal probability density: 
$\tilde{\Pi}(\sigma^{z},h,h^{\prime})=\sum_{ \{ d,d^{\prime} \} }\Pi(\sigma^{z},h,h^{\prime},d,d^{\prime})$
or $\tilde{\Pi}'(\sigma^{z},d,d^{\prime})=\sum_{ \{ h,h^{\prime} \} }\Pi(\sigma^{z},h,h^{\prime},d,d^{\prime})$. 
Defining $\tilde{P} ( \sigma^z, h )$ and $\tilde{P}' (\sigma^z , d)$ as 
\begin{eqnarray}
\tilde{P} ( \sigma^z, h )& =&  \sum_{  \{ d  \}  } P_1(\sigma^z , h) P_2(h ,d ) 
    = 
     e^{\sum_i   a_i \sigma_i^z + \sum_{ij} \sigma_i^z  h_j W_{ij}  +\sum_j  b_j h_j } 
    \times 
     \prod_{k}   2 \cosh \Bigl (c_k  + \sum_k h_j W'_{jk}  \Bigr )   
\end{eqnarray}
and 
\begin{eqnarray}
\tilde{P} '( \sigma^z, d )& =&  \sum_{ \{ h  \} } P_1(\sigma^z , h) P_2(h ,d ) 
    = 
    \prod_{j}   2 \cosh \Bigl (b_j  + \sum_i \sigma^z_i W_{ij} + \sum_k d_k W'_{jk}  \Bigr )   
   \times    e^{\sum_i a_i \sigma_i^z + \sum_k c_k d_k },  
\end{eqnarray}
respectively, the marginal probability densities are given by 
\begin{eqnarray}
\tilde{\Pi}(\sigma^{z},h,h^{\prime}) &=&  \sum_{ \{ d,d^{\prime} \}    }\Pi(\sigma^{z},h,h^{\prime},d,d^{\prime}) =  
 \tilde{P} ( \sigma^z, h ) \tilde{P}^{\star}  ( \sigma^z, h' ),   \\
\tilde{\Pi}'(\sigma^{z},d,d^{\prime}) &=&  \sum_{ \{ h,h^{\prime} \} }  \Pi(\sigma^{z},h,h^{\prime},d,d^{\prime}) =  
  \tilde{P}^{\prime} ( \sigma^z, d ) \tilde{P}^{\prime \star}  ( \sigma^z, d' ).
\end {eqnarray}

With these marginal probability densities, we perform the Metropolis sampling to measure physical quantities. 
The expectation value of a quantum operator ${\mathcal O}$ is given by 
\begin{eqnarray}
\langle\mathcal{O}\rangle  =  \frac{\sum_{ \{ \sigma^z,h,h^{\prime} \}  } \tilde{\Pi} (\sigma^z,h,h^{\prime}) \tilde{O}_{\mathrm{loc}}(\sigma^z,h,h^{\prime})}{\sum_{\{ \sigma^z,h,h^{\prime} \} } \tilde{\Pi} (\sigma^z,h,h^{\prime})} 
=  \frac{\sum_{ \{ \sigma^z,d,d^{\prime} \} } \tilde{\Pi} '(\sigma^z,d,d^{\prime}) \tilde{O}'_{\mathrm{loc}}(\sigma^z,d,d^{\prime})}{\sum_{ \{ \sigma^z,d,d^{\prime} \} } \tilde{\Pi}' (\sigma^z,d,d^{\prime})} 
\label{eq:expdbm_metropolis}
\end{eqnarray}
with 
\begin{eqnarray}
\tilde{O}_{\mathrm{loc}}(\sigma^z,h,h^{\prime})&=&\frac{1}{2}\sum_{\{ \sigma^{\prime z} \} }\left\langle \sigma^z\right|\mathcal{O}\left|\sigma^{\prime z}\right\rangle
 \left(\frac{\tilde{P} (\sigma^{\prime z},h)}{ \tilde{P}(\sigma^z ,h)}+\frac{\tilde{P}(\sigma^{\prime z},h^{\prime})^{\star}}
 {\tilde{P} (\sigma^z ,h^{\prime})^{\star}} \right),  \\
\tilde{O}'_{\mathrm{loc}}(\sigma^z,d,d^{\prime}) &=& \frac{1}{2}\sum_{\{ \sigma^{\prime z} \} }\left\langle \sigma^z\right|\mathcal{O}\left|\sigma^{\prime z}\right\rangle
 \left(\frac{\tilde{P} '(\sigma^{\prime z},d)}{ \tilde{P}'(\sigma^z ,d)}+\frac{\tilde{P}'(\sigma^{\prime z},d^{\prime})^{\star}}
 {\tilde{P}' (\sigma^z ,d^{\prime})^{\star}} \right).
\end{eqnarray}

\subsubsection{Phase problem in the Metropolis scheme}

An advantage of choosing the marginal probability density is that 
by taking the summation over $h$ and $d$, the sign problem can sometimes be avoided
even if the DBM has complex parameters. 
An example is to take the summation over the hidden variables $h$ analytically 
in the three DBM constructions for the Heisenberg models presented in Sec. ~\ref{section_dbm_Heisenberg}. 
In all the three cases, only those $W$ and $W'$ couplings used to enforce the constraints are complex-valued, and the summation over $h$ eliminates the negative weight. 
For example, in the case of the 2d-4h representation in Sec.~\ref{section-two-four-rep},  
though each sample may have a finite imaginary part as in each term of Eq. (\ref{DBM_summary}), 
the total weight becomes real and nonnegative, after the explicit summation over the $h$ degrees of freedom is performed as in Eq. (\ref{DBM_summary2}).

When the lattice is not bipartite, we can still write down the DBM solutions to exactly follow the imaginary time evolutions. 
However, in this case, we will have imaginary $W$ and $W'$ parameters even for the units not involved in enforcing the constraints. 
In this case, the sampling may suffer from sign problem. 
However, as we discuss in the main text, in contrast to the conventional quantum Monte Carlo simulations, we can make the number of imaginary time step to reach the ground state short by starting the analytical DBM time evolution [Eq.~(\ref{eq:psitau})]
from a good stating point $| \Psi_0 \rangle$. 
For example, numerically optimized RBM wave functions can be used for $| \Psi_0 \rangle$, or more generally, 
$| \Psi_0 \rangle$ can be wave functions used in the conventional wave function techniques. 
In this case, before we suffer from a severe sign problems, we might be able to reach the ground state with good statistical accuracy.

\subsubsection{Overall scheme}
We sample over $\sigma^z$, $h$, $h'$ [or $\sigma^z$, $d$, $d'$] with the marginal probability density $\tilde{\Pi}(\sigma^{z},h,h^{\prime})$
[ $\tilde{\Pi}'(\sigma^{z},d,d^{\prime})$ ]. 
The physical quantities are measured following Eq.~(\ref{eq:expdbm_metropolis}).
In the case of Heisenberg model, after tracing out the $h$ spins, 
we have constraints over the values of $\sigma^{z}$, $d$, $d^{\prime}$.  
In that case, a cluster update rather than a local update will be more efficient. 
In particular, in the 2d-6h representation,
 since the imaginary-time evolution of the DBM is equivalent to the path-integral formalism, 
we can apply an efficient cluster update used in the conventional quantum Monte Carlo method, such as so called loop update~\cite{PhysRevLett.70.875}.


\begin{thebibliography}{10}
\expandafter\ifx\csname url\endcsname\relax
  \def\url#1{\texttt{#1}}\fi
\expandafter\ifx\csname urlprefix\endcsname\relax\def\urlprefix{URL }\fi
\providecommand{\bibinfo}[2]{#2}
\providecommand{\eprint}[2][]{\url{#2}}

\bibitem{feynman1948spacetime}
\bibinfo{author}{Feynman, R.~P.}
\newblock \bibinfo{title}{Space-{Time} {Approach} to {Non}-{Relativistic}
  {Quantum} {Mechanics}}.
\newblock \emph{\bibinfo{journal}{Reviews of Modern Physics}}
  \textbf{\bibinfo{volume}{20}}, \bibinfo{pages}{367--387}
  (\bibinfo{year}{1948}).
\newblock \urlprefix\url{https://link.aps.org/doi/10.1103/RevModPhys.20.367}.

\bibitem{dyson1949thes}
\bibinfo{author}{Dyson, F.~J.}
\newblock \bibinfo{title}{The {S} {Matrix} in {Quantum} {Electrodynamics}}.
\newblock \emph{\bibinfo{journal}{Physical Review}}
  \textbf{\bibinfo{volume}{75}}, \bibinfo{pages}{1736--1755}
  (\bibinfo{year}{1949}).
\newblock \urlprefix\url{https://link.aps.org/doi/10.1103/PhysRev.75.1736}.

\bibitem{hubbard1959calculation}
\bibinfo{author}{Hubbard, J.}
\newblock \bibinfo{title}{Calculation of {Partition} {Functions}}.
\newblock \emph{\bibinfo{journal}{Physical Review Letters}}
  \textbf{\bibinfo{volume}{3}}, \bibinfo{pages}{77--78} (\bibinfo{year}{1959}).
\newblock \urlprefix\url{https://link.aps.org/doi/10.1103/PhysRevLett.3.77}.

\bibitem{stratonovich1957ona}
\bibinfo{author}{Stratonovich, R.~L.}
\newblock \bibinfo{title}{On a {Method} of {Calculating} {Quantum}
  {Distribution} {Functions}}.
\newblock \emph{\bibinfo{journal}{Soviet Physics Doklady}}
  \textbf{\bibinfo{volume}{2}}, \bibinfo{pages}{416} (\bibinfo{year}{1957}).
\newblock \urlprefix\url{http://adsabs.harvard.edu/abs/1957SPhD....2..416S}.

\bibitem{abrikosov1975methods}
\bibinfo{author}{Abrikosov, A.~A.}
\newblock \emph{\bibinfo{title}{Methods of {Quantum} {Field} {Theory} in
  {Statistical} {Physics}}} (\bibinfo{publisher}{Dover Publications},
  \bibinfo{address}{New York}, \bibinfo{year}{1975}), \bibinfo{edition}{revised
  edition} edn.

\bibitem{binder}
\bibinfo{author}{Binder, K.}
\newblock \emph{\bibinfo{title}{Applications of the Monte Carlo Method in
  Statistical Physics}} (\bibinfo{publisher}{Springer Verlag},
  \bibinfo{address}{Berlin}, \bibinfo{year}{1984}).

\bibitem{takahashi1984}
\bibinfo{author}{Takahashi, M.} \& \bibinfo{author}{Imada, M.}
\newblock \bibinfo{title}{Monte carlo calculation of quantum systems}.
\newblock \emph{\bibinfo{journal}{J. Phys. Soc. Jpn.}}
  \textbf{\bibinfo{volume}{53}}, \bibinfo{pages}{963} (\bibinfo{year}{1984}).

\bibitem{takahashi1984-2}
\bibinfo{author}{Takahashi, M.} \& \bibinfo{author}{Imada, M.}
\newblock \bibinfo{title}{Monte carlo calculation of quantum systems. ii.
  higher order correction}.
\newblock \emph{\bibinfo{journal}{J. Phys. Soc. Jpn.}}
  \textbf{\bibinfo{volume}{53}}, \bibinfo{pages}{3765} (\bibinfo{year}{1984}).

\bibitem{ceperley1995pathintegrals}
\bibinfo{author}{Ceperley, D.}
\newblock \bibinfo{title}{Path-{Integrals} in the {Theory} of {Condensed}
  {Helium}}.
\newblock \emph{\bibinfo{journal}{Reviews of Modern Physics}}
  \textbf{\bibinfo{volume}{67}}, \bibinfo{pages}{279--355}
  (\bibinfo{year}{1995}).

\bibitem{suzuki}
\bibinfo{author}{Suzuki, M.}
\newblock \bibinfo{title}{Relationship between d-dimensional quantal spin
  systems and (d+1)-dimensional ising systems: Equivalence, critical exponents
  and systematic approximants of the partition function and spin correlations}.
\newblock \emph{\bibinfo{journal}{Prog. Theor. Phys.}}
  \textbf{\bibinfo{volume}{56}}, \bibinfo{pages}{1454} (\bibinfo{year}{1976}).

\bibitem{hirsch}
\bibinfo{author}{Hirsch, J.~E.}, \bibinfo{author}{Sugar, R.},
  \bibinfo{author}{Scalapino, D.} \& \bibinfo{author}{Blankenbecler, R.}
\newblock \bibinfo{title}{Monte carlo simulations of one-dimensional fermion
  systems}.
\newblock \emph{\bibinfo{journal}{Phys. Rev. B}} \textbf{\bibinfo{volume}{26}},
  \bibinfo{pages}{5033} (\bibinfo{year}{1982}).

\bibitem{wiese1996}
\bibinfo{author}{Beard, B.} \& \bibinfo{author}{Wiese, U.-J.}
\newblock \bibinfo{title}{Simulations of discrete quantum systems in continuous
  euclidean time}.
\newblock \emph{\bibinfo{journal}{J.Phys. Rev. Lett.}}
  \textbf{\bibinfo{volume}{77}}, \bibinfo{pages}{5130} (\bibinfo{year}{1996}).

\bibitem{sandvik1999stochastic}
\bibinfo{author}{Sandvik, A.~W.}
\newblock \bibinfo{title}{Stochastic series expansion method with operator-loop
  update}.
\newblock \emph{\bibinfo{journal}{Physical Review B}}
  \textbf{\bibinfo{volume}{59}}, \bibinfo{pages}{R14157--R14160}
  (\bibinfo{year}{1999}).
\newblock \urlprefix\url{http://arxiv.org/abs/cond-mat/9902226}.
\newblock \bibinfo{note}{ArXiv: cond-mat/9902226}.

\bibitem{prokofev2007bolddiagrammatic}
\bibinfo{author}{Prokof'ev, N.} \& \bibinfo{author}{Svistunov, B.}
\newblock \bibinfo{title}{Bold {Diagrammatic} {Monte} {Carlo}: {When} {Sign}
  {Problem} is {Welcome}}.
\newblock \emph{\bibinfo{journal}{Physical Review Letters}}
  \textbf{\bibinfo{volume}{99}} (\bibinfo{year}{2007}).
\newblock \urlprefix\url{http://arxiv.org/abs/cond-mat/0702555}.
\newblock \bibinfo{note}{ArXiv: cond-mat/0702555}.

\bibitem{feynman}
\bibinfo{author}{Feynman, R.~P.}
\newblock \bibinfo{title}{Atomic theory of the two-fluid model of liquid
  helium}.
\newblock \emph{\bibinfo{journal}{Physical Review}}
  \textbf{\bibinfo{volume}{94}}, \bibinfo{pages}{262} (\bibinfo{year}{1954}).

\bibitem{gros}
\bibinfo{author}{Gros, C.}
\newblock \bibinfo{title}{Physics of projected wavefunctions}.
\newblock \emph{\bibinfo{journal}{Ann. Phys.}} \textbf{\bibinfo{volume}{189}},
  \bibinfo{pages}{53} (\bibinfo{year}{1989}).

\bibitem{kashima2001}
\bibinfo{author}{Kashima, T.} \& \bibinfo{author}{Imada, M.}
\newblock \bibinfo{title}{Path-integral renormalization group method for
  numerical study on ground states of strongly correlated electronic systems}.
\newblock \emph{\bibinfo{journal}{J. Phys. Soc. Jpn.}}
  \textbf{\bibinfo{volume}{70}}, \bibinfo{pages}{2287} (\bibinfo{year}{2001}).

\bibitem{tahara}
\bibinfo{author}{Tahara, D.} \& \bibinfo{author}{Imada, M.}
\newblock \bibinfo{title}{Variational monte carlo method combined with
  quantum-number projection and multi-variable optimization}.
\newblock \emph{\bibinfo{journal}{J. Phys. Soc. Jpn.}}
  \textbf{\bibinfo{volume}{77}}, \bibinfo{pages}{114701}
  (\bibinfo{year}{2008}).

\bibitem{becca2017quantum}
\bibinfo{author}{Becca, F.} \& \bibinfo{author}{Sorella, S.}
\newblock \emph{\bibinfo{title}{Quantum {Monte} {Carlo} {Approaches} for
  {Correlated} {Systems}}} (\bibinfo{publisher}{Cambridge University Press},
  \bibinfo{address}{Cambridge, United Kingdom ; New York, NY},
  \bibinfo{year}{2017}).

\bibitem{white}
\bibinfo{author}{White, S.~R.}
\newblock \bibinfo{title}{Density-matrix algorithms for quantum renormalization
  groups}.
\newblock \emph{\bibinfo{journal}{Physical Review B}}
  \textbf{\bibinfo{volume}{48}}, \bibinfo{pages}{10345} (\bibinfo{year}{1993}).

\bibitem{orus}
\bibinfo{author}{Or\'{u}s, R.}
\newblock \bibinfo{title}{A practical introduction to tensor networks: Matrix
  product states and projected entangled pair states}.
\newblock \emph{\bibinfo{journal}{Ann. Phys.}} \textbf{\bibinfo{volume}{349}},
  \bibinfo{pages}{117} (\bibinfo{year}{2014}).

\bibitem{carleo2017solving}
\bibinfo{author}{Carleo, G.} \& \bibinfo{author}{Troyer, M.}
\newblock \bibinfo{title}{Solving the quantum many-body problem with artificial
  neural networks}.
\newblock \emph{\bibinfo{journal}{Science}} \textbf{\bibinfo{volume}{355}},
  \bibinfo{pages}{602--606} (\bibinfo{year}{2017}).
\newblock \urlprefix\url{http://science.sciencemag.org/content/355/6325/602}.

\bibitem{torlai2017manybody}
\bibinfo{author}{Torlai, G.} \emph{et~al.}
\newblock \bibinfo{title}{Many-body quantum state tomography with neural
  networks}.
\newblock \emph{\bibinfo{journal}{arXiv:1703.05334}}  (\bibinfo{year}{2017}).
\newblock \urlprefix\url{http://arxiv.org/abs/1703.05334}.
\newblock \bibinfo{note}{ArXiv: 1703.05334}.

\bibitem{nomura2017restrictedboltzmannmachine}
\bibinfo{author}{Nomura, Y.}, \bibinfo{author}{Darmawan, A.~S.},
  \bibinfo{author}{Yamaji, Y.} \& \bibinfo{author}{Imada, M.}
\newblock \bibinfo{title}{Restricted boltzmann machine learning for solving
  strongly correlated quantum systems}.
\newblock \emph{\bibinfo{journal}{Phys. Rev. B}} \textbf{\bibinfo{volume}{96}},
  \bibinfo{pages}{205152} (\bibinfo{year}{2017}).

\bibitem{deng2017quantum}
\bibinfo{author}{Deng, D.-L.}, \bibinfo{author}{Li, X.} \&
  \bibinfo{author}{Das~Sarma, S.}
\newblock \bibinfo{title}{Quantum {Entanglement} in {Neural} {Network}
  {States}}.
\newblock \emph{\bibinfo{journal}{Physical Review X}}
  \textbf{\bibinfo{volume}{7}}, \bibinfo{pages}{021021} (\bibinfo{year}{2017}).
\newblock \urlprefix\url{https://link.aps.org/doi/10.1103/PhysRevX.7.021021}.

\bibitem{rocchetto2017learning}
\bibinfo{author}{Rocchetto, A.}, \bibinfo{author}{Grant, E.},
  \bibinfo{author}{Strelchuk, S.}, \bibinfo{author}{Carleo, G.} \&
  \bibinfo{author}{Severini, S.}
\newblock \bibinfo{title}{Learning hard quantum distributions with variational
  autoencoders}.
\newblock \emph{\bibinfo{journal}{arXiv:1710.00725 [quant-ph, stat]}}
  (\bibinfo{year}{2017}).
\newblock \urlprefix\url{http://arxiv.org/abs/1710.00725}.
\newblock \bibinfo{note}{ArXiv: 1710.00725}.

\bibitem{glasser2017neuralnetworks}
\bibinfo{author}{Glasser, I.}, \bibinfo{author}{Pancotti, N.},
  \bibinfo{author}{August, M.}, \bibinfo{author}{Rodriguez, I.~D.} \&
  \bibinfo{author}{Cirac, J.~I.}
\newblock \bibinfo{title}{Neural {Networks} {Quantum} {States}, {String}-{Bond}
  {States} and chiral topological states}.
\newblock \emph{\bibinfo{journal}{arXiv:1710.04045 [cond-mat, physics:quant-ph,
  stat]}}  (\bibinfo{year}{2017}).
\newblock \urlprefix\url{http://arxiv.org/abs/1710.04045}.
\newblock \bibinfo{note}{ArXiv: 1710.04045}.

\bibitem{kaubruegger2017chiraltopological}
\bibinfo{author}{Kaubruegger, R.}, \bibinfo{author}{Pastori, L.} \&
  \bibinfo{author}{Budich, J.~C.}
\newblock \bibinfo{title}{Chiral {Topological} {Phases} from {Artificial}
  {Neural} {Networks}}.
\newblock \emph{\bibinfo{journal}{arXiv:1710.04713 [cond-mat,
  physics:quant-ph]}}  (\bibinfo{year}{2017}).
\newblock \urlprefix\url{http://arxiv.org/abs/1710.04713}.
\newblock \bibinfo{note}{ArXiv: 1710.04713}.

\bibitem{cai2017approximating}
\bibinfo{author}{Cai, Z.}
\newblock \bibinfo{title}{Approximating quantum many-body wave-functions using
  artificial neural networks}.
\newblock \emph{\bibinfo{journal}{arXiv:1704.05148 [cond-mat]}}
  (\bibinfo{year}{2017}).
\newblock \urlprefix\url{http://arxiv.org/abs/1704.05148}.
\newblock \bibinfo{note}{ArXiv: 1704.05148}.

\bibitem{saito2017machine}
\bibinfo{author}{Saito, H.} \& \bibinfo{author}{Kato, M.}
\newblock \bibinfo{title}{Machine {Learning} {Technique} to {Find} {Quantum}
  {Many}-{Body} {Ground} {States} of {Bosons} on a {Lattice}}.
\newblock \emph{\bibinfo{journal}{Journal of the Physical Society of Japan}}
  \textbf{\bibinfo{volume}{87}}, \bibinfo{pages}{014001}
  (\bibinfo{year}{2017}).
\newblock \urlprefix\url{http://journals.jps.jp/doi/10.7566/JPSJ.87.014001}.

\bibitem{saito2017solving}
\bibinfo{author}{Saito, H.}
\newblock \bibinfo{title}{Solving the {Bose}{\textendash}{Hubbard} {Model} with
  {Machine} {Learning}}.
\newblock \emph{\bibinfo{journal}{Journal of the Physical Society of Japan}}
  \textbf{\bibinfo{volume}{86}}, \bibinfo{pages}{093001}
  (\bibinfo{year}{2017}).
\newblock \urlprefix\url{http://journals.jps.jp/doi/10.7566/JPSJ.86.093001}.

\bibitem{chen2018equivalence}
\bibinfo{author}{Chen, J.}, \bibinfo{author}{Cheng, S.}, \bibinfo{author}{Xie,
  H.}, \bibinfo{author}{Wang, L.} \& \bibinfo{author}{Xiang, T.}
\newblock \bibinfo{title}{Equivalence of restricted {Boltzmann} machines and
  tensor network states}.
\newblock \emph{\bibinfo{journal}{Physical Review B}}
  \textbf{\bibinfo{volume}{97}}, \bibinfo{pages}{085104}
  (\bibinfo{year}{2018}).
\newblock \urlprefix\url{https://link.aps.org/doi/10.1103/PhysRevB.97.085104}.

\bibitem{clark2017unifying}
\bibinfo{author}{Clark, S.~R.}
\newblock \bibinfo{title}{Unifying {Neural}-network {Quantum} {States} and
  {Correlator} {Product} {States} via {Tensor} {Networks}}.
\newblock \emph{\bibinfo{journal}{arXiv:1710.03545 [cond-mat,
  physics:quant-ph]}}  (\bibinfo{year}{2017}).
\newblock \urlprefix\url{http://arxiv.org/abs/1710.03545}.
\newblock \bibinfo{note}{ArXiv: 1710.03545}.

\bibitem{deng2017machine}
\bibinfo{author}{Deng, D.-L.}, \bibinfo{author}{Li, X.} \&
  \bibinfo{author}{Das~Sarma, S.}
\newblock \bibinfo{title}{Machine learning topological states}.
\newblock \emph{\bibinfo{journal}{Physical Review B}}
  \textbf{\bibinfo{volume}{96}}, \bibinfo{pages}{195145}
  (\bibinfo{year}{2017}).
\newblock \urlprefix\url{https://link.aps.org/doi/10.1103/PhysRevB.96.195145}.

\bibitem{gao_efficient_2017}
\bibinfo{author}{Gao, X.} \& \bibinfo{author}{Duan, L.-M.}
\newblock \bibinfo{title}{Efficient representation of quantum many-body states
  with deep neural networks}.
\newblock \emph{\bibinfo{journal}{Nature Communications}}
  \textbf{\bibinfo{volume}{8}}, \bibinfo{pages}{662} (\bibinfo{year}{2017}).
\newblock \urlprefix\url{https://www.nature.com/articles/s41467-017-00705-2}.

\bibitem{huang2017neuralnetwork}
\bibinfo{author}{Huang, Y.} \& \bibinfo{author}{Moore, J.~E.}
\newblock \bibinfo{title}{Neural network representation of tensor network and
  chiral states}.
\newblock \emph{\bibinfo{journal}{arXiv:1701.06246}}  (\bibinfo{year}{2017}).
\newblock \urlprefix\url{http://arxiv.org/abs/1701.06246}.
\newblock \bibinfo{note}{ArXiv: 1701.06246}.

\bibitem{salakhutdinov2009deepboltzmann}
\bibinfo{author}{Salakhutdinov, R.} \& \bibinfo{author}{Hinton, G.}
\newblock \bibinfo{title}{Deep {Boltzmann} {Machines}}.
\newblock In \emph{\bibinfo{booktitle}{{PMLR}}}, \bibinfo{pages}{448--455}
  (\bibinfo{year}{2009}).
\newblock
  \urlprefix\url{http://proceedings.mlr.press/v5/salakhutdinov09a.html}.

\bibitem{trotter}
\bibinfo{author}{Trotter, H.~F.}
\newblock \bibinfo{title}{On the product of semi-groups of operators}.
\newblock \emph{\bibinfo{journal}{Proc. Amer. Math. Soc.}}
  \textbf{\bibinfo{volume}{10}}, \bibinfo{pages}{545} (\bibinfo{year}{1959}).

\bibitem{salakhutdinov2012anefficient}
\bibinfo{author}{Salakhutdinov, R.} \& \bibinfo{author}{Hinton, G.}
\newblock \bibinfo{title}{An efficient learning procedure for deep {Boltzmann}
  machines}.
\newblock \emph{\bibinfo{journal}{Neural Computation}}
  \textbf{\bibinfo{volume}{24}}, \bibinfo{pages}{1967--2006}
  (\bibinfo{year}{2012}).

\bibitem{PhysRevLett.70.875}
\bibinfo{author}{Evertz, H.~G.}, \bibinfo{author}{Lana, G.} \&
  \bibinfo{author}{Marcu, M.}
\newblock \bibinfo{title}{Cluster algorithm for vertex models}.
\newblock \emph{\bibinfo{journal}{Phys. Rev. Lett.}}
  \textbf{\bibinfo{volume}{70}}, \bibinfo{pages}{875--879}
  (\bibinfo{year}{1993}).

\bibitem{ceperley-alder}
\bibinfo{author}{Ceperley, D.~M.} \& \bibinfo{author}{Alder, J.}
\newblock \bibinfo{title}{Ground state of the electron gas by a stochastic
  method}.
\newblock \emph{\bibinfo{journal}{Phys. Rev. Lett.}}
  \textbf{\bibinfo{volume}{45}}, \bibinfo{pages}{566} (\bibinfo{year}{1980}).

\bibitem{doi:10.1143/PTP.58.1377}
\bibinfo{author}{Suzuki, M.}, \bibinfo{author}{Miyashita, S.} \&
  \bibinfo{author}{Kuroda, A.}
\newblock \bibinfo{title}{Monte carlo simulation of quantum spin systems. i}.
\newblock \emph{\bibinfo{journal}{Prog. Theor. Phys.}}
  \textbf{\bibinfo{volume}{58}}, \bibinfo{pages}{1377--1387}
  (\bibinfo{year}{1977}).

\end{thebibliography}
\end{document}